\documentclass[preprint]{aastex}

\def\kms{~km~s$^{-1}$\ }
\def\kmsc{~km~s$^{-1}$}

\def\arcs{\char'175\ }
\def\arcsec{\char'175 }

\def\etal{et~al.\ }

\def\hub{\ifmmode H_\circ\else H$_\circ$\fi}

\shorttitle{Star Formation Histories of Early-Type Galaxies}
\shortauthors{Caldwell, Rose \& Concannon}

\begin{document}

\title{Star Formation Histories of Early-Type Galaxies. 
I: Higher Order Balmer Lines as Age Indicators}
\author{Nelson Caldwell}
\affil{Smithsonian Astrophysical Observatory, 60 Garden Street, Cambridge, MA 02138}

\and

\author{James A. Rose and Kristi Dendy Concannon}
\affil{Department of Physics and Astronomy, CB \#3255, University of North Carolina,
       Chapel Hill, NC 27599}
\email{caldwell@cfa.harvard.edu, jim@physics.unc.edu, kdconcan@kings.edu}
      
\begin{abstract}

We have obtained blue integrated spectra of 175 nearby early-type
galaxies, covering a wide range in galaxy velocity
dispersion, and emphasizing those with $\sigma$$<$100 \kmsc.
Galaxies have been observed both in the Virgo cluster and in
lower-density environments.  The main goals are the evaluation
of higher order Balmer lines as age indicators, and differences
in stellar populations as a function of mass, environment and morphology.
In this first paper
our emphasis is on presenting the methods used to characterize the behavior of the
Balmer lines through evolutionary population synthesis models.
Lower-$\sigma$ galaxies exhibit a substantially
greater intrinsic scatter, in a variety of line strength indicators, than do
higher-$\sigma$ galaxies, with the large intrinsic scatter setting in below a
$\sigma$ of 100 \kmsc. Moreover, a greater contrast in scatter is present in
the Balmer lines than in the lines of metal features.
Evolutionary synthesis modeling of the observed spectral
indices indicates that the strong Balmer lines found primarily among the
low-$\sigma$ galaxies are caused by young age, rather than by low metallicity.
Thus we find a trend between the population age and the central
velocity dispersion, such that low $\sigma$ galaxies have younger
luminosity-weighted mean ages.  We have repeated this analysis
using several different Balmer lines, and find consistent results from one
spectral indicator to another.

\end{abstract}


\section{INTRODUCTION}

With the exception of galaxies within a few Mpc, efforts to
determine the star formation and
chemical enrichment histories of galaxies must be based on
observations of their integrated starlight, and of the 
modeling of that integrated spectrum.
Unfortunately, establishing reliable luminosity-weighted mean ages and
chemical compositions for early-type galaxies has been difficult to
achieve, due primarily to the degeneracy between age and metallicity on the
integrated spectra of galaxies (e.g., Worthey 1994), and, secondarily, to the
non-solar abundance ratios in most early-type galaxies
(e.g., O'Connell 1976, Worthey \etal 1992, Vazdekis \etal 1997, Trager \etal
2000a, Kuntschner \etal 2001).  Nevertheless,  recently 
improved population synthesis models and higher quality
of spectra of galaxies
have led to a surge of results in this field
(Trager \etal 2000a,b, Kuntschner 2000, Kuntschner \etal
2001, Vazdekis \etal 2001, Terlevich \& Forbes 2002, Proctor \& Sansom 2002,
Kuntschner \etal 2002).

The need for accurate information on galaxy ages and metallicities has been
underscored by rapid developments in cosmological simulations of
the evolution of structure in the universe.
Large-scale numerical simulations of structure formation within the context
of the $\Lambda$CDM (i.e., cold dark matter plus substantial contribution
from a cosmological constant) scenario, when coupled with parametrized
baryonic physics, now make clear predictions as to the formation histories of
early-type galaxies as a function of mass and environment
\citep{ka96,ks98,sp99,be01}.  Hence a comprehensive database of mean ages and
metallicities of early-type galaxies covering a range in mass and
environment can provide crucial constraints on the
hierarchical evolution scenario that is a key component of the
$\Lambda$CDM picture.

We have recently completed a large database of integrated spectra of 175
early-type galaxies  which serves two main objectives.  The first
goal is to observe early-type galaxies covering a large range in
mass, environment, and early-type morphology.  The hierarchical galaxy formation
models now make predictions for the star formation and chemical 
enrichment histories of galaxies as a function of their mass and environment.
In addition, there is  an important distinction to be made between elliptical
galaxies (those galaxies without apparent ongoing star formation and without a
stellar disk) and lenticulars (S0s: those galaxies without apparent ongoing star
formation, but which do have stellar disks).  Whether these two types of 
galaxies have had similar origins and star formation histories, and whether
there is a continuum of properties between them, is a subject of longstanding
debate.  A second goal is to investigate the blue spectral
region, which covers the higher order Balmer lines, as a promising spectral
region for obtaining better decoupling between age and metallicity effects.

There are two aspects in which our survey departs from previous work.
The first is an emphasis on low-mass galaxies, as characterized by their low
central velocity dispersion, $\sigma$.  While previous surveys have tended to
emphasize galaxies with $\sigma$$>$100 \kmsc, our survey contains a substantial
emphasis on galaxies with lower $\sigma$.  Second, we have obtained wavelength
coverage down to $\lambda$3600 \AA, at the relatively  high spectral
resolution of 3.1 \AA \ FWHM, which allows us to evaluate the usefulness
of higher order Balmer lines for age and metallicity determinations.

This paper is the first in a series that will evaluate the ages and chemical
compositions of early-type galaxies as a function of both mass and environment.
In this paper the primary emphasis is on establishing the
techniques that we use to extract luminosity-weighted mean ages and 
metallicities from the galaxy integrated spectra, particularly in regard to
the higher order Balmer lines.  We present an extensive comparison
between ages derived from different Balmer line measurements, to test whether
robust results are obtained from galaxy population synthesis.
The main scientific results presented here
relate to the dependence of the mean age and metallicity of early-type galaxies 
on the central velocity dispersion.  We reserve more extensive discussions of
the influence of environment and morphology for subsequent papers in the series.
A brief
description of the sample of galaxies and the observational data is given in
\S2, while a discussion of the spectral indices used in the study is given in
\S3.  The techniques used to model the integrated spectra of galaxies are
described in \S4.  In \S5 we discuss the dependence of galaxy age and
metallicity on velocity dispersion, and in \S6 we evaluate the results obtained
from different Balmer and metal features.

\section{Observational Data}

To fully investigate the star formation histories of early-type
galaxies, we have chosen our sample to cover a range in mass, morphological
classification, and environment.   Our data sample consists of 175 early-type
galaxies, 59 of which are in the Virgo cluster and 116 of which are in the 
field or other low-density environments.  The galaxies cover a central velocity dispersion
range of 50 $<\sigma<$ 320 km/s, with an emphasis on low-$\sigma$ galaxies,
and have been previously
classified as either E, S0, dS0, or dE as indicated by their ground-based 
morphology.  A more detailed discussion of the sample selection is given in
Paper II, for the Virgo galaxies, and in Paper III, for the field sample.

Long slit spectra of all galaxies in the sample have been obtained at the F.\ L.\ Whipple
Observatory (FLWO) with the 1.5m Tillinghast telescope, the FAST spectrograph and
a Loral 512 x 2688 pixel CCD~\citep{fa98} during twelve 
observing runs between 1997 September and 2002 May in seeing conditions of 
2\arcsec\  or better.  The instrumental setup was identical on all but two occasions.
The dispersing element is the 600 line/mm grating which produces a spectral
coverage of $\sim$2000\AA\  from 3500\AA\ to 5500\AA\ at 0.75\AA \ pixel$^{-1}$ dispersion. 
Using a 3\arcsec\ slit, four pixel resolution is typically achieved, hence the data have
a spectral resolution of 3.1\AA\ FWHM.  On two of the observing runs,
the above setup differed in that a 2\arcsec\ slit was used, resulting in a
spectral resolution of 2.14\AA\ FWHM.  These observations have been smoothed to the
same intrinsic resolution as the remainder of the data.  

Spectra were extracted using a variance weighting scheme which included data 
along the slit 
down to 10\% of the peak flux at a given wavelength.  Thus the resultant spectra
represent the global spectra and refer to an isophotal area, but with 
larger weight given to the brighter, central regions.  Discussion of radial variations
in the derived stellar populations will be taken up in Paper II.
The data were then fluxed using 
standard stars observed during the same nights. 

Radial velocities and velocity 
dispersions were determined for each galaxy, using the IRAF package FXCOR
and K0III template stars.
These are used to deredshift the spectra
and to broaden them to a common resolution. 
The velocities and errors are listed in Table~\ref{tab:log}, where the
errors are the internal errors produced by FXCOR, and averaged from the
multiple observations.  A comparison
with the literature values of Smith et al. (2000) for 25 galaxies in common
gives a mean difference of 13 \kms and rms of 58 \kms.
Our measured velocity dispersions 
are in good accord with the literature values compiled by 
McElroy (1995).  The dispersion errors contained in  Table~\ref{tab:log} are the
rms errors among repeated observations, but with a lower limit of 
7 \kms enforced.

For the low $\sigma$ galaxies ($\sigma$$<$80~\kms), our values tend to be
systematically higher by about 50\% than the recent values determined by Simien \& Prugniel
(2002), Pedraz \etal (2002), and Geha \etal (2002).  
The discrepancy is due  in general to our lower velocity 
resolution than the cited studies ($\sim 80$ \kms versus $\sim 25$ \kms),
and in certain specific cases because of our inclusion in the extracted
spectra of outer parts of galaxies that are rotating, leading to 
broadened absorption lines.  The cited works measured central broadenings,
whereas we measured global broadenings.  For most of our
purpose, the discrepancy does not matter, since we
smooth all spectra to a common velocity broadening of 230 \kms, and
instrumental/observational effects below this value are of no
concern. High resolution dispersions for all the smaller galaxies would
be desirable for the study of relations between dispersions and our derived
stellar population parameters, but are not yet available.
Table~\ref{tab:log} gives the log of the galaxy observations as well
as basic derived parameters.

Although the galaxies were chosen to be free of emission contamination as based 
on the low S/N spectra of the CfA~I redshift survey (Huchra \etal 1983), 
a significant fraction of the
absorption line spectra were found to suffer from the effects of emission in the
Balmer lines. The contamination is not completely unexpected since previous surveys
have shown that 40-50\% of E/S0 galaxies have at least small
amounts of ionized gas~\citep{cald84,ph86}, but we do need to correct for this
emission.  To determine the quantitative impact of the fill-in on the spectral 
line strengths, we obtained spectra in the red, centered near H$\alpha$,
for 40\% of the galaxies in the sample (24 Virgo, 45 Field). The red spectra 
were taken during observing runs in 2000 February-April and 2000 September-October
using a similar setup as the blue spectra, with a 3\arcs slit, covering a spectral range of 5500-7500\AA\
and overlapping the blue spectra by $\sim$50\AA.  Typically, 1-2 exposures,
each 700-1200 sec in duration, were taken for each galaxy.  The red spectra
were reduced in the same manner as the blue spectra.

The galaxies observed in the red are so indicated in column 9 of Table~\ref{tab:log} and
were chosen from two subsets of our original data -- (i) those
galaxies with suspected emission at [OII] $\lambda$3727 and (ii) those galaxies
whose spectra appear to be emission-free.  An 
emission-free spectrum can be normalized and subtracted from an 
emission-contaminated spectrum to yield the flux in H$\alpha$ and in the neighboring
[NII] lines.
The H$\alpha$ information that is gained from the red spectrum can then be
used to correct the bluer Balmer line indices employed in the age-dating 
analysis. A discussion of the method used to correct for emission contamination
is given in Appendix A.

\section{Spectral Indices}

In studying the integrated light of galaxies, our goal is to determine
the distribution in age and chemical composition of the stellar populations
present.  Given that each population of a given age and metallicity itself is
a composite of stars covering a wide range in the HR diagram, the interpretation
of integrated spectra has been plagued by problems of non-uniqueness.
In addressing this, observers have
naturally been driven in two conflicting directions.  One is to observe the
entire spectral energy distribution, hence at low resolution and with a wide
slit, to best sample the contributions from all types of stars.  The other is
to work at higher spectral resolution and a narrow slit, to measure the 
strengths of specific absorption features which, by having primary sensitivity
to one particular atmospheric parameter (e.g., T$_{eff}$), can help solve
the non-uniqueness problem. Both
methods supply important constraints to the analysis of integrated light.  In
what follows we exclusively rely on higher resolution spectroscopy, with
emphasis on specific spectral features relevant to resolving 
non-uniqueness issues.

\subsection{Index Definitions}

In measuring the strengths of specific absorption features there is again a
tradeoff between low and high resolution approaches.  To obtain a robust 
result for the equivalent width of an absorption line requires a determination 
of the continuum at a sufficient distance from the line center so that the wings
of the feature are excluded, while the measurement of the feature itself must 
include both the line core and wings.
Such an approach forms the basis of the Lick index
system (e.g., \cite{fab85} and described in detail in \cite{wet94}), in
which indices are defined by measuring the flux in
the line bandpass and subtracting it from a continuum formed by connecting a
straight line between the average flux in red and blue sidebands.  While the
traditional Lick indices primarily concentrate on prominent metal features
in the red and on H$\beta$ (see \cite{wet94} for a complete
list),
several new Lick-style indices were defined in \cite{worott}
which concentrate on the higher order Balmer lines of H$\gamma$ and H$\delta$.

The disadvantage of the relatively broad bandpass system is the contamination
of both the line and continuum bandpasses with unwanted features which
can complicate the interpretation of the resultant spectral index.  Hence 
an alternative approach is to narrow both line and continuum bandpasses
sufficiently to obtain the cleanest possible measurements of both line and
(pseudo)-continuum.  This approach has been utilized in \citet{ro94} and
\citet{jw95} in defining indices that are narrowly focussed on,
e.g., H$\gamma$ or the Fe I$\lambda$4045 iron feature.  While minimizing the
contribution from unwanted features, this method has two disadvantages.  First,
by narrowing the line measurement, only the core is included, hence the
method loses sensitivity if significant wings are present.  Moreover, a
decrease in spectral resolution, due to instrumental or velocity broadening
effects, will redistribute the flux deficit in the line core out into the
wings, where the pseudocontinuum ``sidebands'' are defined.  The result is that
slight changes in spectral resolution can lead to large changes in the 
measured strength of the line feature, thereby making it critical to 
determine and account for this spectral resolution effect.  More recently,
\citet{va99} and Vazdekis \etal (2001a,b) have addressed this
problem in the case of a narrow band H$\gamma$ index in which the location
of the line and pseudocontinuum bandpasses have been carefully selected so as
to minimize the effect of spectral resolution on the index measurement.  
The result is
a more robust definition of H$\gamma$, but which also requires high precision in
the wavelength scale so as to measure the correct line and continuum intervals.

A third approach to the problem, originally formulated in \citet{ro84}, and
used subsequently in, e.g., \citet{ro85} and \citet{ro94}, is to maximize 
the contribution from the feature of interest by measuring the residual central
intensity in the line, and that in a neighboring reference line, and forming 
the ratio of these two quantities.  By using the deepest part of the spectral
lines in the two features, the maximum
contrast with other contaminating absorption lines is achieved.  In addition,
the line ratio index is defined without reference to the (pseudo)-continuum
level, which is generally problematic to locate, especially at bluer
wavelengths where line crowding is such a problem.  Moreover, since both line
centers tend to be similarly affected by a change in spectral resolution, the
line ratio indices are not strongly sensitive to resolution effects.  A 
disadvantage to this method is that the index loses sensitivity if one of the
lines saturates in the core, a problem that is a particular issue in younger
populations, where the Balmer lines become strong.  A further problem is that
the indices all consist of relative measurements of one feature against 
another, and thus cannot provide the equivalent width measurement of a single
feature that is ideal for resolving the degeneracy between age and metallicity
effects.

In this paper we utilize the well-studied Lick index system for the red 
spectral region, in particular for characterizing the strength of H$\beta$.
Although we use the standard Lick index definitions (Worthey \etal 1994), our 
index measurements \emph{have not been calibrated to the Lick/IDS system}, 
since that would require further smoothing of our data as well as
transformation onto the Lick/IDS instrumental response system.
For the higher order Balmer lines in the blue, where the line crowding is
more extreme, we switch primarily to the line ratio indices of \citet{ro94}.
In particular, we make extensive use of the Hn/Fe 
index\footnote{$Hn/Fe=<H\delta$/Fe4045+$H\gamma$/Fe4325+H8/Fe3859$>$} which
has the advantage of averaging three higher-order Balmer line to nearby
iron line ratios, and which was first shown to be an effective age indicator
in~\cite{crc00}.  Altogether, 
we measure a total of 30 spectral indices -- 15 Lick equivalent width indices, 
3 high-resolution equivalent width indices, and 12 line ratio indices.  
We do not utilize
the \cite{va01a}  H$\gamma_{\sigma<130}$ index in this paper, due to the
high velocity dispersion of many of the galaxies in our sample, but reserve it
for a future paper, in which we examine the low velocity dispersion Virgo 
galaxies.

To readily compare the observed galaxy spectral indices with those produced
by the evolutionary synthesis models, we broadened all of the galaxy spectra to
a common effective velocity dispersion of 230 \kmsc.  Consequently, we have
excised from the discussion in this paper all galaxies with $\sigma$$>$230 \kms
and leave them for a future discussion.  This somewhat arbitrary cutoff was
arrived at as a compromise between desiring to preserve as much dynamic range as
possible in the blue indices while also including as many galaxies as possible 
in the discussion.  The removal of of the high $\sigma$ galaxies reduces our
sample to 56 Virgo galaxies and 97 field galaxies.

\subsection{Errors in the Spectral Indices}

The spectral indices discussed above are all measured with a single, convenient
FORTRAN program written by Alexandre Vazdekis and publicly available on his 
website\footnote{http://www.iac.es/galeria/vazdekis/models.html}.  To determine 
the uncertainties in each spectral index, we use a weighted combination of the
rms in the individual galaxy measurements as outlined in Paper I; a brief
review of the process is given here.  First, the spectral index is measured for
each of the individual galaxy exposures, typically 2-4 for each object, and the
standard deviation of these values is determined.  Since the rms among individual
exposures is particularly unreliable for those galaxies which have only two observations,
we utilize the standard deviations in all indices for all galaxies in the following 
manner.  For a particular galaxy, \emph{i}, the error $\Lambda_{ij}$ in index \emph{j},
where there are a total of \emph{m} indices and \emph{n} galaxies, is given by:

\begin{equation}
\Lambda_{ij} = \frac{\sum_{i=1}^n \varepsilon_{ij} \sum_{j=1}^m 
\varepsilon_{ij}}{\sum_{i,j=1}^{m,n} \varepsilon_{ij}}
\end{equation}

\vskip 0.6cm
\noindent where $\sum_{i=1}^n \varepsilon_{ij}$ is the average over all galaxies of the
standard deviation in the $j^{th}$ index, $\sum_{j=1}^m \varepsilon_{ij}$ is the average 
of the standard deviations of all indices of the same type (i.e, all Lick indices or all 
line ratios indices) for the $i^{th}$ galaxy,
and $\sum_{i,j=1}^{n,m} \varepsilon_{ij}$ is the average standard deviation in all 
indices of the same type for all galaxies in the sample. Thus, in effect, by using 
all of the spectral index error information, we have found the average 
error for each index, and weighted those errors by the relative mean error in all
indices for a particular galaxy compared to the mean error
over the whole sample.  For the two galaxies in our sample for which we have only 
one observation, the above process is still followed, except that the standard 
deviation in each index is taken to be the mean rms in that index for galaxies with 
comparable signal-to-noise ratios. 

\subsection{Correcting for Emission Contamination}\label{sec:emiss-cont}

Early-type galaxies normally contain much less ionized gas and dust than spiral 
galaxies. However, 
spectroscopic surveys have revealed that nearly 40-50\% of ellipticals do
have at least some weak
optical emission~\citep{cald84,ph86}.  If emission lines \emph{are} present in the
galaxy spectrum -- from either ionized gas in HII regions, active galactic nuclei (AGN)
and/or planetary nebulae -- the measured absorption line indices may be severely affected. 
In the case of the Balmer lines, emission fill-in can weaken the
line strength and lead to older derived ages.  Thus, a potential source of systematic
error in the measurement of the spectral line indices comes from the contamination
of the absorption feature by fill-in from emission lines.  The method that we
follow to detect and remove contamination of the absorption features from
superimposed emission is discussed in Appendix A.

\subsection{Compilation of Spectral Index Data}

In Table~\ref{tab:listind} we list all of the spectral indices measured by us
for each galaxy.  In the first column the spectral index is given, and in the
second column a reference is given to the paper in which the index is defined.
The full table of spectral indices, and the errors in those indices, is listed
electronically in Table~\ref{tab:indices}.  To clarify matters, in the third 
column of Table~\ref{tab:listind} we list the column number in 
Table~\ref{tab:indices} in which a particular index appears.
Each index in Table~\ref{tab:indices} is followed by its $\pm$1$\sigma$ error.
Many of the indices in Table~\ref{tab:indices} have been corrected
for emission contamination (in the Balmer lines) and/or for the effects of 
non-solar abundance ratios (see \S\ref{sec:nsar}). 
As a guide to the amount of correction that has been
applied, in the second to last column in Table~\ref{tab:indices} we give the
correction that has been made to the Lick H$\beta$ index for emission fill-in,
and in the last column we give the correction applied to the H$\beta$ index to 
account for non-solar abundance ratios.

\section{Evolutionary Synthesis Models}

The information extracted from the spectral indices defined in the previous
Section must still be translated into mean ages and metallicities of the
integrated stellar population through the use of population synthesis models.
In these models, a theoretical stellar evolutionary
isochrone is computed for a given population representing a unique age and metallicity,
hereafter referred to as a simple stellar population (SSP).  Each point
along the isochrone represents a specific star of known temperature, luminosity and
surface gravity which produces a specific set of absorption lines.  The
spectral line strengths for each point along the isochrone are weighted 
by the stellar luminosity and the number of stars expected at that point
(from the computed luminosity function)
and then added to derive the total line strength for the integrated population.
By computing the expected integrated line strength for SSPs covering a 
range of ages and metallicities, the spectral indices from the galaxies can be
compared to the predicted values, and, the luminosity-weighted mean
age and metallicity of the galaxy can be determined.  

While the sophistication of the evolutionary population synthesis models that
supply the framework for interpreting
the integrated light of galaxies has increased substantially
in the last few years, the current generation of models still has its 
uncertainties.  There are a variety of issues including: uncertainties in the
model isochrones (with inadequate understanding of convection and of mass loss
being the chief culprits), uncertainties in transforming from the theoretical
H-R diagram into an observational plane (with the T$_{eff}$ scale providing a
major problem), inadequate coverage in empirical stellar libraries, 
uncertainties in the accuracy of model stellar atmospheres and synthetic
spectra, and uncertainties in the treatment of non-solar abundance ratios.
Additionally, the population models assume a single age and metallicity for the 
entire galaxy stellar population while a galaxy is most likely a composite of
different populations.  As a result of the uncertainties in the models, we
place primary emphasis on using the models to determine the \emph{relative}
ages of galaxies, rather than their \emph{absolute} ages.
In the present section, the components and 
implementation of the population models are described, followed by a comparison 
of the models used in this work to other available synthesis models.

\subsection{The Worthey Population Models}

For our analysis, we use a modified version of the population synthesis models
originally introduced in~\cite{wo94}.  The Worthey (1994, hereafter W94) models
depend on two adjustable parameters -- the metallicity and single-burst age
of the stellar population -- and one fixed parameter, the IMF exponent, here chosen to
be the Salpeter value.  The basic ideas behind the models are reviewed briefly, and
the reader is referred to W94 for more details.  

To assemble the theoretical integrated spectrum, the models incorporate three 
ingredients: stellar evolutionary isochrones, a stellar SED library, and a library
of absorption line strengths. For the evolutionary information, the modified W94 
models use the isochrones of the Padova group~\citep{padova} which have been 
computed for six metallicities in the range -1.7 $<$ [Z/H] $<$ +0.4 and for 29 ages 
from 0.004 to 19 Gyr.  The stellar SED library was constructed using the model
atmospheres and synthetic spectra of~\citet{kur93} for stars hotter than 3750K and the
model SEDs of~\citet{bess91} and observed  SEDs from~\citet{gs83} for cooler
M giants. Ideally, the spectral line strengths should be determined from a 
single observed library of stellar spectra which fully sample the entire range
of atmospheric
parameters (T$_{eff}$, log~g, and [Z/H]).  Since the Galactic star formation
and chemical enrichment history excludes some regions of the atmospheric 
parameter space, especially metal-poor high-gravity stars, we have chosen to 
combine libraries of both empirical and synthetic spectra to fully cover the
line strengths along each isochrone.

\subsubsection{The Empirical Stellar Library}

For the empirical spectral library, we use a set of high-resolution stellar 
spectra compiled by L.\ Jones (1999; see also Leitherer \etal 1996),
which are well-suited for stellar population 
analysis.  The library consists of two spectral regions --- one in the blue covering 
the wavelength range $\lambda$3820-4509\AA\  and one in the red covering 
$\lambda$4780-5469\AA\ --- for each of 684 stars.  
The spectra\footnote{The complete Coud\'e Feed spectral library is publicly
available from the NOAO ftp archive at ftp://ftp.noao.edu/catalogs/coudelib/.}
were taken with the KPNO Coud\'e Feed telescope and spectrograph and have a pixel
sampling of 0.62\AA\ pixel$^{-1}$ with a spectral resolution of 1.8\AA\
FWHM.  Because the stellar library is primarily drawn from the local solar neighborhood,
all areas of the log~g-[Fe/H]-log T$_{eff}$ grid are not equally covered.  Specifically,
the grid is relatively well-populated for cool stars (T$_{eff} <$ 6000), but at higher
temperatures, it has a much more limited coverage due to the lack of hot, metal-poor and
hot, low-gravity stars.
To ensure adequate sampling throughout the parameter space, we restrict our use of 
the empirical library to stars in the temperature range 3750-6300K.  Further stars
were eliminated from the library if they are observed to have chromospheric emission
(which contaminates the absorption line indices) or if their atmospheric parameters
are either undetermined or are in striking disagreement with their measured 
spectral indices.  In all, 547 stars are used from the empirical library.

The empirical spectral library is incorporated into the population synthesis models
through polynomial fitting functions which describe how the absorption line strength
varies as a function of the effective temperature, gravity and metallicity of the
stars. To produce the fitting functions, we employed a FORTRAN program kindly provided by
G. Worthey.  The program uses a series of least-squares regressions to fit a third order
polynomial in $\Theta$=5040/T$_{eff}$, [Fe/H], and log~g. Through an iterative procedure, the
terms in the polynomials are either included or excluded until no systematic trends
appear in the residuals of atmospheric parameters versus the fit. Stars with very 
deviant index values are excluded from the fits, unless they are in regions of 
particularly poor coverage (e.g., very low metallicity).  Each
absorption line used in the spectral indices is fitted with one polynomial across the 
entire temperature range, 3500-6300K.  

\subsubsection{The Synthetic Spectral Library}

Due to the lack of hot, unevolved metal-poor stars in the Galaxy, the
Coud\'e Feed spectral library (CFSL) only has adequate metallicity coverage for
temperatures less than 7000K.
To compensate for this
shortcoming in the Coud\'e Feed library, the empirical spectra are augmented
with 2103 synthetic spectra generated with the R. Kurucz SYNTHE program using
the Kurucz ATLAS model atmospheres~\citep{kur93}, both kindly provided by 
R. Kurucz.
A complete description of the construction of the synthetic
spectra is given in~\citet{lr02}.  Briefly stated, the SYNTHE program 
generates 
a synthetic spectrum at high resolution (0.1\AA\ pix$^{-1}$) from a given model atmosphere and line list.  The
SYNTHE spectrum is then rebinned and smoothed to match that of the 
observations.
The synthetic library covers the wavelength
range $\lambda$3500-5500\AA\  and covers the 3-dimensional parameter space of
log~g, [Fe/H] and T$_{eff}$ (see Table~4 of \citet{lr02} for the specific
coverage).  Because the synthetic spectra
are subject neither to observational errors nor to uncertainties in the atmospheric
parameters, the calculated line strengths of the SYNTHE stars evolve systematically
with the atmospheric parameters, and there is no need to fit them with polynomial
functions. Instead, the synthetic spectra are incorporated into the population
synthesis models through linear interpolation between the grid points for a
given set of atmospheric parameters. For points outside the SYNTHE parameter
grid, the line strength corresponding to the nearest grid point is used.

\begin{figure}
\plotone{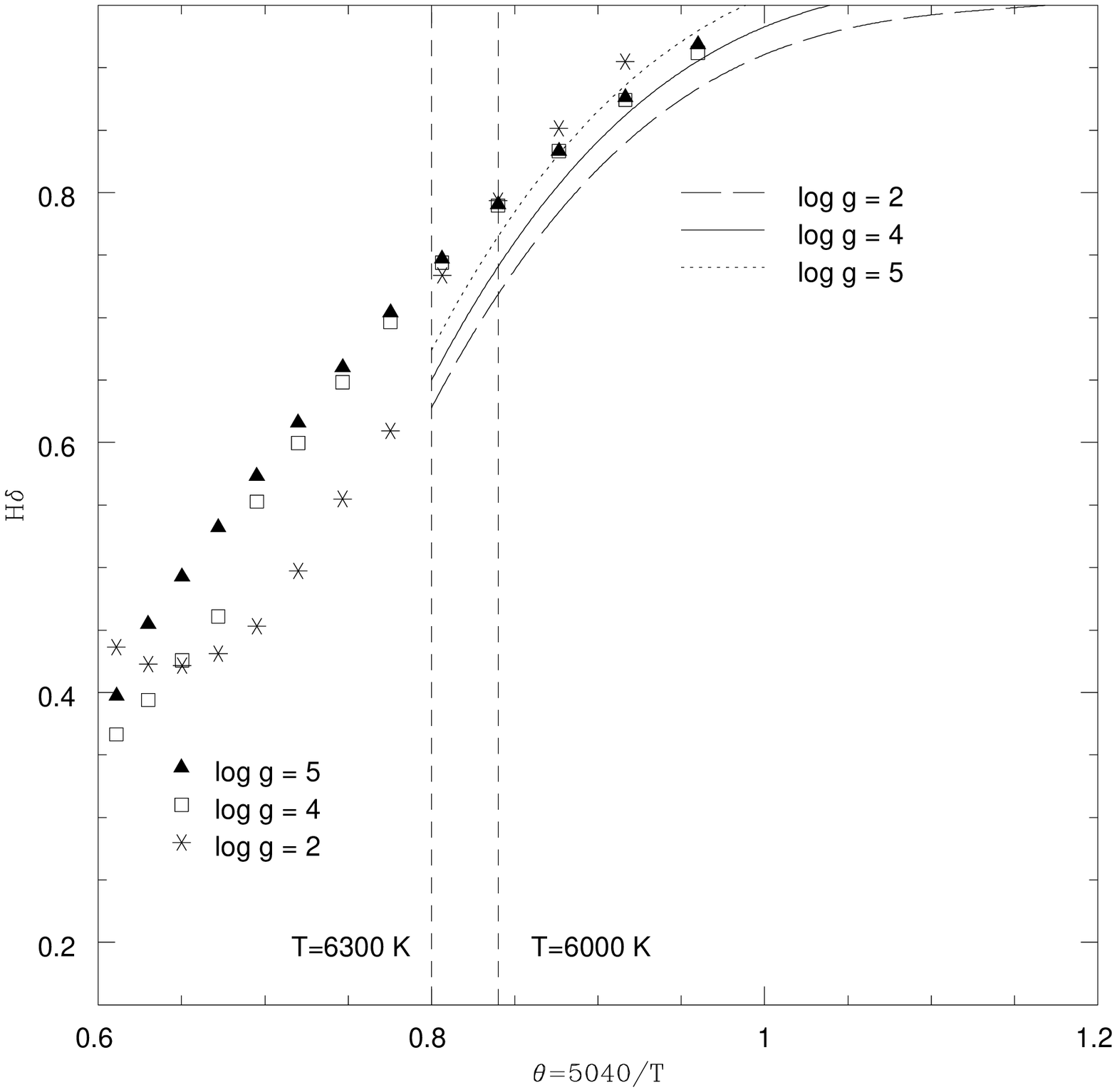}
\caption{Comparison between the empirical H$\delta$ line strength fitting function 
(curves) and the synthetic library stars (symbols) for solar metallicity and three 
different surface gravities.  The region of overlap is indicated by the vertical dashed 
lines. The fitting functions are calculated for T$<$6300 K and used for T$<$6000K.}
\label{coude-syn}
\end{figure}

It is, of course, important that there be no strong discontinuity in line
strength across the 6000K boundary in temperature between the empirical cool
star spectra and the synthetic hot star spectra.  The agreement between the
empirical and the synthetic data is in general reasonably good in
the region of overlapping temperature.  We plot an example of the interface
between synthetic and empirical spectra for the H$\delta$ feature in
Figure~\ref{coude-syn}.

\subsection{Putting It All Together}

Before using the population models to interpret the integrated light of the galaxies,
it is important to ensure that all of the model ingredients are smoothed to the
same spectral broadening as the galaxy data.  The spectra in the Coud\'e Feed library
were observed at an intrinsic spectral resolution of $\sim$1.8 \AA\ FWHM whereas the FLWO galaxy
spectra are at a resolution of $\sim$3.1 \AA.  To put the two datasets on the same scale, 
the Coud\'e spectra are convolved with a Gaussian of $\sigma$ = 0.97 \AA.
To determine the amount of broadening required to match the SYNTHE and FLWO spectra,
a synthetic spectrum was chosen with the appropriate atmospheric parameters of a 
star observed with the same configuration as the galaxy spectra.  The SYNTHE spectrum was then
broadened by various Gaussians ranging from 0.2 to 6.0 pixels, each of the broadened 
spectra were cross-correlated with the original spectrum, and the relationship between 
the FWHM of the cross-correlation peak and the amount of smoothing was determined.
Using this relationship, the amount of broadening required to match the SYNTHE 
spectra with the FLWO spectra was found
to be $\sigma$ = 0.24 \AA.  After both stellar libraries were smoothed to the same
intrinsic spectral resolution as the galaxy data, they were then smoothed by an
additional amount to match the effective galaxy broadening of 230 \kms (recall
that all final galaxy spectra are broadened to a common velocity dispersion of
230 \kmsc).  At this point, both the empirical and synthetic libraries and the
galaxy data are all at a common spectral resolution.

Upon implementation, the population synthesis models combine the information from
the Padova isochrones, the Kurucz SED fluxes and the library line strengths to create a
theoretical integrated population of a specified age and metallicity.  To assemble
the integrated spectrum, the absorption line strengths are weighted by the number
and luminosity of stars at that isochrone point and are superposed on the 
continuum that has been formed by the appropriate weighting of each SED.  As the 
models integrate along the distribution of stars, the line strengths for isochrone points 
cooler than 6000K are calculated from the empirical fitting functions (where the
Coud\'e Feed library coverage is complete) and from interpolating in the SYNTHE
line strength grid for stars hotter than 6000K.  

\subsection{Comparison with Other Models}

The ultimate results of the W94 population models are integrated absorption line
strengths, which can be translated into the various spectral indices described in $\S$3.
For each spectral index we calculate a grid of model values at a series of
ages and metallicities.
By choosing an appropriate pair of indices, the galaxy data can be over-plotted
on the index-index model grids, and the age and metallicity of the stellar population can be 
determined. Since the derived ages depend on the model grids against 
which they are measured, an important gauge is to compare the index-index grids produced 
by different synthesis models.  Thus, as a check of our implementation of
the W94 models, and hence our interpretation of the early-type galaxy spectra, we
compare our model grids to those produced by the publicly available\footnote{
http://www.iac.es/galeria/vazdekis/models.html} population synthesis models 
of A. Vazdekis (Vazdekis 1999, V99).  The Vazdekis models are similar to 
the updated W94 models in that they both use the set of evolutionary isochrones produced 
by the Padova group (e.g, Bertelli et al.\ 1994),
although the V99 models supplement these isochrones with the stellar tracks of
Pols et al.\ (1995) for low mass stars (M$<$0.6M$_\odot$).  Both models make
use of the CFSL in constructing the integrated stellar population. However, an important
difference in the two models is the method in which the stellar spectra are 
utilized in the computation of the integrated line strengths.  As described in the 
previous section, the W94 line strengths are formed by interpolation in the synthetic 
spectral grid for hot stars, and through the use of empirical fitting functions for the
cooler CFSL stars (T$_{eff}<$6000K).  The V99 models, on the other hand, rely
solely on the CFSL (with slightly modified stellar parameters), and for each 
point in the H-R diagram choose a CFSL stellar spectrum with closely matching 
atmospheric parameters.

Figure~\ref{fig:guyvaz} shows a comparison of the two sets of models for four
different spectral index pairs that are used extensively in this paper.  M32
is once again shown alone in all the plots as a visual aid. For
clarity, we plot only a limited number of discrete ages, and only the two 
chemical compositions [Fe/H]=0.0 and [Fe/H]=-0.4.  
Figure~\ref{fig:guyvaz} indicates that there is good agreement between
the predictions of the W94 models and the V99 models in the H$\beta$ versus
Fe 4383 diagram. Agreement is satisfactory, as well, in the Hn/Fe versus 
Fe 4383 diagram.   For the Mg~$b$ versus H$\beta$ and the Mg~$b$ versus Hn/Fe
plots, however, the agreement is less satisfactory.
The discrepancies highlight the problems
inherent in determining the absolute mean ages of the galaxies.  On the other
hand, the {\it relative} ages of the galaxies, which are the primary concern of this
paper, are less vulnerable to modeling issues.  A more detailed
comparison of the two models has been carried out by \citet{ku00} 
and \citet{tf02}.

\subsection{Determining Galaxy Ages and Metallicities}

\begin{figure}
\plotone{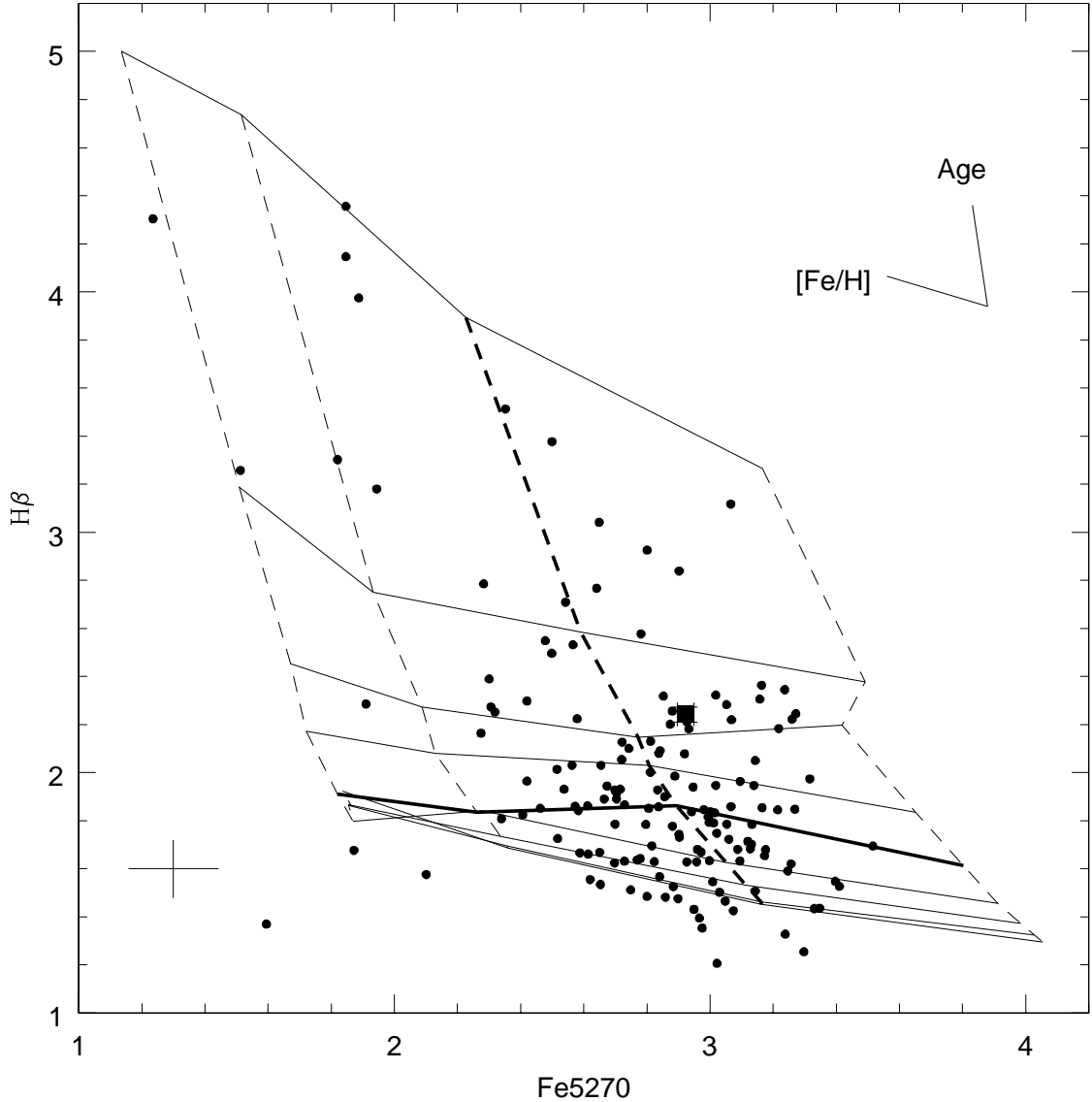}
\caption{A sample of the galaxy distribution (filled circles) on an iron versus 
Balmer line index grid.  The Fe~5270 indices have been corrected for NSAR, while
the H$\beta $ indices have been corrected for NSAR and emission.  
Tracks of constant age are denoted by solid lines, 
with ages, increasing from top to bottom, of 1.00, 2.00, 3.16, 5.01, 7.94 (shown in bold), 
12.02, 15.13, 17.38, and 19.05 Gyr.  Tracks of constant metallicity are denoted 
by dashed lines, with [Fe/H], increasing from left to right, of -0.7, -0.4, 0.0 (shown in
bold),
and +0,4.  The average $\pm$1$\sigma$ error bars for the observed indices are 
shown in the lower left corner.  M32 is plotted for reference as a solid 
square. The units of the axes, and indeed all Lick indices in subsequent plots,
are in \AA . Line ratios used in other plots are of course unitless.}
\label{fig:Hb-Fe}
\end{figure}

\begin{figure}
\plotone{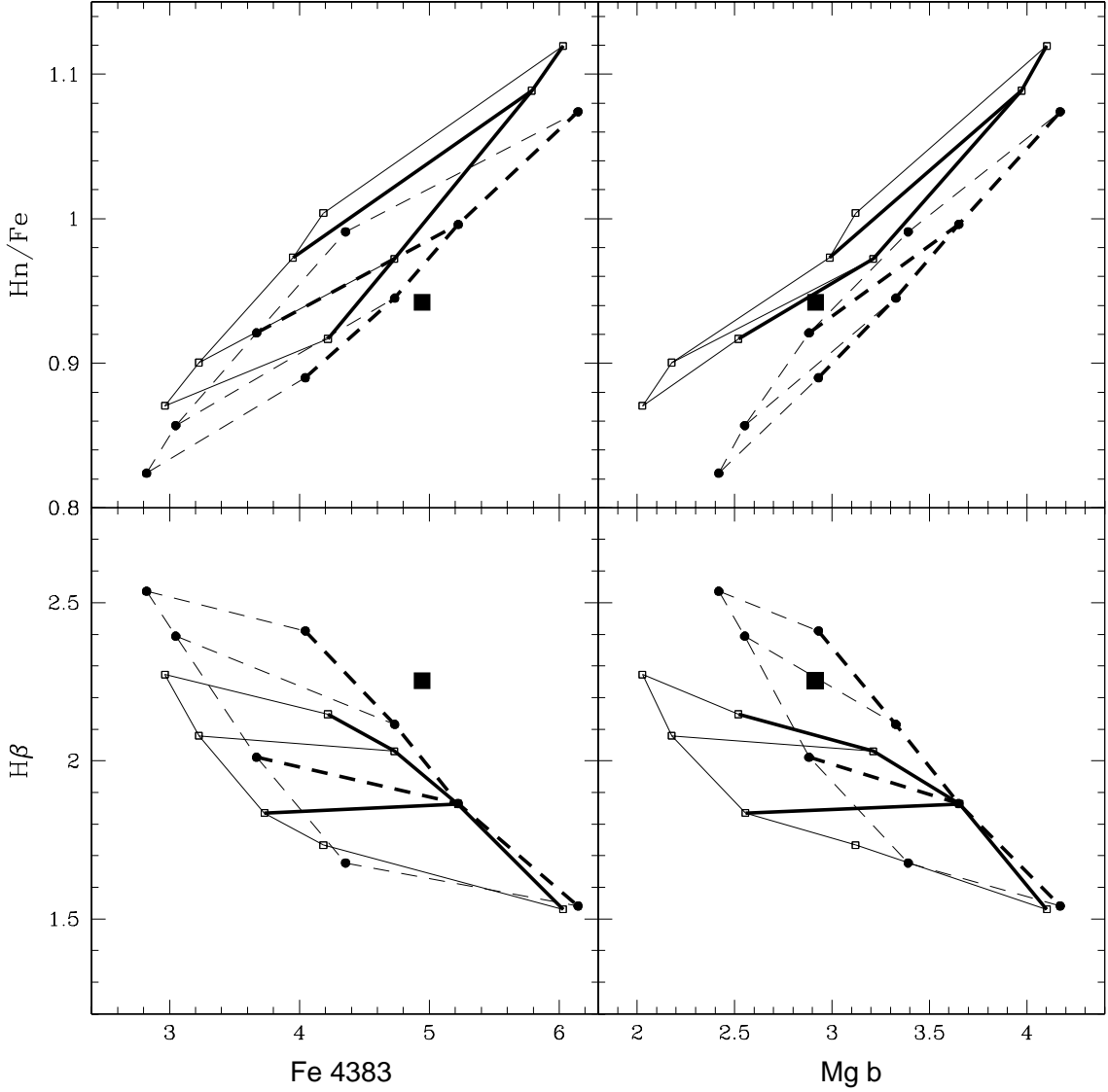}
\caption{A comparison between the grids formed from the W94 models used here
(solid lines), and the Vazdekis \etal 2001b models (dashed lines). Shown are grids
for 4 line indices: Hn/Fe, H$\beta$, Mg~$b$, and Fe~4383. 
 The grid points in age
occur at 15.13, 7.94, 5.01 and 3.12 Gyr; in metallicity they occur at -0.4 and 0.0.
Bold lines show the solar abundance line, and the 7.94 Gyr age line. The index
values for M32 are plotted as  squares. 
The left two panels show good agreement, while
the panels on the right with Mg~$b$ have some discord. 
  }
\label{fig:guyvaz}
\end{figure}

The goal of the population synthesis modeling is to compare the grid of 
computed model indices to those of our galaxy sample, and thereby to derive 
ages and metallicities of the galaxies in our data set.  To do so, a primarily
metallicity sensitive index (e.g., Fe5270) is plotted against a primarily age sensitive
index (e.g., H$\beta$) for a range of SSP model predictions producing a largely
orthogonal grid of ages and metallicities.  By over-plotting the galaxy
data, the SSP equivalent age and metallicity can be determined (Figure~\ref{fig:Hb-Fe})
through interpolation in the index grid.  However, because the age-metallicity
degeneracy is not completely broken, the lines of the model grids are not
perpendicular. This fact complicates the reliable extraction of age and 
metallicity information when observational errors are included.
An additional complication in deriving the galaxy ages and 
metallicities is that, as seen in Figure~\ref{fig:Hb-Fe}, not all galaxies lie 
within the model predictions.  This may be due to observational errors,
residual emission contamination in the Balmer line indices (see 
\S\ref{sec:emiss-cont}), differing
abundance ratios in the galaxies and the model components (addressed in 
\S\ref{sec:nsar}),
and/or zero-point offsets in the population models.

To extract the age and metallicity of a galaxy from a particular index-index
diagram, the four model grid points which form a box encompassing the data
point are found.  Using the four corners of the surrounding box, the lines of
constant age and metallicity which pass through the data point are determined,
and the age and metallicity values are interpolated along
the constant metallicity and age lines.  For galaxies
which lie outside the model grids, the population values are extrapolated only
in the cases where the the extrapolation is small and reliable.  
Because the metallicity
tracks tend to be evenly spaced, extrapolation of the galaxy metallicity
is relatively stable. On the other hand, the constant age tracks are not evenly
spaced and are, in fact, packed closer together for older ages.  This makes a 
linear
extrapolation of the galaxy age uncertain for data points which lie far from the
 model
grids, particularly at old ages.  Thus we do not attempt an age extrapolation
for those galaxies which lie beyond the oldest age line in the diagram.

\subsection{Non-Solar Abundance Ratios}
\label{sec:nsar}

In the previous section we have seen how spectral line-strength indices and 
population synthesis models can be used to circumvent the age-metallicity
degeneracy, allowing us to derive the SSP age and metallicity for a galaxy. 
However, there is an additional complication to consider in making these
determinations.
Since the late 1970s, evidence has mounted that the abundance ratios in early-type
galaxies are often non-solar and that the Mg/Fe abundance ratio, in particular, 
is larger
in more luminous early-type galaxies than in solar-neighborhood stars (O'Connell 1976,
Peletier 1989; Worthey, Faber \& Gonzalez 1992; Vazdekis et al 1997).  The
population synthesis models are built using spectral databases of Solar
Neighborhood stars, which
reflect the specific pattern of the chemical enrichment of our Galaxy.  In 
addition, the theoretical isochrones that are used as the backbone of 
evolutionary synthesis modeling are based on solar abundance ratios.
Hence for galaxies with elemental abundance ratios deviating from that found in the Solar
Neighborhood, there are additional corrections to be made in modeling their 
luminosity-weighted mean ages and chemical compositions.  Recent population
synthesis modeling is now beginning to simulate the effect of non-solar
abundance ratios (NSAR) \citep{tr00a,ku01,va01a,ps02}, based on theoretical
isochrones incorporating non-solar abundance ratios \citep{sw98,vdb00,sa00} and
on the synthetic spectral modeling of key spectral indicators carried out by
\cite{tb95}.
In Appendix B we describe how the effects of NSAR are accounted for, following
the prescriptions developed by \cite{tr00a} and \cite{ku01}. 

\section{Results on Ages of Early-Type Galaxies}
\label{sec:syseffects}

We are now in a position to combine the galaxy spectral index data, described 
in \S3, with the modeling procedures, described in \S4, to determine 
luminosity-weighted mean ages and metallicities for our early-type galaxy 
sample.  There will be two main goals for the discussion in this Section and
in \S6.
First, in this Section we demonstrate that there is a trend in the
mean ages of early-type galaxies with velocity dispersion, in the sense that 
low-$\sigma$ galaxies have younger mean ages than their higher-$\sigma$ (i.e.,
more massive) counterparts.  In establishing this result we will evaluate 
several potential sources of systematic error in age determinations, namely,
``contamination'' from a metal-poor population and/or from a red horizontal 
branch (RHB) population, ``contamination'' from emission lines, and the
effect of non-solar abundance ratios. The second goal, discussed in \S6, is to 
compare results
obtained from different Balmer and metal line indicators, and to evaluate whether
robust and consistent ages and abundances emerge when different diagnostic
spectral features are employed.

\subsection{Spectral Variations as a Function of Velocity Dispersion}

\begin{figure}
\plotone{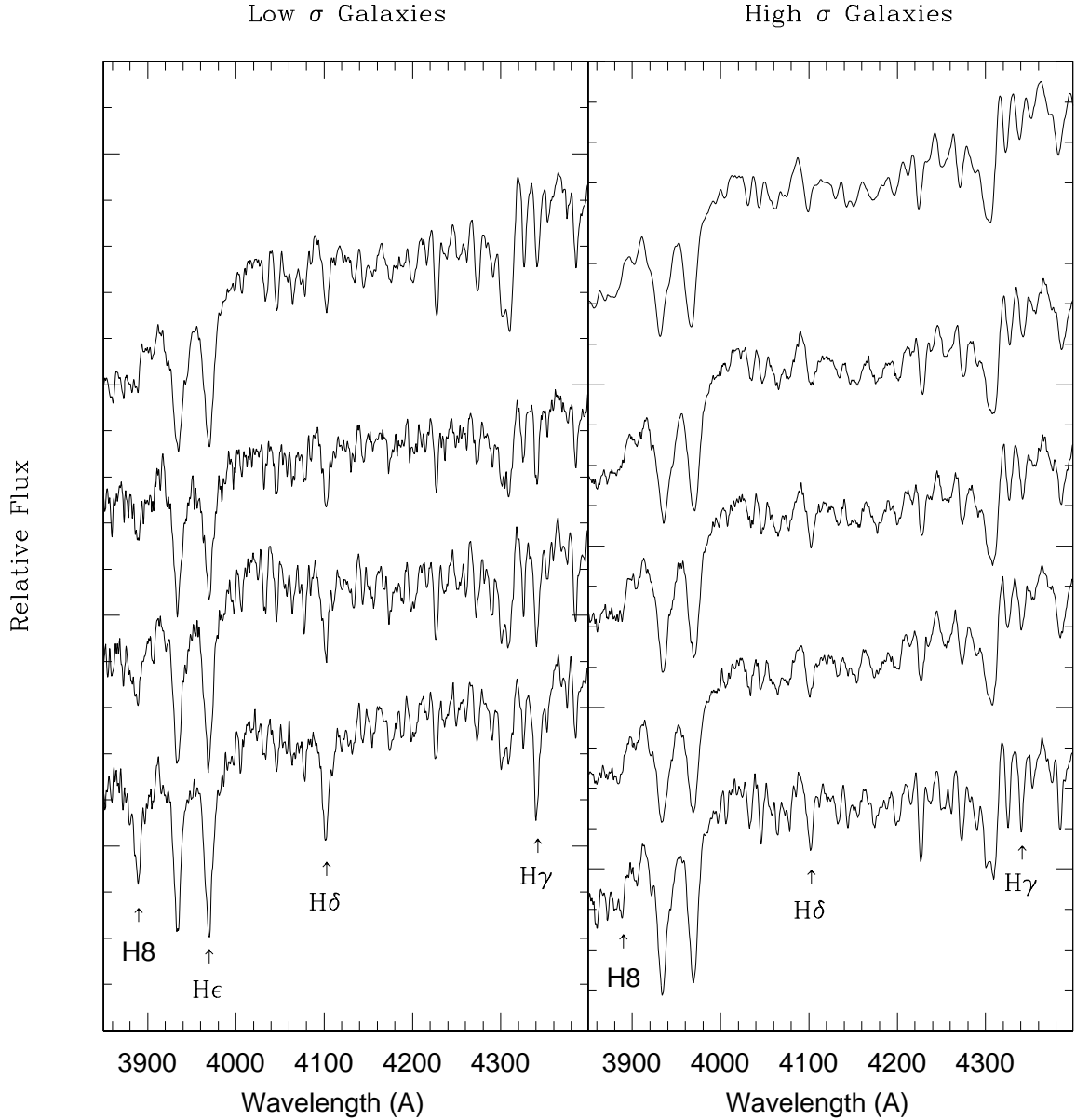}
\caption{Representative spectra of low-$\sigma$ (left panel) and high-$\sigma$
(right panel) galaxies are plotted, at their observed broadenings.  The effect of increasing Balmer line
strength can be seen by looking from top to bottom in the panels. The effect of
increased line broadening can be seen by comparing the lowest two galaxies in
the right panel.  The increase
in Balmer strength is more pronounced in the low-$\sigma$ galaxies.  The 
spectra plotted in the left hand panel are of, from top to bottom, NGC~770,
VCC140, VCC523, and VCC1912.  In the right hand panel the spectra are of, from
top to bottom, VCC685=NGC 4350, NGC~821, NGC~938, NGC~2954, and VCC1475.}
\label{fig:lines}
\end{figure}

We begin by illustrating the large range in spectral behavior that is evident
in our galaxy sample.  
In Fig.~\ref{fig:lines}, we plot representative examples of low-$\sigma$ and
high-$\sigma$ galaxies, concentrating on the spectral region containing
H$\delta$ at 4101 \AA \ and H$\gamma$ at 4340 \AA, as well as H8 at 3888 \AA.  
These spectra are shown at their observed spectral broadenings.
From their appearance, it
is clear that the low $\sigma$ galaxies have a much wider range in Balmer line
strength, and that the increase in Balmer line strength in the galaxies from
top to bottom in the figure is systematic across all Balmer lines.

\begin{figure}
\plotone{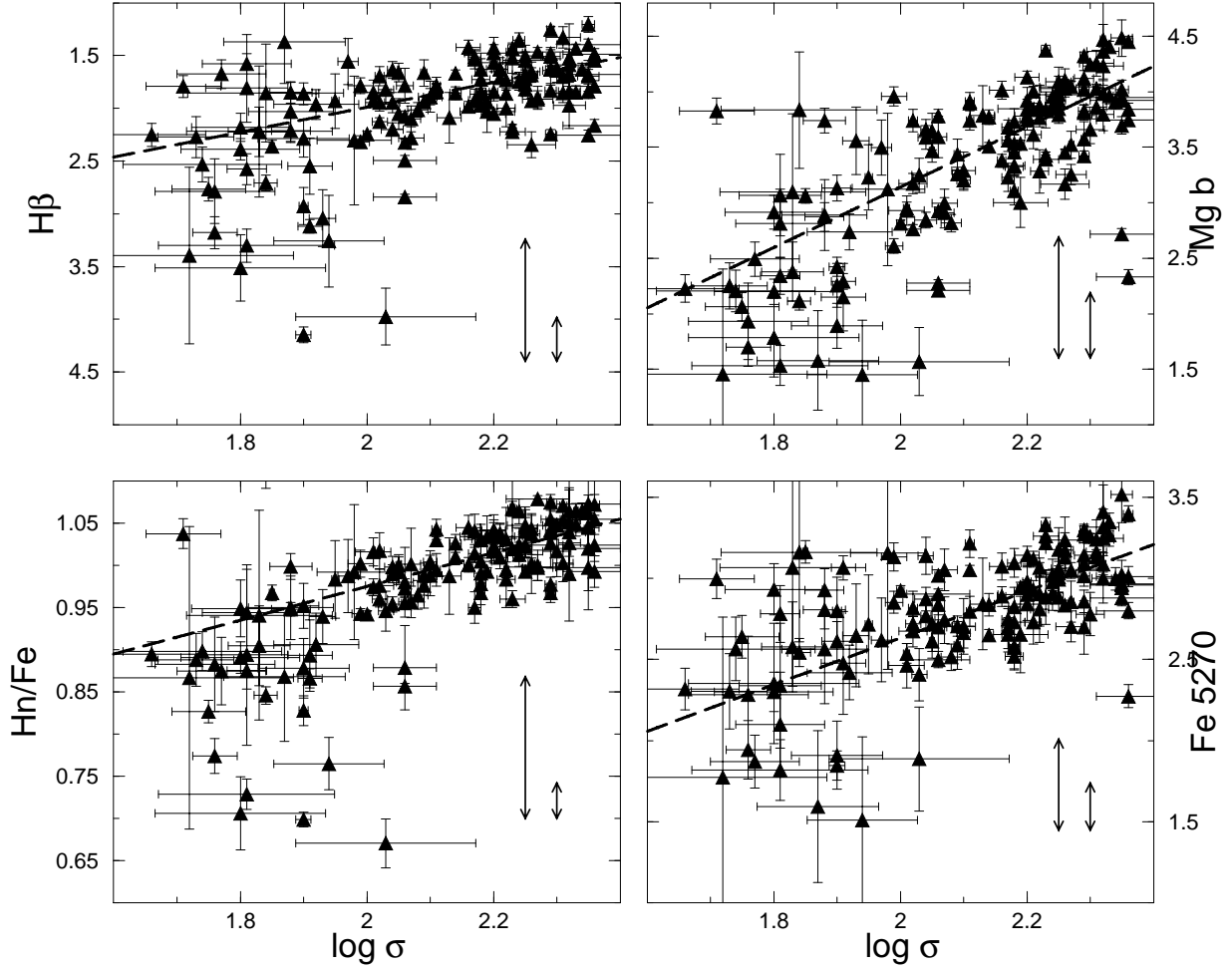}
\caption{Four spectral indices are plotted versus log $\sigma$. All indices 
have been emission corrected, and all except the Hn/Fe index have
been NSAR corrected. Note that the
vertical axis of the H$\beta$ plot has been inverted to mirror the relation in
the other three indices.  The long-dashed line is the BCES regression fit to
the data with log $\sigma$ $>$ 2.0, excluding outliers.   The vertical arrows 
in the lower right corner of each panel is a representation of the measured
intrinsic scatter in the high $\sigma$ galaxies (log $\sigma$ $>$ 2.0, right) 
and the low $\sigma$ galaxies (log $\sigma$ $<$ 2.0, left).}
\label{bces}
\end{figure}

Hereinafter, the data we discuss, either in the form of spectra or indices,
have been smoothed to a common velocity dispersion of 230 \kms, emission corrected
(see appendix A),
and corrected for NSAR where applicable (see appendix B).
A quantitative representation of the variation in line strength among our
early-type galaxy spectra is given in Fig.~\ref{bces}, where the behavior of
four spectral indicators as a function of galaxy velocity dispersion is
plotted.  The two right hand panels characterize the behavior of the metal
features Mg$~b$ and Fe5270 from the Lick system.  The upper and lower left
hand panels plot the behavior of the Balmer-sensitive indices H$\beta$ and
Hn/Fe.  Note that for the two Balmer line strength indices
the Balmer strength is plotted so as to increase from top to bottom.
As defined in \S3, the Hn/Fe index is an average value for the line ratios
of H$\gamma$, H$\delta$, and H8 (at 3888 \AA ) to neighboring metal lines.
Plotted with the galaxy data in Fig.~\ref{bces} are the regression line
fits to the indices as a function of log $\sigma$.  The fits were determined
using the BCES (bivariate correlated errors and intrinsic scatter) method of 
Akritas \& Bershady (1996) for the higher $\sigma$ galaxies (with
log $\sigma$ $>$ 2.0, the outliers were exluded from the fit), 
and then extrapolated to lower $\sigma$.  The application
of the BCES method is further described in \citet{crc00}.

There are two key features to note in Fig.~\ref{bces}.  First, while the
metal-sensitive Mg$~b$ and Fe 5270 indices exhibit a relatively constant and 
symmetric scatter at all velocity dispersions, the two Balmer line indices
show a striking increase in scatter for log $\sigma$ $<$ 2. Second, the large
scatter at low log $\sigma$ is asymmetric, with many galaxies having high 
Balmer line strengths.  As is previously described in
\citet{crc00}, we have extracted the {\it intrinsic}
scatter in the indices from the {\it observed} scatter, and the intrinsic
scatter at high and low log $\sigma$ is denoted by the two vertical arrows
in the lower right corner of each plot.  The results of
the intrinsic scatter measurements are summarized in Table~\ref{tab:inscat}.
There we have divided the galaxy sample into low- and high-$\sigma$ bins. For
each index we give the derived intrinsic scatter in the index as well as the
$\pm$1$\sigma$ uncertainty in the intrinsic scatter, first for the uncorrected 
indices, then for emission-corrected indices, and finally for indices corrected
both for emission and for NSAR.  The number in parentheses below the 
uncorrected value represents the median observational error for the index.
Since different indices are measured in different ways, the ratio in intrinsic
scatter between low- and high-$\sigma$ samples provides the clearest assessment
of the dependence of spectral index behavior on velocity dispersion.  While that
ratio is only 1.8 and 2.0 for Mg$~b$ and Fe5270, respectively, in the final 
(i.e., emission-corrected and NSAR-corrected) indices, it is 2.8
and 4.0 for H$\beta$ and Hn/Fe, thus underscoring that the large intrinsic
scatter in low-$\sigma$ galaxies is more pronounced in the Balmer lines.  While
the scatter in indices at low $\sigma$ is large, a clearly measurable intrinsic
scatter is also evident at high $\sigma$.  We return to this fact in \S\ref{sec:inscatter}.

We have recalculated the statistics using the lower velocity dispersions
from the literature, where available (\S 2.), 
and found no significant change. Perhaps
when all of these lower mass galaxies have precise dispersions measured some
of the details of the scatter calculations may change, but the main results that
the scatter is due to age variations will likely not change.
\citet{crc00} proposed that the large intrinsic scatter 
in the Balmer lines is due primarily to a range in age that
is particularly pronounced among the lower $\sigma$ galaxies.  With the population
synthesis models and technique established in the previous sections, we now evaluate
that assertion.

\subsection{Variations in Age as a Function of Velocity Dispersion}
\label{sec:agevar}

\begin{figure}
\plotone{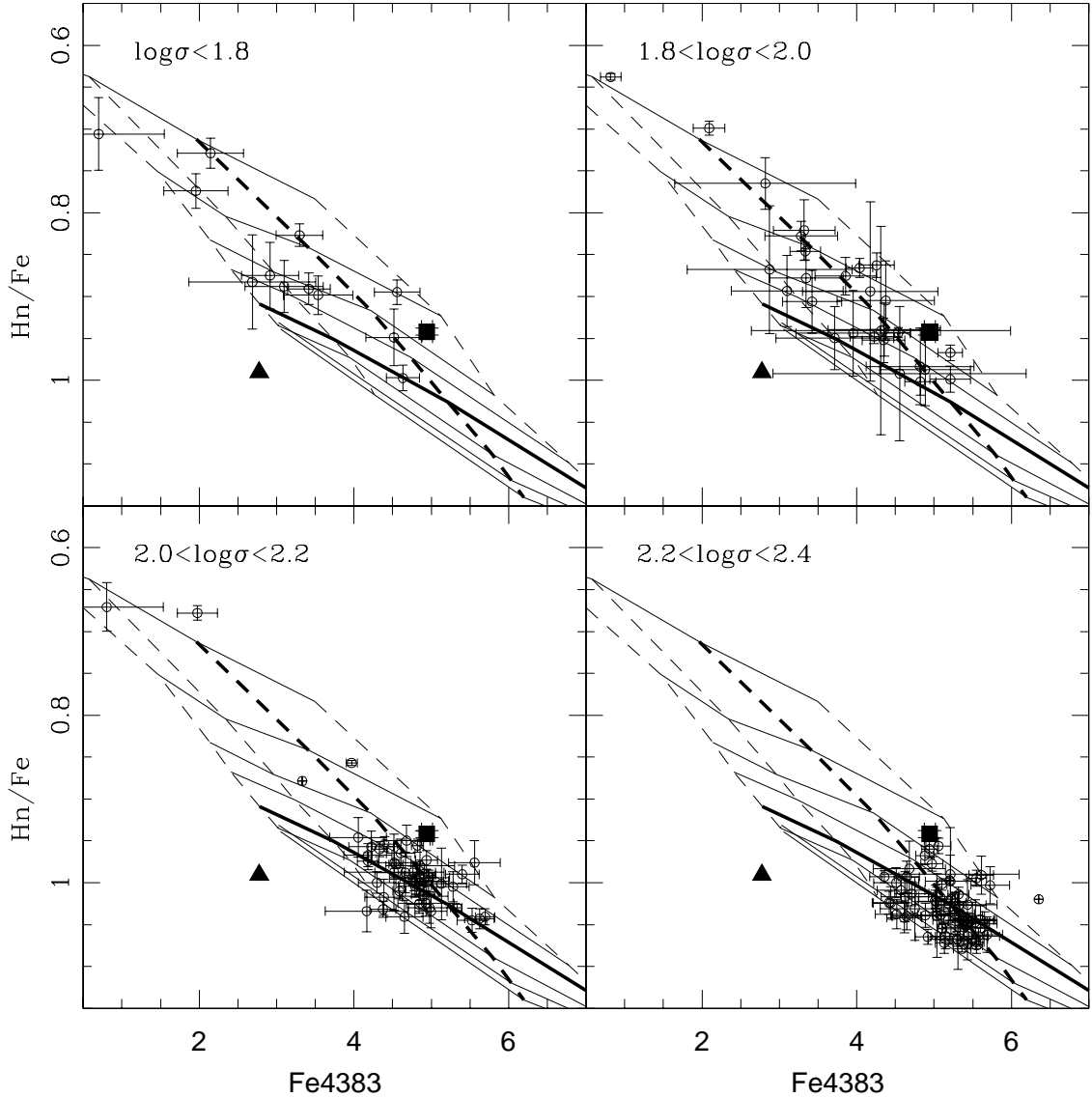}
\caption{The Lick Fe4383 index is plotted versus the Hn/Fe index for all galaxies in our
sample with $\sigma\le$230 \kmsc.  Note that {\it low} values of Hn/Fe indicate
strong Balmer line strength relative to the neighboring metal lines.  
The galaxies have been subdivided into four different velocity dispersion bins
in the four panels of the Figure.  The large square denotes the elliptical 
galaxy M32, while the large triangle denotes the globular cluster 47 Tuc.  
The grid lines connect the simple stellar populations
from the Worthey models of various ages and metallicities.  Solid lines connect
models of constant age; the ages (in Gyr), from top to bottom, are: 1.00, 2.00, 
3.16, 5.01, 7.94 (shown in bold), 12.02, 15.13, 17.38..  Dashed lines connect models of
constant metal-abundance; the [Fe/H] values, from left to right, are: +0.4, 0.0 (shown in bold),
-0.4, and -0.7.  The galaxy data for this plot have been corrected for emission only. 
}
\label{fig:Hn4383_4plot}
\end{figure}

The distribution of galaxies in luminosity-weighted mean age 
and metallicity can be assessed by plotting the combination of a primarily age-sensitive 
Balmer index versus a primarily metal-sensitive index.  To begin, we plot the
metal-sensitive Fe4383 index from the Lick system versus the Balmer-sensitive 
Hn/Fe index in Fig.~\ref{fig:Hn4383_4plot}.  The galaxy data have been corrected for
emission, and have been subdivided into four different velocity dispersion bins.
The data are superposed on grid lines of
SSPs of constant age and metallicity as determined from the Worthey models
described in \S4.  For comparison with more familiar plots of this kind, 
Fig.~\ref{fig:Hn4383_4plot} is similar in nature to the log H$\beta$ versus 
log $<$Fe$>$ plot in, e.g.,  Figure 1(b) of Trager et al.\ (2000b), except that 
here the Balmer index is on the x-axis.  A striking
feature of Fig.~\ref{fig:Hn4383_4plot} is that the two groups with low $\sigma$ 
have a lower mean
age than does the highest $\sigma$ group (log $\sigma\ge$2.2),
while the galaxies with 2.0$<$log $\sigma$$<$2.2 show an intermediate
distribution in age.  In short, the increased scatter in Hn/Fe seen among the
low $\sigma$ galaxies (in Fig.~\ref{fig:Hn4383_4plot})
is indeed reflected in an increased trend towards younger ages.  In 
Table~\ref{tab:agesum} we summarize data on the mean age, and dispersion in age,
for each of the four log~$\sigma$ groups.  The age dispersions
listed there have not been corrected for observational error.  Note that the
rms scatter in age is actually {\it lower} among the low $\sigma$ galaxies
than for the high $\sigma$ galaxies, which may seem surprising, given the greatly
{\it increased} rms scatter in Balmer line strength for the low $\sigma$ galaxies.  
This result simply reflects the highly non-linear dependence of Balmer line
strength on age, coupled with the pronounced trend to younger mean ages among
the low $\sigma$ galaxies.

Another important aspect of Fig.~\ref{fig:Hn4383_4plot} is the fact that 47 Tuc
lies separated from virtually all galaxies in the diagram.  Clearly, no
galaxy in our sample is similar to 47 Tuc, which supplies a
well-defined empirical anchor point at an age of $\sim$12-13 Gyr 
\citep{lc00,zo01,gsa02} and [Fe/H]$\sim$-0.7 - -0.8 (\citep{bw92,cg97}).  
The large discrepancy between isochrone-based
SSP models and the location of 47 Tuc has been pointed out in previous work (Gibson et al.
1999).  Through a combination of the inclusion of $\alpha$-element enhancement and
diffusion to the models \citep{va01a}, along with adjustments to the 
T$_{eff}$ scale for metal-poor giants and, most importantly, to the failure of
model isochrones to produce sufficient numbers of stars above the HB level (Schiavon et al.
2002a,b), the most recent models now reproduce 
the position of 47 Tuc in many diagnostic diagrams.  However, the key point is
that, independent of zero-point problems with the models, no galaxies are
found with the old age and low metallicity of 47 Tuc.  Despite the
fact that 47 Tuc shares the same B-V color as a typical E/S0 galaxy (e.g., M32) 
their mean ages
and chemical compositions are quite distinct.  The systematic
differences in absorption line indices between 47 Tuc 
and M32 have been previously analyzed by \citet{ro94} and \citet{rd99}.  

While the mean age and the scatter in age vary substantially with velocity
dispersion, there is, on the other hand, relatively little change in metallicity,
as designated by the iron-sensitive Fe5270 feature.  It is also worthy of note
that M32 (designated by the large square) is found among the highest
[Fe/H] objects in the sample.  One should bear in mind that in M32 we are
sampling only the central $\sim$10 pc of the galaxy, thus it is not clear how
relevant the comparison is with the more widespread central region spectra of
the other galaxies in the sample.

To quantify the appearance of different mean ages for the four log~$\sigma$
groups in Fig.~\ref{fig:Hn4383_4plot}, we have applied the
Kolmogorov-Smirnoff two-sample test 
to various combinations of the four log $\sigma$ bins. The results are summarized in
Table~\ref{tab:ks_table}.  The two discrepant data points in the $2.0 < $ log~$\sigma$ $< 2.2$
group have been excluded from the test.
 With the one exception of the two lowest $\sigma$ groups, the 
likelihood that the age distributions of any two log $\sigma$ groups are drawn from the
same parent population is small.  In the case of the two lowest $\sigma$ bins, the 
probability is high that their age distributions are similar.

\begin{figure}
\plotone{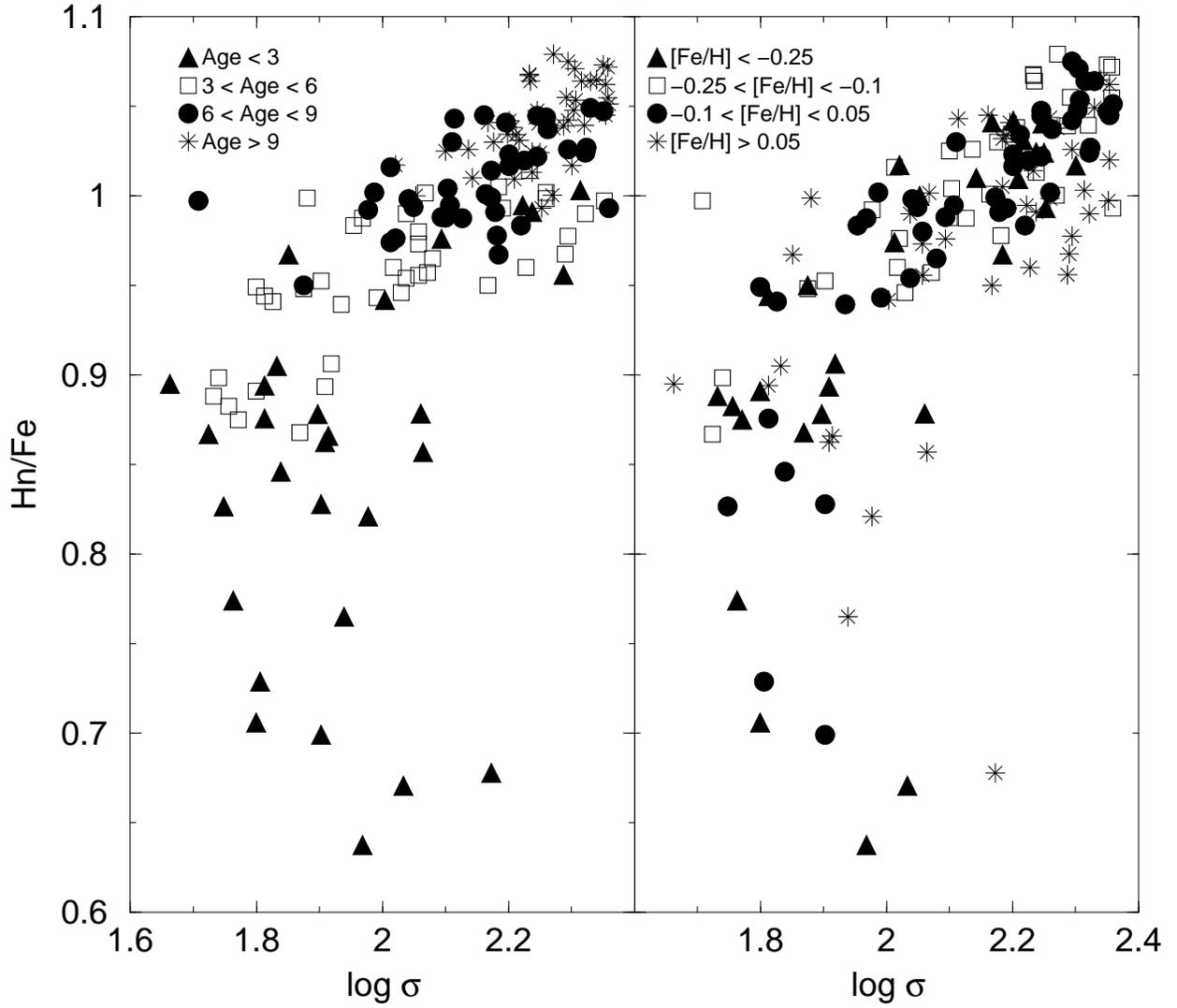}
\caption{The derived ages of galaxies in the Hn/Fe-log$\sigma$ diagram.
The left panel shows galaxies coded by age intervals.  In the
right panel, the coding is by [Fe/H].}
\label{fig:HnFe-agemet}
\end{figure}

\begin{figure}
\plotone{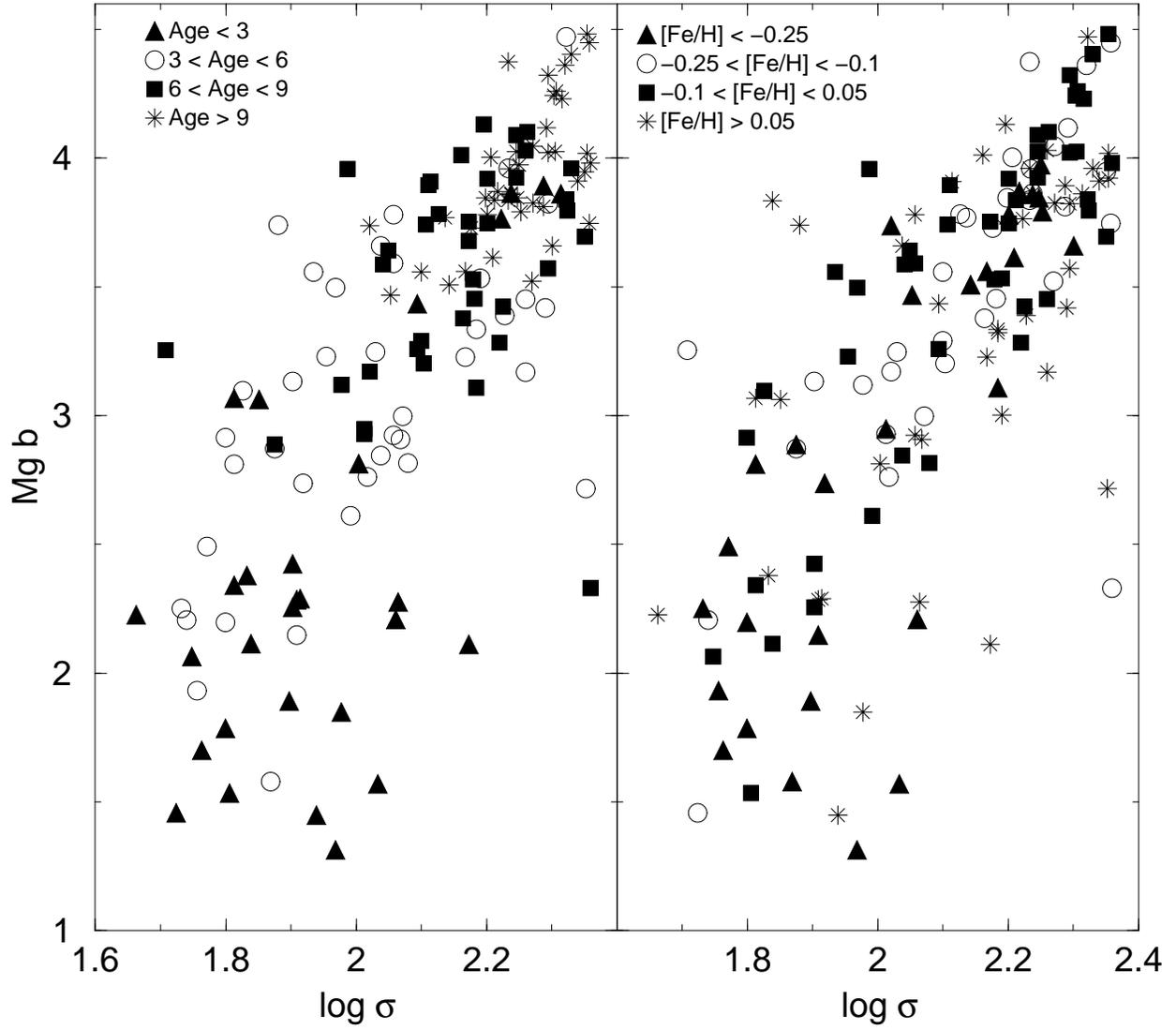}
\caption{The derived ages of galaxies in the Mg~$b$-log$\sigma$ diagram.
The symbols are the same as in Fig.~\ref{fig:HnFe-agemet}}
\label{fig:Mgb-agemet}
\end{figure}

To further emphasize that the intrinsic scatter in Balmer line strength found
in Fig.~\ref{bces}, especially at low log $\sigma$, is primarily due to a
spread to younger ages,
we have replotted the correlation between Hn/Fe and log $\sigma$ and that 
between Mg~$b$ and log $\sigma$ in Figs.~\ref{fig:HnFe-agemet} and 
\ref{fig:Mgb-agemet},
respectively. In the left hand panels we have subdivided the galaxies into 
different age bins, where the ages come directly from interpolating each galaxy
data point between the age-metallicity grid lines in the Fe4383 versus Hn/Fe
diagram (Fig.~\ref{fig:Hn4383_4plot}).  The right hand panels show the galaxies
coded by metallicity, where the [Fe/H] values also come from interpolation in
Fig.~\ref{fig:Hn4383_4plot}.  
It is certainly evident from Fig.~\ref{fig:HnFe-agemet} that the low $\sigma$ 
galaxies with
strong Balmer lines (low Hn/Fe) have young ages as well.  In contrast, there
is no clear trend with metallicity.  In the case of the Mg~$b$ index, plotted in
Fig.~\ref{fig:Mgb-agemet}, a trend still exists in the sense that at a given 
log $\sigma$, a lower Mg~$b$ correlates with lower age. However, the effect is 
now subtler for this metallicity feature than for the more age-sensitive Balmer
index. Although we do not show them here, similar results are found for plots
 of both
H$\beta$ and Fe5270 versus log $\sigma$, in that H$\beta$ closely follows the 
trends seen in Hn/Fe while the behavior of Fe5270 is similar to that of Mg~$b$.

While the above discussion strongly indicates that there is a considerable
spread in age among early-type galaxies, in particular a trend to young ages
at low $\sigma$, two 
questions need to be addressed at this point.  First, could some type of
systematic effect lead to a spurious conclusion of young ages in galaxies?
Second, the ages we have presented are based on a single pair of spectral
indices; are consistent results obtained with other spectral indicators?
We first consider
several systematic effects that could, in principle, lead to incorrect age
determinations, and then in \S6 we evaluate the degree to which different
Balmer and metal features produce consistent results.

\subsection{Systematic Effects on the Integrated Spectra of Galaxies}
\label{sec:system}

\subsubsection{A Metal-Poor Component in Early-Type\\ Galaxies?}
\label{sec:metalpoor}

It has been pointed out in the past that the inability to model the integrated
spectrum of early-type galaxies such as M32 with a single old-age SSP does not
necessarily imply the existence of intermediate-age stars.  Instead,
one can argue for multiple metallicity components in these galaxies.  This
approach has been advocated most recently by Maraston \& Thomas (2000), who
propose that the enhanced Balmer absorption lines found in many early-type
galaxies can be reproduced by adding in an old metal-poor population, rather
than resorting to intermediate-age populations.  Since it is indeed crucial
to distinguish between the intermediate-age and metal-poor population scenarios,
we now address that degeneracy.  

As has been described in Rose (1985, 1994), Leonardi \& Rose (1996), Leonardi \&
Worthey (2000), and
\cite{lr02}, the line ratio index composed of the Ca II H +
H$\epsilon$ and Ca II K lines (hereafter referred to as the Ca II index)
provides a sensitive test for the presence of
early-type stars in an integrated spectrum, whether they come from young
upper main sequence stars, blue stragglers, horizontal branch stars, or
main sequence turnoff stars from a metal-poor population.  The key idea is
that for all cool stars the Ca II H and K line cores are saturated and
insensitive to temperature, thus the line ratio is constant in cool stars.
Only in stars hotter than early F does this ratio change, once the line cores
are no longer saturated, and, once H$\epsilon$, whose line center coincides with
Ca II H, also begins to make a significant contribution.
Thus the Ca II index provides a sensitive lever for detecting small amounts
of hot stars in the integrated light.  

\begin{figure}
\plotone{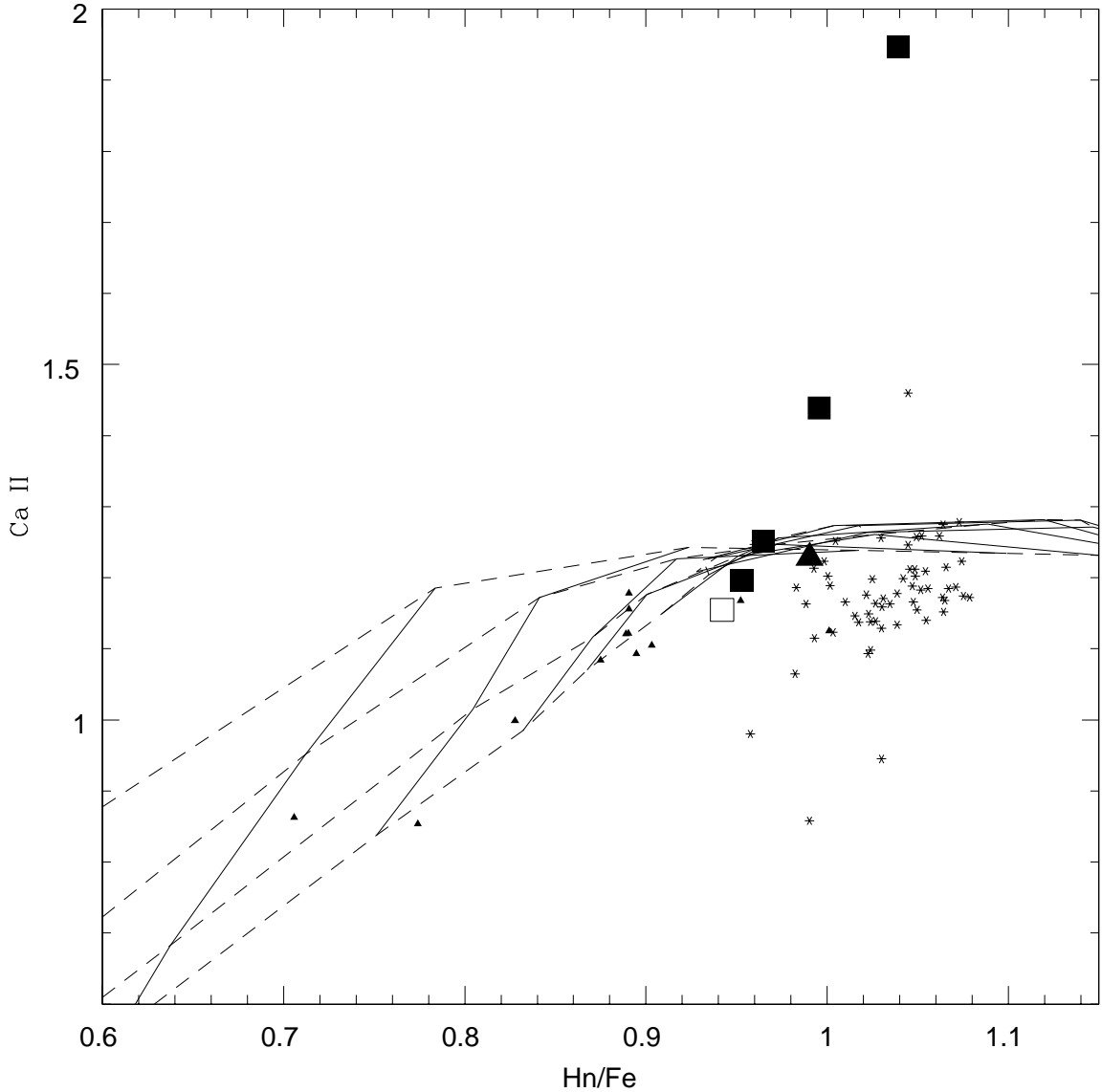}
\caption{The Ca II index is plotted against the Hn/Fe index.  The highest and
lowest 
$\sigma$ galaxies are plotted as small asterisks and filled triangles,
respectively. M32 and 47 Tuc are plotted as a large
open square and a large filled triangle, respectively.  The filled squares, at
increasing distance from M32, represent the resultant indices when the spectrum
of M15 is subtracted from that of M32 with
contributions of 5\%, 10\%, 20\%, and 30\% at 4000 \AA.  See text for details.
The grid lines connect the simple stellar populations
from the Worthey models of various ages and metallicities.  Solid lines connect
models of constant age; while
dashed lines connect models of constant
metal-abundance. Note that both axes place the strong Balmer absorption
at low index values, and are plotted in the same manner as in our previous
papers that used these indices.}
\label{fig:HnFevCaII_M15}
\end{figure}

In Fig.~\ref{fig:HnFevCaII_M15} we plot the Ca II index versus Hn/Fe, for the
highest and lowest $\sigma$ galaxies
in our sample (in the previous references, Ca II has been plotted against 
H$\delta$/Fe4045, rather than Hn/Fe), along with the model age and metallicity
grid lines.   Note that the Worthey models all arrive
at an asymptotic value of $\sim$1.2 for old age populations, independent of 
metallicity.  The globular cluster 47 Tuc indeed lies just slightly below this
old age value, which is consistent with a $\sim$1-2\% contribution from
blue stragglers at 4000 \AA\ predicted by the
color-magnitude diagram of its core region (Howell, Guhathakurta, \& Gilliland
2000; Ferraro et al. 2001).  On average, even the oldest galaxies in our sample
are displaced by $\sim$0.05 below the old-age value, and the previously 
identified young galaxies have low Ca II indices, consistent with ages
below 2 Gyr. 

As was pointed  out by Rose (1985) in the case of M32, the 
$\sim$0.05 discrepancy between the Ca II index of M32 and the cool star value
can be accounted for by introducing either a small component of very young stars, 
blue stragglers, or a metal-poor component.  In the latter case, which appears
to be most plausible, given the inevitable existence of a metal-poor component
in a chemically enriched galaxy, the Ca II discrepancy in M32 can be reproduced
by adding in a $\sim$7.5\% contribution to the integrated light at 4000 \AA \ from
a metal-poor globular cluster spectrum.  In fact, such a contribution is 
expected in the case of a closed box model (Rose 1985) and is in concert with
the near-UV synthesis of M32 performed by Kjaergaard (1987).
We reproduce the effect of this metal-poor contribution in 
Fig.~\ref{fig:HnFevCaII_M15}.  The large squares show
the result of subtracting off varying amounts of light from M32 in the form of
an integrated spectrum of the metal-poor Galactic globular cluster M15,
smoothed to the same resolution as the FAST data.
Specifically, the sequence of squares, moving away from the open square
representing M32, are the results of subtracting 5\%, 10\%, 20\%, and 30\% of
the light respectively.  The key point is that if one tries to weaken the Balmer
lines of M32 by removing the metal-poor component, one very quickly comes into
conflict with the old-age value of the Ca II index.  That is, the Ca II index
is quickly affected by the subtraction of hot stars, because those stars have
unsaturated Ca II lines. 

Removing $\sim$7-8\% of the
light of M32 in the form of an M15 spectrum brings the Ca II index in concert
with that of 47 Tuc, but only reduces the Hn/Fe index by a small amount (0.009).  To
displace M32 in Hn/Fe by the amount required to give it an old age in the
Fe4383 versus Hn/Fe diagram of Fig.~\ref{fig:Hn4383_4plot}, i.e., $\sim$0.08 in
Hn/Fe, requires well in excess of 30\% of the light coming from an M15
component, which is in strong violation of the Ca II criterion.

It has also been proposed recently that the strong Balmer lines in M32 and some other
early-type galaxies can be reproduced by adding in a component of extreme horizontal branch
(HB) stars, other hot stars in advanced evolutionary stages (De Propris 2000),
and/or a population of hot blue stragglers.  
Again, either scenario avoids the necessity for intermediate ages.  While we do not
reproduce the specifics here, Rose (1985) placed upper limits of only a few
percent of the light in M32 coming from $any$ type of hot population, again based
on the observed Ca II index.  In short, the stringent limit on hot star light 
placed by the Ca II index definitely rules out the possibility that the enhanced
Balmer lines in galaxies such as M32 can be ascribed to a population of hot 
stars, rather than to an intermediate age population.

In short, a small metal-poor contribution to the integrated light of early-type
galaxies is indicated by the position of these galaxies in the Ca II versus
Hn/Fe diagram.  Removal of this apparently uniform contribution of $\sim$7\%
at 4000 \AA \ will slightly increase the derived ages for all of the galaxies.
However, the inferred age increase is small, and does not account for the
galaxies, such as M32, with intermediate and young ages.  The amount of age
adjustment, naturally, is a function of the age of the galaxy, since the age
lines crowd together at higher age, and also the contribution of the underlying
metal-poor component will vary with age, due to the systematically brighter 
main-sequence turn off (MSTO) of a younger component.  In the case of M32, the
adjustment amounts to $\sim$1~Gyr.  Because the emphasis in this paper is on the age 
$distribution$, rather than $zero \ point$, and because correcting for a uniform
metal-poor component will result in a systematic age shift, thus hardly
affecting the age distribution, we do not choose to make that correction here.

\subsubsection{Contamination from Red Horizontal Branch Stars?}

\begin{figure}
\plotone{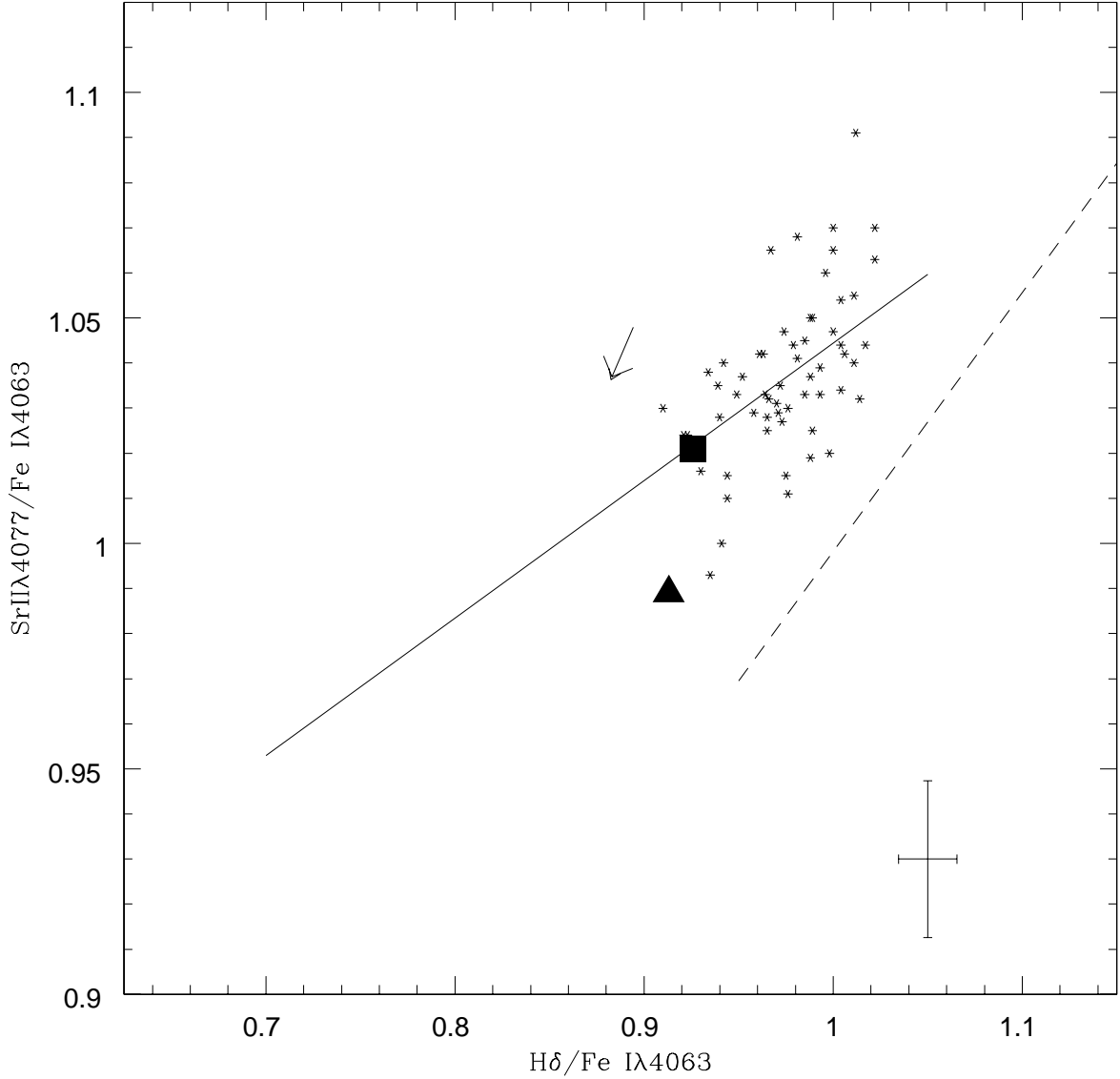}
\caption{The SrII4077/FeI4063 index is plotted versus the H$\delta/4063$ index.
High-$\sigma$ galaxies are plotted as small asterisks, while M32 and 47 Tuc are
plotted as a large filled square and  triangle, respectively.  The solid line
represents the mean relation for dwarf stars, while the dashed line represents
the mean relation for giants.  The left arrow indicates the direction of
correlated errors in the plot.  The error bars in the lower right corner 
indicate the average $\pm$1$\sigma$ errors in the galaxy indices. Note that the
lower values of the Balmer sensitive H$\delta/4063$ index, as well as the gravity 
sensitive index SrII4077/FeI4063 indicate early stellar spectral types. The diagram
is plotted in the same manner as previous papers.}
\label{fig:Hd63vSr63}
\end{figure}

The enhanced Balmer lines which lead to young age estimates for galaxies such as
M32 in metal versus Balmer line diagrams can in principle be due to a population
of red horizontal-branch (RHB) stars.  These stars would be cool enough to have the Ca II 
index of a typical late-type star, but hotter than the MSTO of an old population.  For
example, the bluest RHB stars in 47 Tuc approach these characteristics.  To
assess this possibility, we rely on spectral indices, described extensively
by Rose (1985, 1994), which are sensitive to surface gravity, and hence can be
used to determine the relative contribution of dwarf and evolved stars to the
integrated spectrum of a galaxy.  Specifically, the two spectral indices used
form the ratio of the singly ionized Sr II$\lambda$4077 feature to the
neutral $\lambda$4063 Fe I feature (the SrII$\lambda$4077 index) and the 
ratio of H$\delta$ to Fe I$\lambda$4063 (the H$\delta/\lambda$4063 index).
In Fig.~\ref{fig:Hd63vSr63} we plot a sub-sample
of galaxies in the surface gravity dependent 
SrII$\lambda$4077 versus H$\delta/\lambda$4063 diagram.  The solid and dashed 
lines represent the mean relations for dwarfs and giants respectively, 
determined from the behavior of stars in the Coud\'e Feed spectral library.  
As has been
discussed in Rose (1985, 1994) the separation of dwarfs and giants in this
diagram is independent of metal-abundance, except in the case of the most
metal-deficient stars ([Fe/H]$<$-1.5).  As can be seen in Fig.~\ref{fig:Hd63vSr63},
the Galactic globular cluster 47 Tuc lies between the dwarf and giant sequences, in
fact close enough to the giant sequence to indicate that $\sim$2/3 of its
integrated light at 4000 \AA \ comes from evolved stars, which is in accord with
its observed color-magnitude diagram (Schiavon \etal 2002a,b).  In contrast, M32 lies close to the
dwarf sequence, thereby ruling out the possibility that RHB stars are making a
significant contribution.  Hence the enhanced Balmer lines are indeed due to
an intermediate age population.  It is also evident in Fig.~\ref{fig:Hd63vSr63} that
the other galaxies in the sample are clustered around the dwarf relation, thus
the same conclusion should be drawn that RHB stars are not an important issue
in early-type galaxies.  Note that since the Fe I$\lambda$4063 line is used in
both indices, we can expect the galaxy index errors in the plot to be 
correlated.  The left arrow in the plot shows the expected direction of the
correlated errors.

\subsubsection{The Effect of Emission Corrections}

Contamination of the stellar absorption spectrum by emission is clearly a potential source
of systematic error in age determinations.  The most serious effect of a
superposed emission spectrum is to fill in the Balmer absorption lines, thereby
causing the galaxy age to be overestimated.  Thus, varying amounts of emission
fill-in, if not accounted for, can lead to a spurious spread in derived ages,
in the sense that an emission-induced spread to older ages is produced.  Note
that this effect will only produce younger ages if an {\it overcorrection} is
made for emission.

\begin{figure}
\plotone{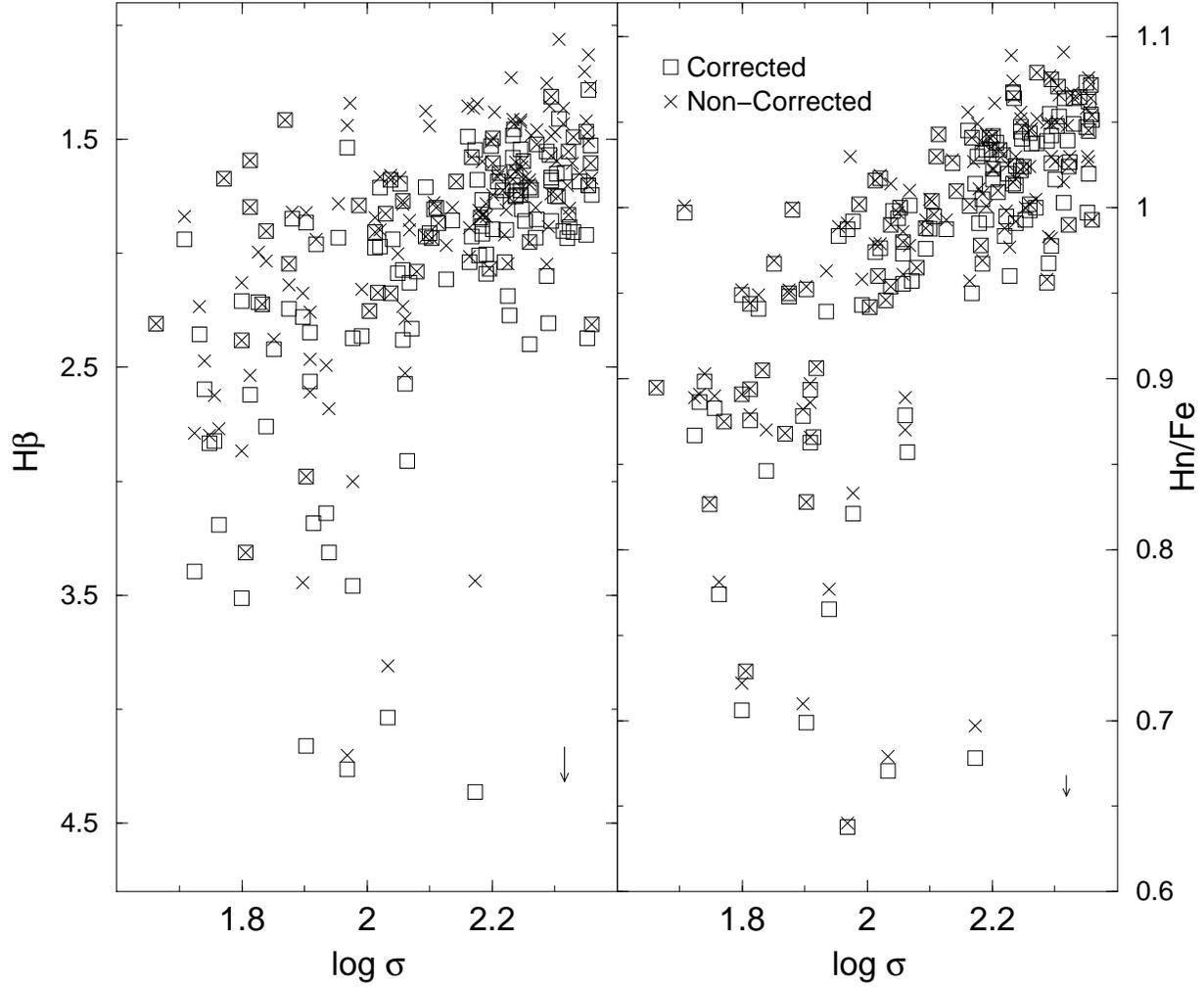}
\caption{The effects of correcting the spectral indices for emission fill-in, 
shown for two Balmer line indices, the Lick H$\beta$ index (left panel, note
that H$\beta$ is plotted so that strong absorption is at the bottom) and the 
Hn/Fe line ratio index (right panel), both plotted versus log $\sigma$.  
The original index values are shown 
as x's and the corrected
values are shown as open squares. The mean index corrections are 0.15 for 
H$\beta$ and 0.012 for Hn/Fe, and are denoted by arrows in the lower right
corners of the panels.}
\label{fig:hbhn.newcor}
\end{figure}

In \S3.3 and in Appendix A we discussed the strategy for identifying and removing the effect of 
emission in our sample of early-type galaxy spectra.  As described there, we
either remove the emission, based on a determination of H$\alpha$ emission
from supplementary red spectra, or use the [OII]$\lambda$3727 line, along with an
assumed [OII]$\lambda$3727/H$\alpha$ ratio, to estimate the emission correction.  The
results of removing emission contamination of the Balmer line indices are
plotted in Fig~\ref{fig:hbhn.newcor}, where the galaxy data is plotted both
before and after emission correction for both H$\beta$ (left panel) and Hn/Fe
(right panel) indices.  As expected from the steep Balmer decrement in emission,
the average emission correction is larger
in H$\beta$ than for the Hn/Fe index, but the corrections are modest in both
cases, and certainly do not explain the young ages seen at low $\sigma$.  An
important point, to which we return in \S\ref{sec:inscatter}, is that at the
high log~$\sigma$ end, both the mean H$\beta$ at a given log~$\sigma$ and the
scatter in H$\beta$ are significantly affected by the emission corrections.
Hence making the H$\beta$ emission corrections accurately is crucial for 
determining the intrinsic scatter in age for the high $\sigma$ galaxies from
that index.

\subsubsection{The Effect of NSAR Corrections}
\label{sec:nsarcor}

\begin{figure}
\plotone{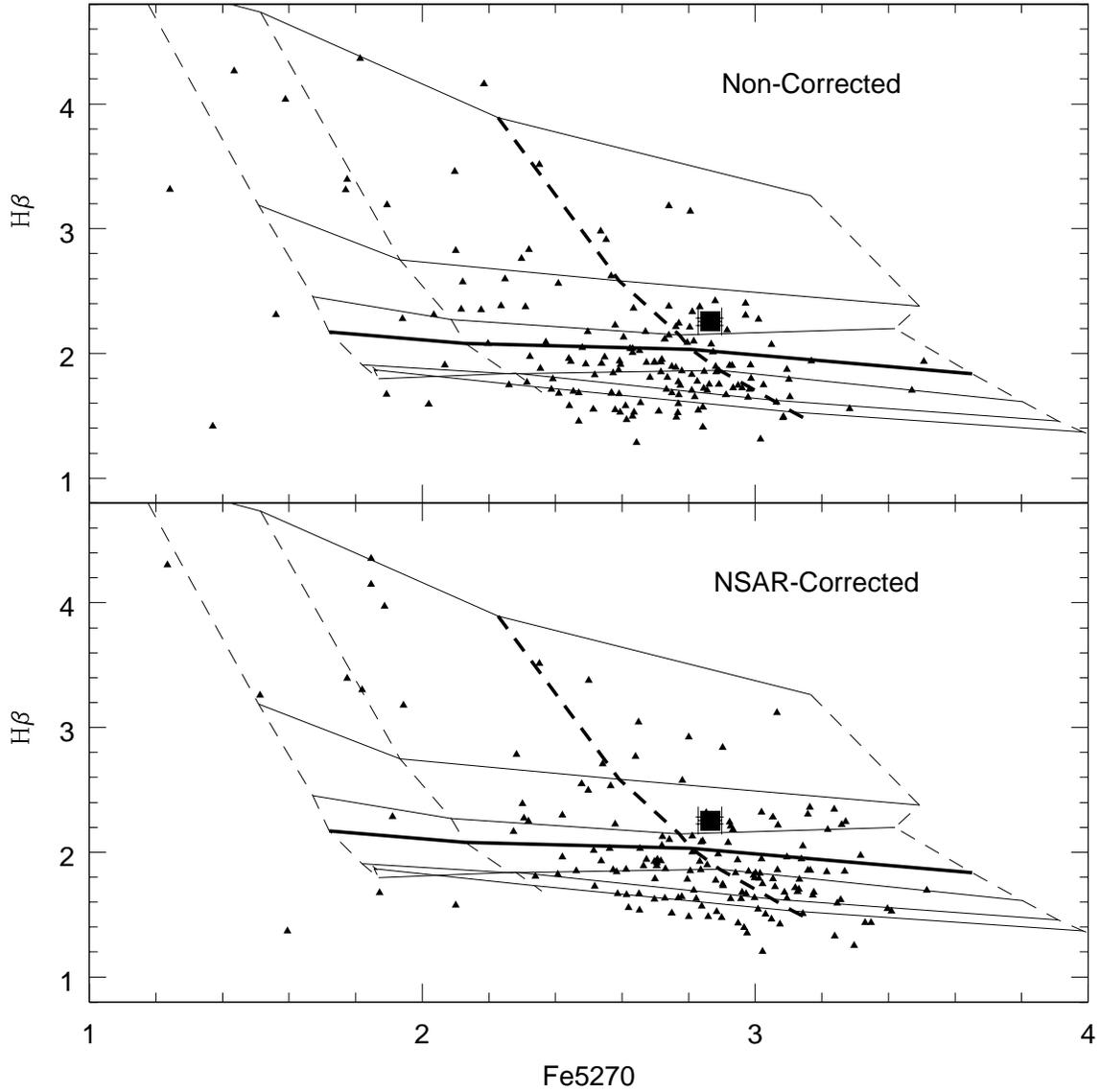}
\caption{The effects of correcting the H$\beta$ and Fe5270 indices for NSAR.
The upper and lower panels show the galaxy data (filled triangles) before and 
after NSAR corrections, respectively.  In both panels the data have already 
been corrected for emission.  The large square with error bars represents M32.
The lines of constant age and metallicity are the same as in 
Fig.~\ref{fig:Hn4383_4plot}.}
\label{fig:nsnew_Hb5270}
\end{figure}

As was mentioned in \S\ref{sec:nsar} and fully described in Appendix B, 
the chemical composition of early-type galaxies deviates
from the element abundance ratios found in the Galactic stars that
comprise our empirical stellar database. The effects of non-solar abundance
ratios (NSAR) on the various
spectral indices used to characterize the galaxy integrated spectrum
must be accounted for if reliable ages and heavy element abundances are to be 
extracted.  In Figure~\ref{fig:nsnew_Hb5270}, the effects of applying NSAR 
corrections to the Fe5270 versus H$\beta$ plot are shown.  We have followed the
prescription described in Appendix B, which follows that of \cite{tr00a}
and \cite{ku01}.  The primary effect,
as expected, is to increase the overall heavy element abundance derived from the
Fe5270 index, since iron is found to be underabundant, relative to magnesium,
in early-type galaxies.   We do not have prescriptions for correcting either the
Fe4383 or Hn/Fe indices for NSAR effects.  For both of these indices, the 
primary metal sensitivity is due to an Fe~I absorption line, hence the derived
abundances from the Fe4383 versus Hn/Fe diagram are primarily [Fe/H].  Therefore, 
we will use this index pair to measure only [Fe/H], and not [Z/H]. On the
other hand, the abundances derived from the NSAR-corrected Fe5270 versus H$\beta$
diagram reflect overall heavy element composition, thus are [Z/H].

Hereinafter, we will explicitly
use the term {\it [Z/H]} (derived from the grid of Fe5270 vs Hn/Fe or H$\beta$)
to refer to the mean abundance of all the heavy elements,
which have the solar abundance pattern.  We will use {\it [Fe/H]} to refer specifically
the iron abundance (derived from the grid of Fe4383 vs Hn/Fe), which 
may not reflect [Z/H] precisely, because of the presence
of NSAR. We will continue to use {\it metallicity} in a looser fashion, to refer
either to [Fe/H] or [Z/H], depending on the context.

In Table~\ref{tab:indcor} we summarize the effects on the derived mean ages and 
metallicities for our galaxy sample of both emission  and NSAR corrections.
For ages derived using the Hn/Fe index, the age shifts due to emission are 
fairly small,
with the mean of the total sample changing from 8.9 to 8.2 Gyr.  For H$\beta$
derived ages, the mean shift in age due to emission correction is somewhat 
larger, from 11.1 to 9.3 Gyr.  The change in the metallicity
determinations due to the emission correction is +0.03 dex in the H$\beta$-Fe5270 grids, 
and is negligible (+0.002 dex) in the Hn/Fe versus Fe4383 grids.
As expected, the NSAR corrections change the mean derived metallicity from the
Fe5270 versus H$\beta$ diagram, from [Fe/H]=-0.15 to [Z/H]=-0.03, while
the derived mean age is essentially unaffected by the NSAR corrections.
In addition, the mean galaxy age extracted from the emission-corrected, but not 
NSAR-corrected, Hn/Fe versus Fe4383 (8.3 Gyr) is quite consistent with that 
derived from the emission and NSAR-corrected H$\beta$ versus Fe5270 (9.3 Gyr). 
Hence we infer that the lack of NSAR correction to Hn/Fe and Fe4383 does not 
produce a large bias in derived ages.

\section{Consistency of Age and Abundance Determinations}

At this point, a case has been made that the luminosity-weighted mean ages of 
early-type galaxies are substantially younger for low-$\sigma$ galaxies, 
particularly those with log $\sigma$ $\leq$ 2.0.  In addition, an intrinsic 
scatter in several key spectral indicators, including those measuring
Balmer line strength, appears to be present at {\it all} $\sigma$.  We have
considered and eliminated a variety of effects, other than mean age variations,
that can explain the large variation in Balmer line strength observed in the low $\sigma$
galaxies.  In this section we take up the issue as to whether consistent age
and metal abundance determinations are obtained from a variety of different
spectral indicators.  We are especially concerned with whether the higher order
Balmer lines (H$\gamma$, H$\delta$, and H8) give similar results to that 
obtained with the Lick H$\beta$ index, whose characteristics as an age indicator
have been well established.

\begin{figure}
\plotone{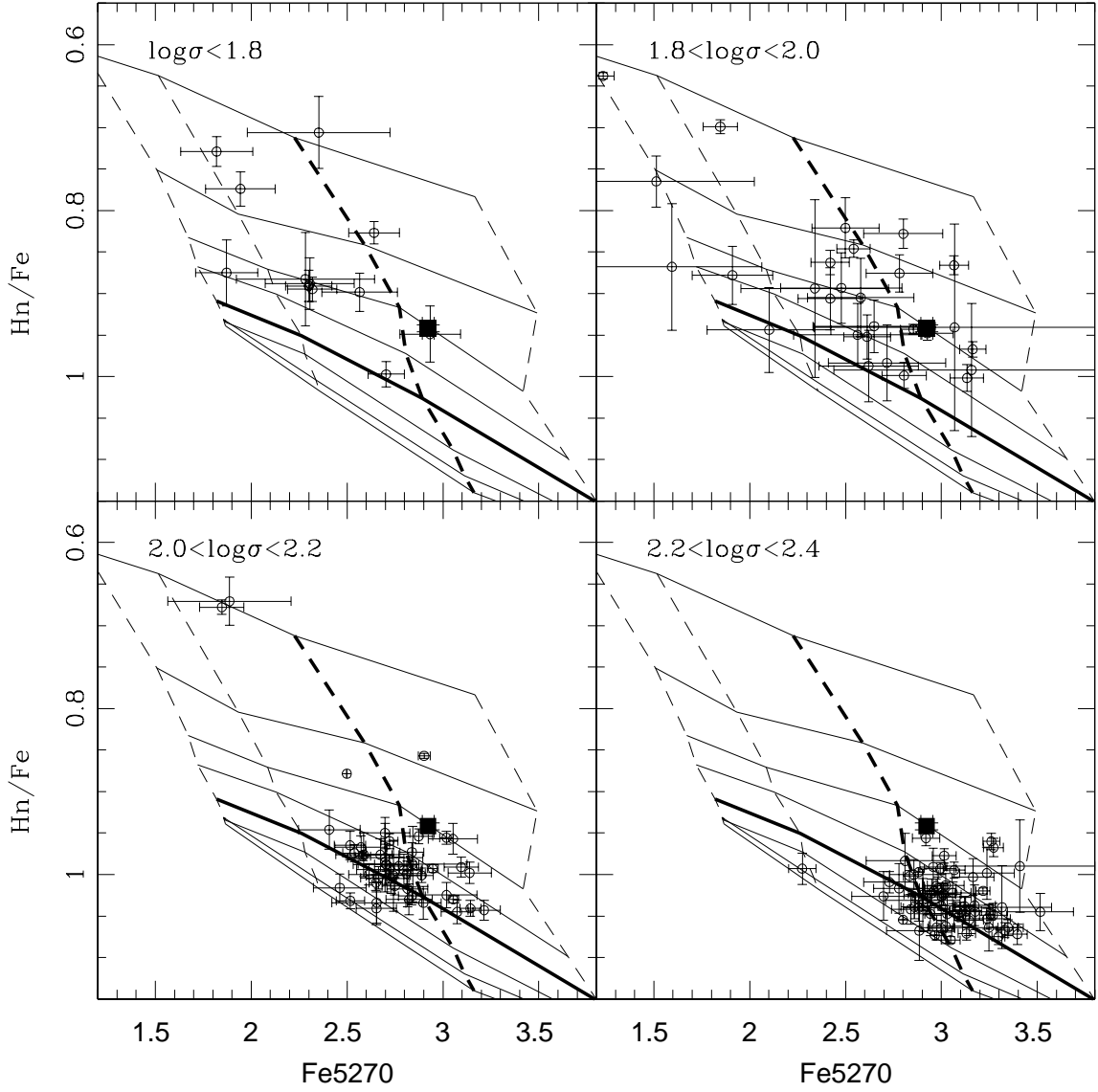}
\caption{The emission corrected
Hn/Fe Balmer line index is plotted versus the Lick Fe5270 index (which 
has been NSAR corrected). 
Symbols and grid lines are the same as in Fig.~\ref{fig:Hn4383_4plot}.}
\label{fig:Hn5270_4plot}
\end{figure}

In the preceding section we showed that early-type galaxies cover a large
range in age when compared with model grids in the Fe4383 versus Hn/Fe diagram.
To demonstrate that the basic results are not strictly dependent on the choice
of metal feature, in Fig.~\ref{fig:Hn5270_4plot} we plot the Lick Fe5270 index 
versus Hn/Fe.  As can be seen by comparing with Fig.~\ref{fig:Hn4383_4plot}, 
the basic characteristic of the Fe4383 versus Hn/Fe diagram, i.e., a trend to
young ages among the lower $\sigma$ galaxies, is reproduced in the F5270 versus
Hn/Fe diagram.  Again, M32 is located among galaxies of the highest abundance.

\subsection{Comparison Between Higher Order Balmer Lines and H$\beta$}

\begin{figure}
\plotone{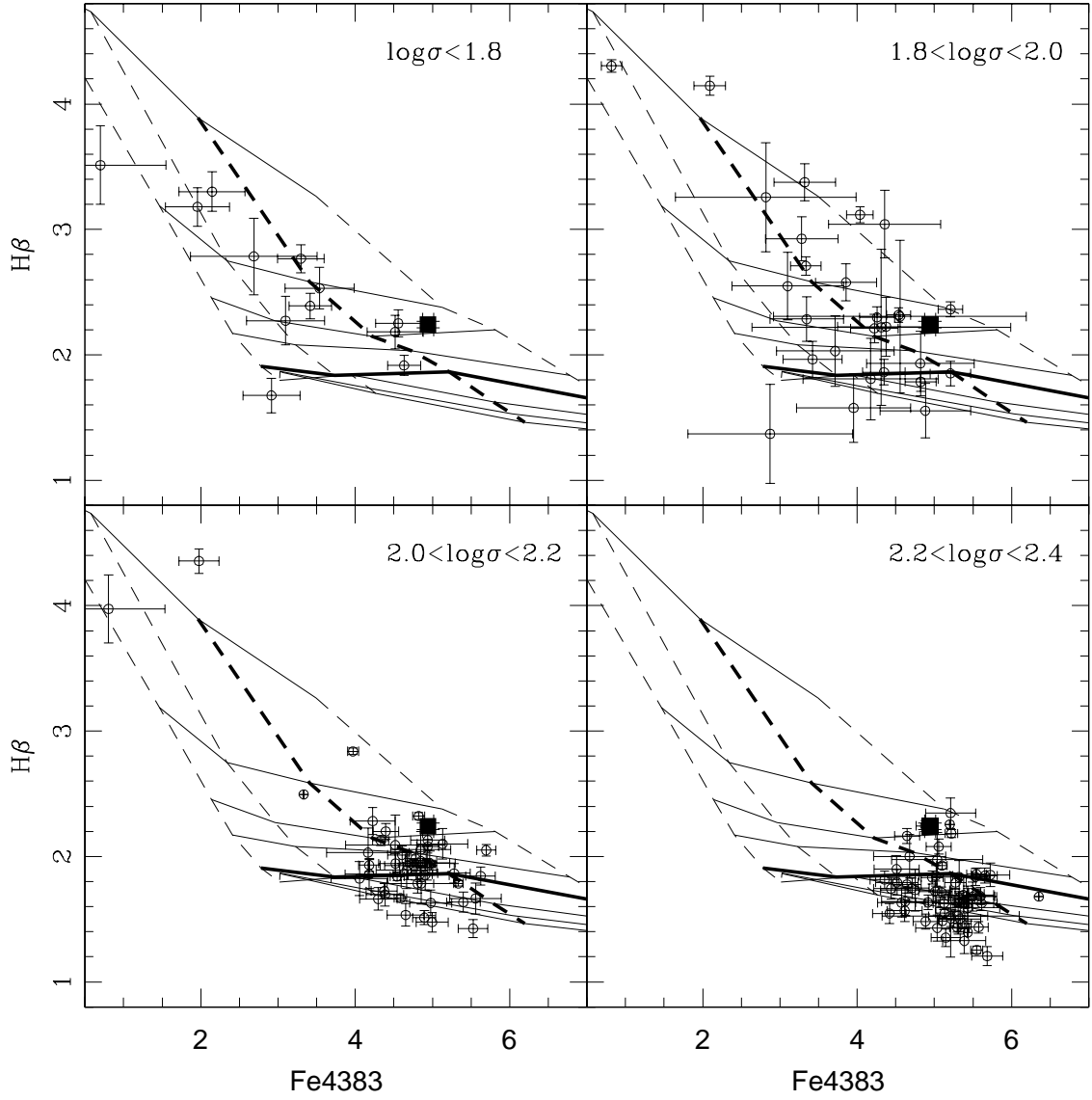}
\caption{The Lick  H$\beta$ index is plotted versus the Lick Fe4383 index.
H$\beta$ has been corrected for NSAR and emission.
Symbols and grid lines are the same as in Fig.~\ref{fig:Hn4383_4plot}.}
\label{fig:Hb4383_4plot}
\end{figure}

\begin{figure}
\plotone{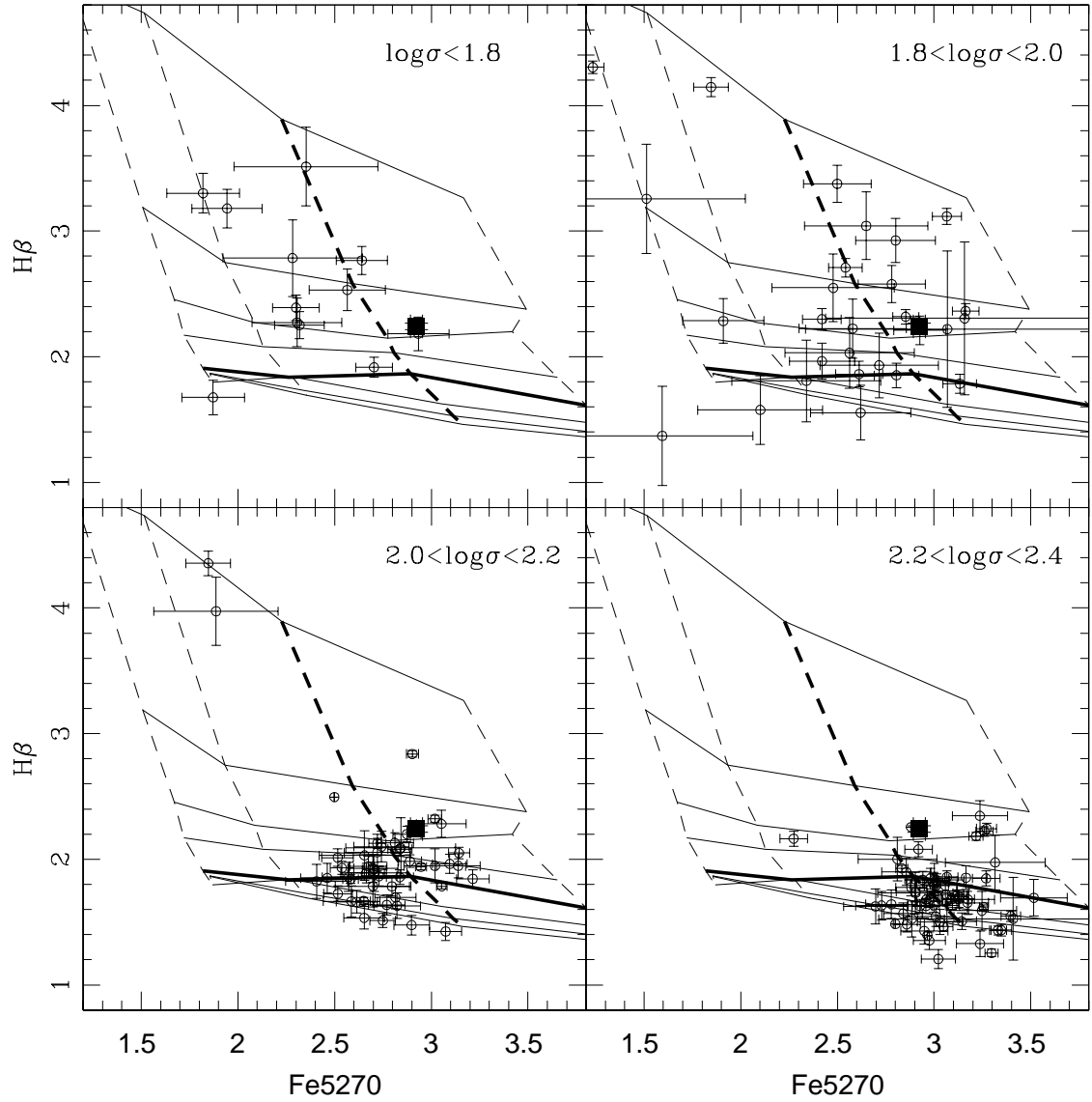}
\caption{The Lick H$\beta$ index is plotted versus the Lick  Fe5270 index.
H$\beta$ has been corrected for emission, and both indices have been
NSAR corrected.
Symbols and grid lines are the same as in Fig.~\ref{fig:Hn4383_4plot}.}
\label{fig:Hb5270_4plot}
\end{figure}

A more stringent test of the age-metallicity determinations is to
repeat the analysis for different Balmer lines, as well as for alternative
measurements of a given Balmer feature.  A natural starting point is to compare
the results obtained from the higher order Balmer lines, as measured by the 
Hn/Fe index, with that obtained from the Lick H$\beta$ index, which has been
extensively modeled and utilized in the literature (e.g., Trager \etal 2000a,b), 
and hence represents a benchmark for all age 
determinations.  The galaxy data is compared with model grids in the Fe4383
versus H$\beta$ and Fe5270 versus H$\beta$ diagrams in 
Figs.~\ref{fig:Hb4383_4plot} and \ref{fig:Hb5270_4plot}, respectively.
The H$\beta$ data has been corrected both for emission and NSAR effects, and
Fe5270 has been corrected for NSAR.
There are several aspects of the H$\beta$ plots which are noteworthy.  First,
the same qualitative features found in the higher order Balmer lines, as
represented by Hn/Fe, are clearly present in both H$\beta$ plots.  Most
important, the general trend to young ages for lower $\sigma$ galaxies is
reproduced.  Second, the model grid lines in H$\beta$ plots, especially in
the Fe5270 versus H$\beta$ diagram, show a better orientation of isochrone
and iso-metallicity lines along the index axes.  This is due to the fact that
the Lick H$\beta$ index is an equivalent width measure of that feature, while
Hn/Fe is a relative measure of a Balmer line depth relative to an Fe line depth.
Hence the H$\beta$ index more directly disentangles age from metallicity 
effects.  A third feature of the H$\beta$ diagrams is that we find a greater
number of galaxies lying outside the model grids lines at the oldest ages,
which is strictly a zero point issue in the modeling.  However, as a result
it is easier to define a complete sample of ages and metallicities
from the Hn/Fe diagrams, especially for 4383 versus Hn/Fe, in which only four
galaxies lie in a region of the diagram for which no age/metallicity 
extrapolation is possible.

From the four combinations involving the two metal lines, Fe4383 and Fe5270,
and the two Balmer line indices, Hn/Fe and H$\beta$, we have produced four
ages and metallicities for each galaxy.  The results of these age and 
metallicity
determinations are compiled in Table~\ref{tab:gal_ageZ}. The first column
gives the galaxy ID, while columns (2) - (5) give the ages determined from
Fe4383 $vs$  Hn/Fe, Fe5270 $vs$ Hn/Fe, Fe4383 $vs$ H$\beta$, and Fe5270 $vs$
H$\beta$, respectively.  Columns (6) - (10) give the metal abundances determined
from those same four combinations of indices.  Errors in the ages and
abundances are listed
for the parameters derived from only the Fe4383 $vs$  Hn/Fe diagram, since
those for the other index combinations will be similar. These errors were derived
by finding the maximum age and metallicity differences implied by the 
index errors \& model grids.

\begin{figure}
\plotone{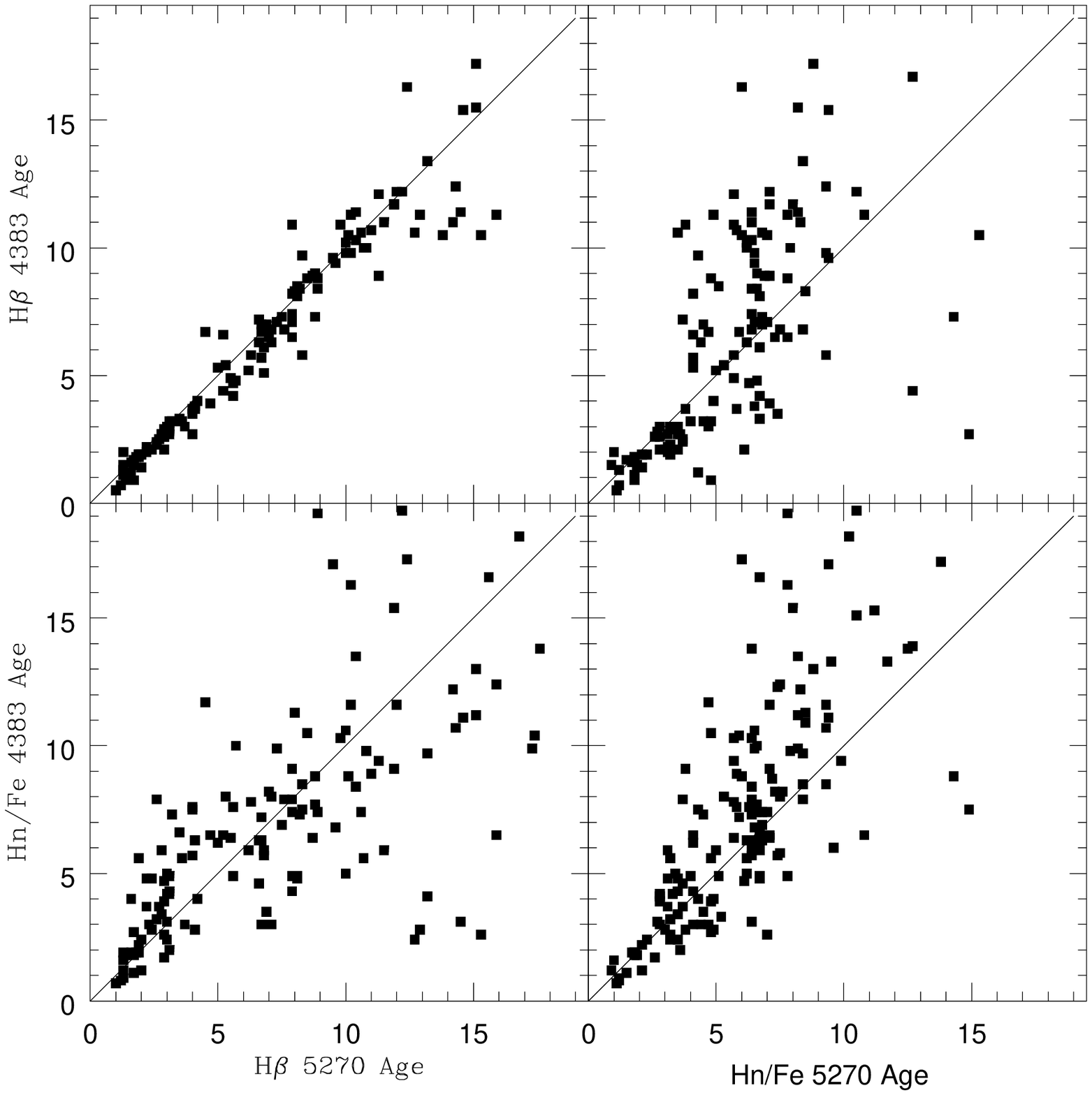}
\caption{A comparison is made between the derived ages for our galaxy sample
from four different model grids. The unity line is shown as a visual aid.}
\label{fig:HbHn.age}
\end{figure}

\begin{figure}
\plotone{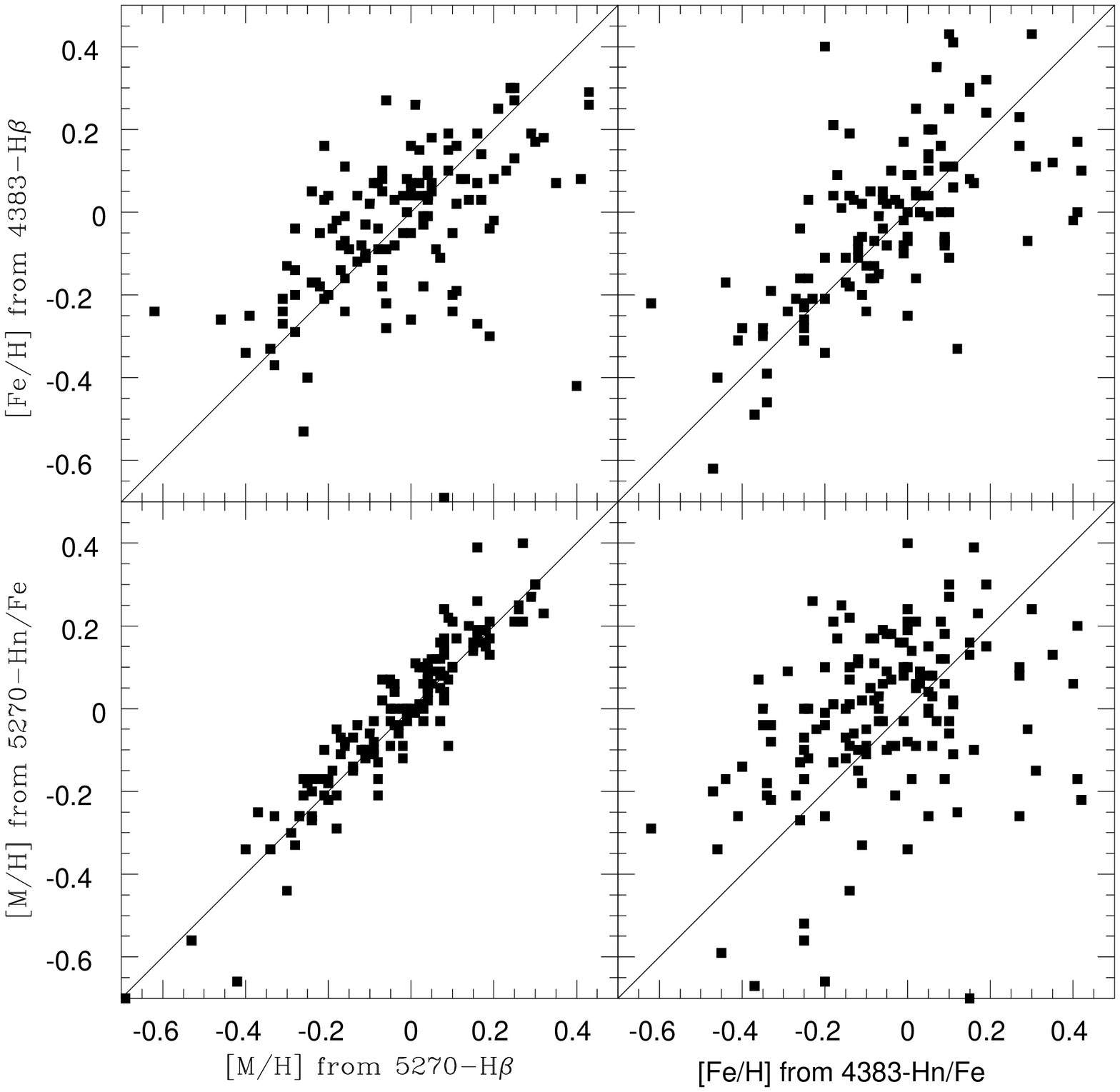}
\caption{A comparison is made between the derived metallicities for our galaxy 
sample from four different model grids. The unity line is shown as a visual aid.}
\label{fig:HbHn.met}
\end{figure}

To illustrate the degree to which
these individual age and metallicity estimates correlate with each other, in
Figs.~\ref{fig:HbHn.age} and \ref{fig:HbHn.met} we show correlation plots for 
several combinations of age and metals determinations.  The very tight 
correlation between ages derived from Fe5270 versus H$\beta$ and those 
determined from Fe4383 versus H$\beta$ again reflect the fact that age and
metallicity are so nearly decoupled in these diagrams.  Thus inconsistencies
between Fe4383 and Fe5270 do not translate into corresponding age uncertainties.
A more interesting comparison can be seen in the correlation between the
H$\beta$-Fe5270 ages and the Hn/Fe-Fe4383 ages, since these ages come from
completely independent sets of spectral indicators, with rather different
measurement techniques.  The correlation between the derived ages is quite good,
with an rms scatter around a straight line fit of 3.4 Gyr.  It is also clear 
that the age scale for Fe5270 versus Hn/Fe determinations is systematically
lower at large age than for the other determinations.  In the metallicity
correlation plots, the best correlation is achieved, not surprisingly, when the
same metal line is plotted versus the different Balmer indices.  Note, however,
that the correlation is still good between the [Fe/H] values extracted from
Fe4383 versus Hn/Fe and the [Z/H] values derived from Fe5270 versus H$\beta$.
In short, we find good overall consistency between age and abundance 
determinations based on the two distinct measures of Balmer line strength.

The previous results have demonstrated that the ages and
chemical compositions of galaxies are reasonably well reproduced from one
Balmer and metal feature to another.  Specifically, we have made a detailed
comparison between ages and metallicities derived from the H$\beta$ and Hn/Fe
Balmer line indices and the Fe4383 and Fe5270 metal line indices.  We have also
carried out a comparison between different methods for measuring specific
Balmer lines.  For instance, in the Rose line ratio index system, H$\gamma$
is characterized by the index ratio H$\gamma$/Fe$\lambda$4325.  On the other
hand, \cite{worott} defined an equivalent width measure of
H$\gamma$, namely H$\gamma$$_F$.  To compare the results obtained from these
and other Balmer indices, we calculated the mean indices for the four 
different log $\sigma$ groups for a variety of different indices.  We then
extracted mean ages and metallicities for these groups for various index
combinations.  The results for a variety of two-index diagrams
are given in Table~\ref{tab:gal_sigs3}.  These results are summarized as follows:

\vskip 0.5cm
\noindent(1) The overall trend of younger to older mean age with increasing
log $\sigma$ is reproduced regardless of which Balmer and metal line
combination is used.  

\vskip 0.5cm
\noindent(2) The mean ages and metallicities are particularly insensitive to
the specific Fe feature used, as can be seen by comparing the Fe4383 results to
the Fe5270 results.

\vskip 0.5cm
\noindent(3) There are significant differences in zero point (and scale) in
the ages that are derived from different Balmer lines, {\it particularly in the 
case of H$\delta$}.  We note that while Schiavon et al. (2002a,b) are able to
obtain consistency between the spectroscopic age for 47 Tuc, as derived from 
its integrated spectrum using both H$\beta$ and H$\gamma$, and the age based
on its observed color-magnitude diagram, they are unable to obtain satisfactory
agreement in the case of H$\delta$.  The problem appears to reside in the
contamination of the red wing of H$\delta$ by the CN 4216 band.  This effect can
be seen in the right panel of Fig. \ref{fig:lines}, where the spectra of several
high $\sigma$ galaxies with relatively old ages are plotted.  Note that the
pseudo-continuum peak redward of H$\delta$ is substantially lower than its
counterpart on the blueward side of the line.  Because the details of the CN
feature are still not fully reproduced at the 1 \AA\ level, it is difficult to
assess to what extent the H$\delta$ line core may be contaminated by CN in the
case of $\alpha$-element enhanced galaxy spectra.  Since the stellar spectra
used in the SSP modeling come from solar neighborhood stars, the population 
models will not reproduce the enhanced CN features that are evident in, e.g.,
47 Tuc and other Galactic globular clusters, and in the integrated spectra of 
globular clusters in M31 (Burstein et al 1984; Tripicco 1989).

\vskip 0.5cm
\noindent(4) The derived ages from Balmer lines are reasonably 
insensitive to the method of quantifying the line strengths.  Note that for
both H$\gamma$ and H$\delta$ the derived ages are quite consistent between
the line ratio indices (H$\gamma/4325$ and H$\delta/4045$) and the 
pseudo-equivalent width indices (H$\gamma_F$ and H$\delta_F$).

\subsection{Comparison with Other Studies}

\begin{figure}
\plotone{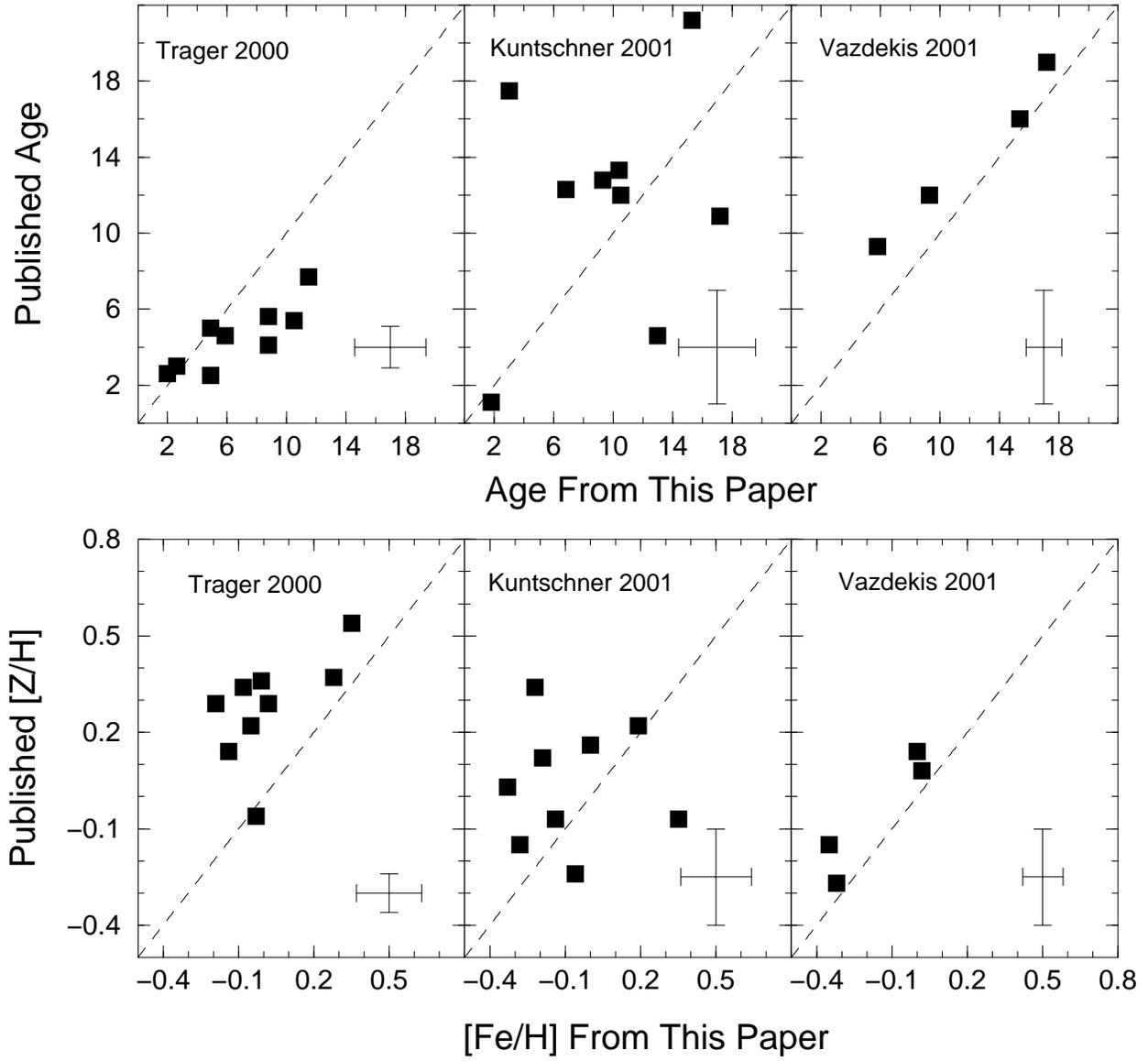}
\caption{A comparison of our derived ages and metallicities to determinations from
the literature.   Mean errors {\it for the specific
data plotted} are shown as an error cross in the lower right corner of each
plot. The unity line appears as a dashed line.
The same age and [Fe/H] scale applies in all panels.}
\label{fig:litage}
\end{figure}

With the large number of galaxies in our sample and the existence of stellar
population studies similar to ours \citep{jor99,tr00a,ku01,tf02},
it is worthwhile to compare our derived ages and metallicities with those
published by other groups.  Unfortunately, because our study has focused
on the lower luminosity galaxies, our overlap with other groups is small.
However, there are three studies for which we have galaxies in common.  For
the Gonzalez sample of primarily field ellipticals analyzed by \cite{tr00a}
we have 9 galaxies in common; from the cluster and field sample of
\cite{ku01}, we have 9 galaxies in common; and from the study of
the Virgo cluster by \cite{va01b} we have 4 galaxies in common.  The
ages and metallicities determined in these studies
are listed in Table~\ref{tab:lit} along with
our values for comparison, and are shown graphically in 
Figure~\ref{fig:litage}.
The agreement is good between our determinations and those published by
\cite{tr00a} and \cite{va01b}, and less so with the
\cite{ku01} study.  There are four galaxies which are in common between
three of the four studies, NGC 2778, NGC 4473, NGC 4478, and NGC 5831. With
the exception of single ``discordant'' values for NGC 2778 and NGC 5831, the
agreement for these multiply determined ages is good.
As expected, the agreement is better in relative ages and metallicities than
in the zero points between one study and another.

On the matter of the enhancement ratios (see appendix B), 
we can make a comparison only
with the published Trager et al.  Recall that we have avoided in this
paper the massive galaxies, which would have the largest enhancement
ratios, and hence the derived average ratios for the nine galaxies we
have in common will not be large.  For those galaxies, the mean 
[E/Fe] ratio (the ratio of all ``enhanced elements'' to iron) 
found by Trager et al. is 0.11, while our value for
[Mg/Fe] is 0.04.

\subsection{Intrinsic Scatter in Age and Metallicity}
\label{sec:inscatter}

While there is little doubt from the previous discussion that the low-$\sigma$ 
galaxies exhibit a pronounced trend towards lower
luminosity-weighted mean age, a more subtle question regards the possibility
of an intrinsic age spread among the higher-$\sigma$ galaxies as well.  
Before considering the question from the perspective of age, i.e., a
modeled parameter, we first revisit the issue as to whether intrinsic scatter 
is present among the Balmer line indices at high $\sigma$.  The results of
Table~\ref{tab:inscat} indicate, at face value, that there is a significant 
intrinsic scatter present in all indices at high $\sigma$.  Given, however, that
emission and NSAR corrections do produce a significant change in the assessment
of the scatter, it is worthwhile to consider the role played by these
corrections.  Of greatest significance is the effect of emission on Balmer line
indices.  Returning to Fig.~\ref{fig:hbhn.newcor} it is apparent that emission 
corrections are important for accurately assessing the high $\sigma$ galaxies 
in H$\beta$, while they are substantially less important for Hn/Fe.  
More specifically, we previously found that the
mean change in age that results from applying emission corrections to H$\beta$
is 1.8 Gyr (and 0.7 Gyr for Hn/Fe).  This figure is actually an underestimate,
since for many galaxies no age determination can be made before the correction
is made, because the emission-contaminated indices lie outside the 
age-metallicity grid lines.  Note that the 
observational error used in extracting the intrinsic scatter in \S5 only
includes the actual uncertainties in the observations, and does not include
the uncertainty introduced in making the emission correction.  Hence part of 
the intrinsic scatter found for the emission-corrected Balmer indices could in 
principle be due to errors in correcting for emission. 

We now take a closer look at the issue of age scatter as determined from
the Hn/Fe index.
An inspection of Fig.~\ref{fig:Hn4383_4plot} indicates that, even 
among the higher-$\sigma$ galaxies, there is a sizable spread in that diagram,
particularly in age.  We discuss two ways to demonstrate that a true age
spread is apparent even in the high $\sigma$ galaxies.  First, we show that the
more extreme examples of younger and older galaxies are indeed reproducibly
different in their spectra, i.e., are not simply due to observational scatter
from low S/N ratio.  We then show that the distribution of 
high $\sigma$ galaxies in the Fe4383 versus Hn/Fe diagram
has a scatter in excess of that expected from the observational errors.

\begin{figure}
\plotone{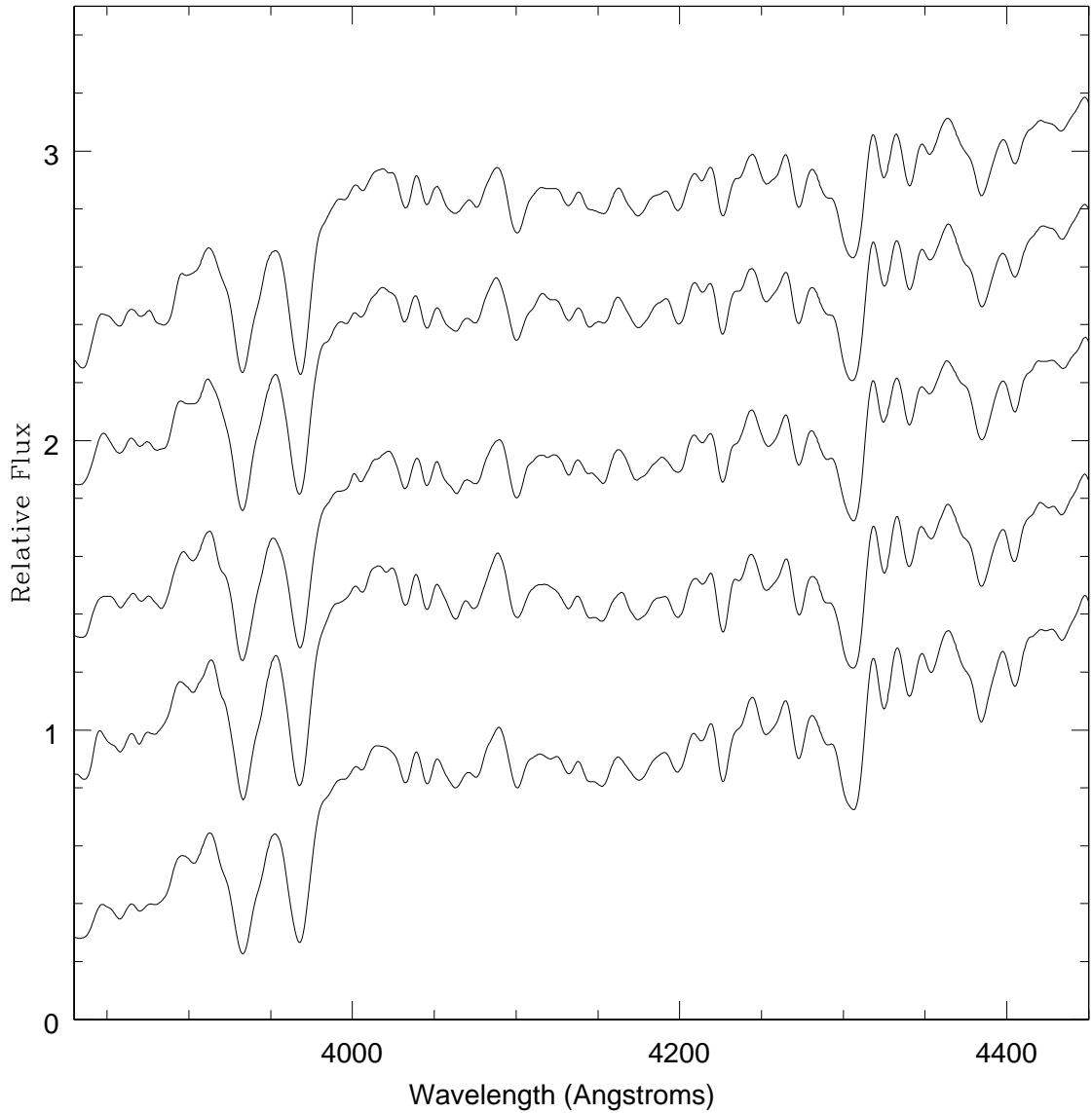}
\caption{The spectra of five high-$\sigma$ galaxies are plotted to show the
correlated trend in all Balmer lines from older to younger galaxies.  Plotted
from top to bottom are: NGC~7611, NGC~7612, NGC~2954, NGC~821, and
NGC~4442 (VCC1062), which have ages of 2.8, 2.8, 3.9, 11.6, and 
12.3 Gyr respectively.}
\label{agespecs}
\end{figure}

To show that there is a systematic change in the spectra from the youngest to 
oldest galaxies in the high $\sigma$ sample, in Fig.~\ref{agespecs} we plot the
spectra of five high-$\sigma$ galaxies, ranging in derived age from 2.8 Gyr to
12.3 Gyr, based on their location in Fig.~\ref{fig:Hn4383_4plot}.  Note that as one
proceeds from the bottom (oldest) spectra, to the top (youngest) spectrum, a
systematic change can be seen in the three higher order Balmer lines, H$\gamma$,
H$\delta$, and H8, relative to their neighboring metal lines.  Specifically,
in the bottom two spectra H$\gamma$ is weaker than Fe4325, while in the top 
spectra it is stronger.  Likewise, H$\delta$ is slightly weaker 
than the neighboring blended feature at 4063 \AA\ in the bottom spectra, while
in the top spectrum it is distinctly stronger than that feature.  Finally, while
H8 is buried in the CN3883 band in the bottom spectrum, it clearly emerges at
3888 \AA\ in the top spectra.  The systematic trend in all three Balmer lines,
from younger to older galaxies, coupled with the reproducible nature of the 
metal lines from one galaxy to another, indicates the reality of the changing
Balmer line strengths along the age sequence.  However, the changes in Balmer
line strength from 12.3 Gyr to 2.8 Gyr is indeed quite subtle, i.e., high
S/N ratio is required to discriminate such age changes from integrated spectra.

\begin{figure}
\plotone{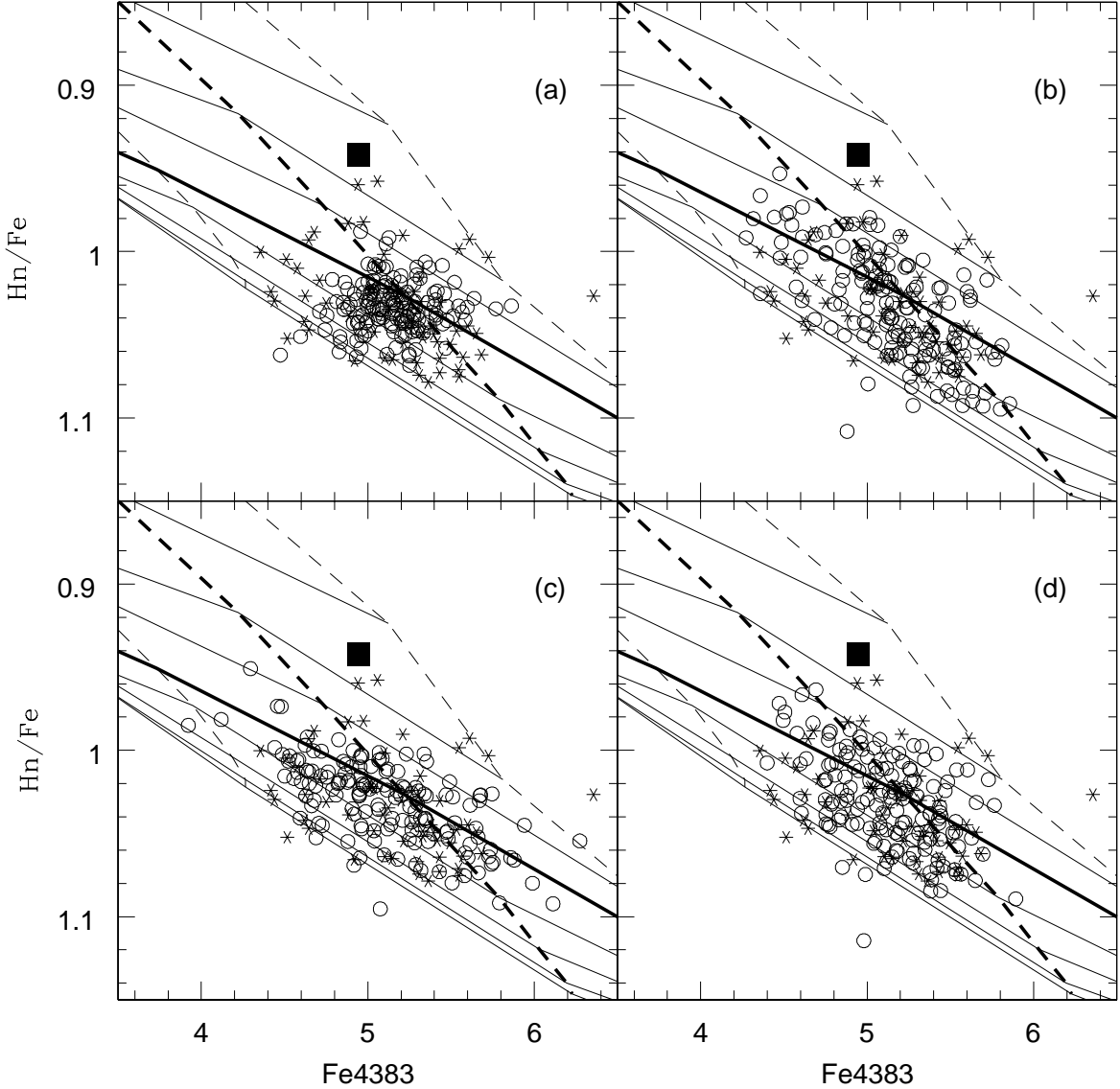}
\caption{The observational data for the high-$\sigma$ galaxies (asterisks) 
are plotted along with simulated observations (open circles)
in the Fe4383 versus Hn/Fe diagram.  The solid and dashed lines indicate the 
grid of 
Worthey models, and the large square represents M32. An additional age
line is shown here, at 19.05 Gyr. In panel (a) we have
only included the known $\pm$1$\sigma$ observational errors in producing the 
simulated galaxy points.  In panels (b) and (c) we have allowed for a 
$\pm$1$\sigma$ scatter of 2 Gyr in age and 0.1 in [Fe/H], respectively, along
with the observational errors.  In panel (d) we have simulated correlated
errors in age and [Fe/H].  See text for details.}
\label{fig:HnFev4383_sim}
\end{figure}

To further demonstrate the degree of intrinsic age and metallicity scatter among
the higher $\sigma$ galaxies, we have simulated the expected observational 
scatter in the Fe4383 versus Hn/Fe diagram under different assumptions regarding
the intrinsic age and metallicity scatter in the
galaxies.  The results of these simulations are displayed in 
Fig.~\ref{fig:HnFev4383_sim}.  In the upper left panel, (a), we have assumed
a single mean age and metallicity for all of our simulated galaxies, hence only
observational errors, and interpolated the Hn/Fe and Fe4383 indices
corresponding to that age and [Fe/H]. We then randomly sampled 150 times from 
Gaussian error distributions having rms scatters of $\pm$0.014 and $\pm$0.21,
which correspond to the median observational errors in Hn/Fe and Fe4383,
respectively.  In each case we added the observational errors to the mean 
indices corresponding to the selected mean age and metallicity of 9.5 Gyr and
[Fe/H]=-0.07.  In Fig.~\ref{fig:HnFev4383_sim}(a) we have plotted both the
actual galaxy observations for the high-$\sigma$ galaxies (log $\sigma$$>$2.2),
as asterisks, and the simulated data, as unfilled circles.  The large filled
square represents the observed data for M32.  As usual, we plot the age
and metallicity grid lines determined from the modeling.  It is evident that 
the observed data has a significantly larger scatter than the simulations.  
Hence we conclude that some intrinsic scatter in age and/or metallicity must be
present in the high-$\sigma$ galaxies.  

In Fig.~\ref{fig:HnFev4383_sim}(b) we have included an intrinsic scatter in age 
of $\pm$2 Gyr.  To simulate such an age spread, we first randomly selected 150
times from an age distribution with an rms dispersion of $\pm$2 Gyr and a mean
of 9.5 Gyr.  In all cases the metallicity was held fixed at [Fe/H]=-0.07.
For each realization, we interpolated the Hn/Fe and Fe4383 indices for the given
age and [Fe/H].  We then randomly selected observational errors in Hn/Fe and
Fe4383 for each trial galaxy following the prescription above.  It is
evident that the simulated galaxy data more closely follows the observed data
than in the case of Fig.~\ref{fig:HnFev4383_sim}(a).  As a further test, we held
the age fixed at 9.5 Gyr and randomly selected in [Fe/H] with an rms dispersion
of $\pm$0.1 around a mean value of -0.07.  After adding in observational error,
we find the distribution of points plotted in Fig.~\ref{fig:HnFev4383_sim}(c).
The correspondence with the actual galaxy data is noticeably better in the case
of Fig.~\ref{fig:HnFev4383_sim}(b), where an age scatter is introduced, rather
than a metallicity scatter.  Finally, in Fig.~\ref{fig:HnFev4383_sim}(d) we have
introduced a correlated age and metallicity scatter.  We first randomly selected 150
ages with an rms scatter of 2 Gyr around the mean age of 9.5 Gyr.  Then for each
age we calculated a metallicity according to the prescription:

     [Fe/H] = $-0.3 log(\tau/9.5) -0.07,$

\noindent where $\tau$ is the age, in Gyr, of the simulated galaxy.  Then we
perturbed the metallicity by randomly sampling from a distribution with an rms
of $\pm$0.05, and finally added in a randomly sampled observational error.  The
results of this simulation can be compared to the observed galaxy data in
Fig.~\ref{fig:HnFev4383_sim}(d).  This correlated age-metallicity prescription
appears to provide a closer representation to the observed galaxy data than
either a scatter in age or metallicity alone.  

In short, while the higher-$\sigma$ galaxies clearly form a systematically older
sample in 
age than their lower-$\sigma$ counterparts, there are signs of an intrinsic 
spread in age even among these more massive galaxies.  We return to this issue
in more detail in Papers II and III.

\section{Discussion \& Conclusions}

\begin{figure}
\plotone{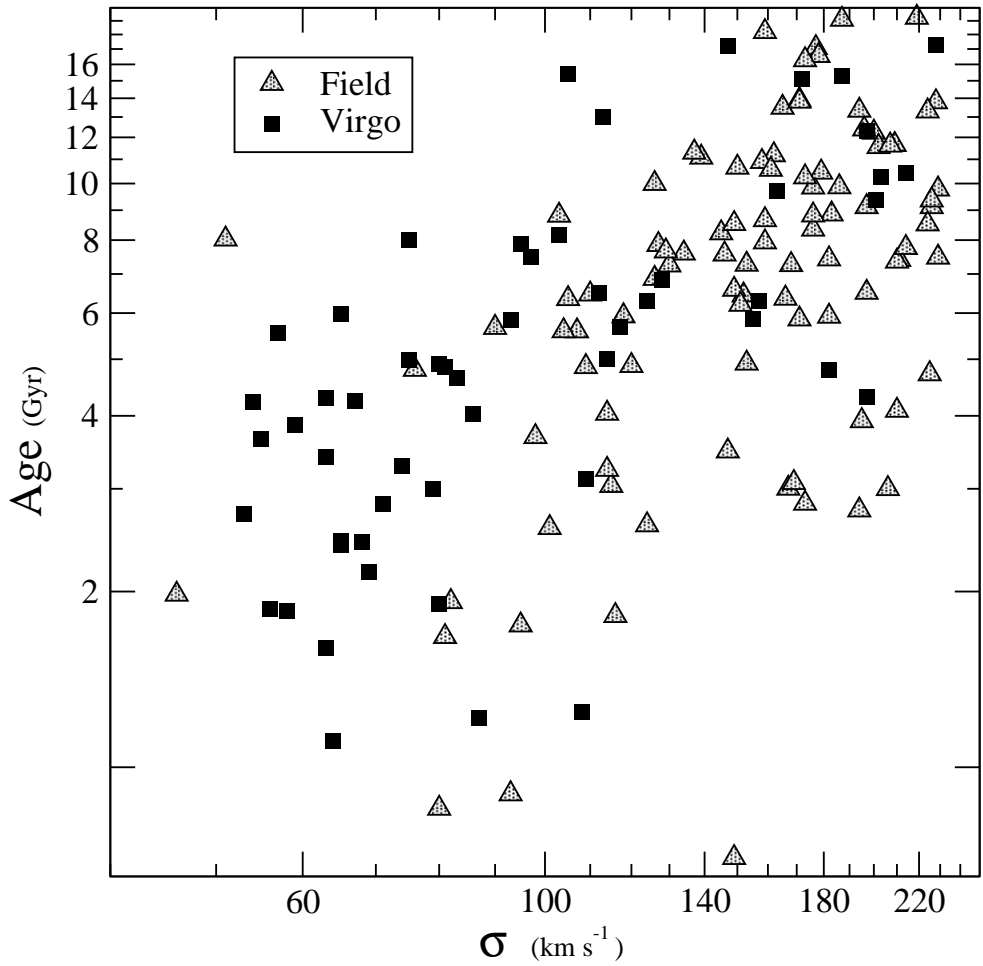}
\caption{The age-$\sigma$ diagram for all galaxies in the sample.  The 
squares denote the Virgo galaxies; the triangles denote field galaxies.}
\label{fig:age-mass}
\end{figure}

While the previous sections have focused on establishing
the reliability of the ages and metallicities derived in our study, here we
emphasize our principal results for the E/S0 stellar populations
and then briefly examine the
implications for hierarchical models of galaxy evolution.  We begin by 
showing the trend of mean age in our sample as a function of the measured
velocity dispersion in Fig.~\ref{fig:age-mass}.  
Different symbols distinguish the field and Virgo samples
(discussion of those samples will be taken up in the succeeding papers).
In addition to the scatter seen at all velocity dispersions discussed above,
a clear trend is seen, such that lower $\sigma$ galaxies have
younger ages.
We have again used our velocity dispersions for consistency
even though some values for low mass galaxies are higher than literature values.
The overall sense of this figure does not change if we use the smaller
dispersions.  The ages in this figure come from the Hn/Fe-Fe~4383 diagram; if
we use the H$\beta$-Fe~5270 diagram, which is corrected for NSAR, the same
basic character of the data is found.

\begin{figure}
\plotone{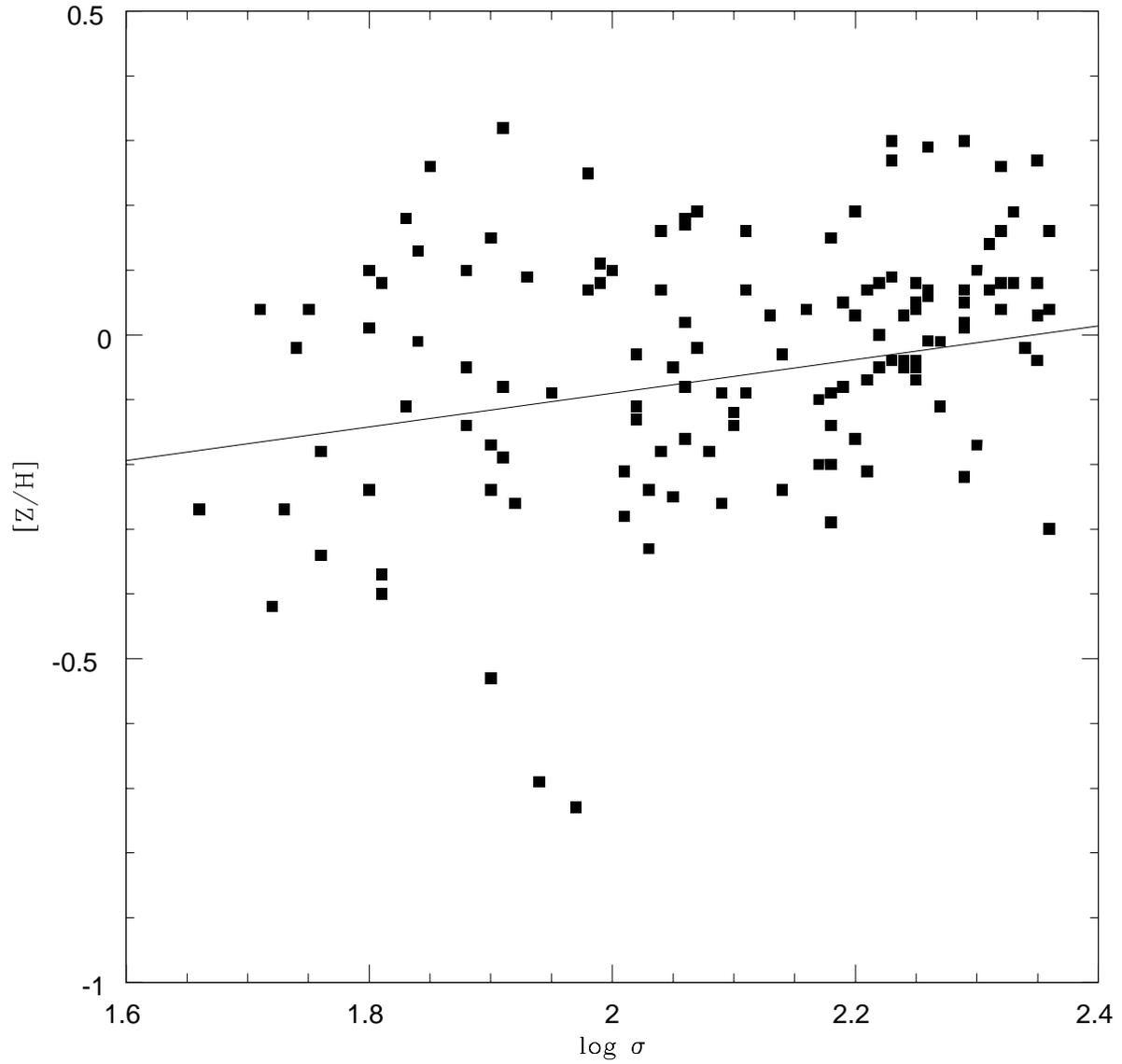}
\caption{The [Z/H]-log $\sigma$ correlation for all galaxies in the sample.  The
[Z/H] values have been determined from the Fe5270 versus H$\beta$ diagram, 
where both indices have been corrected for NSAR effects.}
\label{fig:metsvssigma}
\end{figure}

There are then three principal conclusions to arise from the work presented in
this paper.
First, there is a large intrinsic spread in Balmer line strength at low $\sigma$
(Table~\ref{tab:inscat}), and
it appears that a small intrinsic spread is present at high $\sigma$ as well.
This spread in Balmer line strength has
been extensively modeled and shown to be due primarily to a spread in age.
Second, the mean age of galaxies increases with
increasing velocity dispersion, leading to an age-$\sigma$ trend, shown
in Figure~\ref{fig:age-mass}.  Although no age-$\sigma$ trend was observed
in the samples of Kuntschner (2000) or Terlevich and Forbes (2001), our observed
age-$\sigma$ trend is in accordance with the results of 
Trager \etal (2002a,b), who suggest that large Es are older, on average,
than smaller Es.  It is not clear from
Figure~\ref{fig:age-mass} whether there exists a continuous age-$\sigma$\ 
correlation, or whether there are
two distinct populations --- a sample of younger, low-$\sigma$ galaxies (log $\sigma \le
\sim$2.1) and a sample of older, high-$\sigma$ galaxies.  With the available information, 
we cannot reliably distinguish between these two scenarios.  Third, the trend in
[Z/H] with $\sigma$ is modest, as seen in Fig.~\ref{fig:metsvssigma}, where the
[Z/H] values derived from the NSAR-corrected Fe5270 and H$\beta$ indices are
plotted versus log~$\sigma$.  A fit to the relation in 
Fig.~\ref{fig:metsvssigma} yields a slope of 0.32, such that [Z/H] increases
by only $\sim$0.2 over the observed range of 1.6$<$log~$\sigma$$<$2.4.  

Can our results place key constraints on age distributions predicted by
numerical simulations of structure formation in the universe?  
The most extensively modeled scenario is the cold dark matter
picture, in which structure evolves through a hierarchical sequence of mergers
(e.g., Davis \etal 1985).  Two basic features of all hierarchical models provide
testable predictions.  First, it is expected that the merger process will lead
to a significant scatter in the ages of assembly of the present epoch dark 
matter halos.  Second, the merger sequence naturally leads to an earlier 
assembly epoch (and hence older age) of the dark matter halos in lower mass
galaxies, while the assembly of the higher mass galaxies occurs more recently,
on average.  Consequently, the hierarchical models tend to predict that
more massive early-type galaxies should have younger mean ages than low mass
galaxies \citep{ka96,ks98,sp99}.  Our results do indeed indicate a substantial scatter in the ages of
early-type galaxies, consistent with a hierarchical
scenario.  On the other hand, we find the lower mass galaxies to be younger,
which reverses the trend of the hierarchical predictions. We reserve a more
extensive discussion of this apparent discrepancy to Paper III, but refer the
reader to a discussion of this point in Terlevich \& Forbes (2002). 

The emphasis of this
first paper has been to describe the techniques used to measure key spectral
indicators and to model them through evolutionary population synthesis.
We have demonstrated that spectral indicators which measure Balmer
lines from H$\beta$ through H8, and measured using different techniques, give
consistent results for the luminosity-weighted mean ages of early-type
galaxies.  Our principal conclusion then is that a significant trend is present 
between galaxy mean age and mass in the sense that galaxies with velocity
dispersions below 100 \kms are younger on average than their higher $\sigma$
counterparts.

\acknowledgements

We wish to thank G. Worthey for generously providing his evolutionary synthesis
code, which has been resident at UNC for many years now.  We are also grateful to R. 
Kurucz for making available his stellar atmosphere models and SYNTHE code.  In
addition, we thank A. Leonardi for updating the Worthey models by producing
and incorporating the Kurucz synthetic spectra for hotter stars and for other
updates to the original Worthey models.  H. Kuntschner kindly
supplied us with the ages and abundances of the galaxies in his sample.
We thank the Mount Hopkins observers, Perry Berlind and Michael 
Calkins for obtaining some of the data reported here.  Lastly, we thank
the referee for comments that helped us to improve the presentation of
this paper.  This 
research was partially supported 
by NSF grant AST-9900720 to the University of North Carolina.

\appendix

\section{Emission Corrections}

The detection of an emission spectrum superimposed on an absorption spectrum is
a non-trivial task.
In the case of the age-sensitive Balmer lines, the task is particularly
problematic, since the extracted emission line strength will depend strongly on
the assumed character of the underlying absorption spectrum.  Naturally, the
properties of the absorption spectrum are themselves strongly dependent on 
the presumed age and metallicity of the galaxy, yet these are the
quantities that we wish to determine.  The problem is less severe for most of the
forbidden metal lines, which tend to be superimposed on regions of the absorption spectrum
that are less sensitive to assumptions about the stellar population.  However, 
unless the temperature, chemical composition, and emission 
mechanism for the ionized gas can be independently determined, the ratio of 
emission in the forbidden metal lines to that in the Balmer lines is not well
constrained.  
To correct for the effects of emission on the absorption line strengths,
we have obtained spectra in the red, centered near H$\alpha$ for 40\% of the galaxies
in our sample.  A description of those observations is given in 
$\S$2.2.2.
The observed galaxies are indicated in column 9 of Table~\ref{tab:log}, where a
``y'' designation indicates that emission is detected in the galaxy and the
[NII]/H$\alpha$ emission line intensity ratio is reliably measured, a ``n''
indicates that emission was not detected, 
and a ``u'' indicates that emission is probably detected, but no reliable value
can be determined for the [NII]/H$\alpha$ ratio.

\subsection{Measuring the Flux in H$\alpha$}

To evaluate the emission contamination of the Balmer lines, we exploit two
basic advantages of H$\alpha$ for this purpose.  First,
while the equivalent widths of the Balmer absorption lines are
relatively constant along the Balmer sequence, the emission decrement is steep.
Thus at H$\alpha$ we maximize the contribution of emission relative to the
underlying absorption.  Second, at H$\alpha$ the sensitivity to the age and
chemical composition of the underlying absorption spectrum is minimized 
(compared with the higher order Balmer lines).  This is due to the fact that
at H$\alpha$ the light is preferentially dominated by the giant branch, thus
relatively insensitive to age when compared to the higher order Balmer lines
(where the main sequence turnoff light provides a higher contribution).

The emission contamination was detected and removed in the following manner.
Of the 69 galaxies in our sample for which we have spectra at H$\alpha$, 29 of
those have no detected emission in the [OII]$\lambda$3727 line and are 
consequently considered to be emission free.  From this sample of 29 
emission free galaxies
we have measured the relation between the Hn/Fe index (which measures higher 
order Balmer line absorption strength) and the H$\alpha$ absorption. A linear
regression yields the relationship:
\begin{equation}
H\alpha_{absorption} = -2.15(Hn/Fe)+3.94
\end{equation}
with an rms scatter around the mean relation of $\pm$0.14, after eliminating two
discrepant galaxies. 
Since none of the galaxies are plagued by strong emission, we make the 
assumption that the emission is small enough in the H$\gamma$ and higher order
Balmer lines so that the
observed Hn/Fe indices for our galaxy spectra provide a good indicator of the
true underlying absorption spectrum and a reliable predictor of the strength
of H$\alpha$ absorption.  For a given emission-contaminated galaxy, we select
from our sample of 29 emission-free galaxies a galaxy whose Hn/Fe index closely
matches that of the contaminated galaxy and preferably has a lower velocity
dispersion.  From the velocity dispersion information determined from the blue
spectra, we then smooth the red spectra of both the contaminated galaxy and
the emission-free template to a common resolution, normalize the continuum of
the template spectrum to that of the contaminated spectrum in the vicinity of
H$\alpha$ and then subtract off the template.  From the difference spectrum the
flux of $H\alpha$ \emph{emission} is measured, using the IRAF \emph{splot} 
routine.  Then,
assuming the emission follows the expected Case B Balmer decrement 
\citep{ost89}, the expected H$\beta$ emission flux is determined.

To translate the estimated H$\beta$ emission flux into corrected absorption
line indices in the blue, a spectrum of the Balmer emission must be subtracted from 
the contaminated blue spectrum.  We have produced a spectrum of the Balmer decrement
covering H$\beta$ through H8 from a flux-calibrated observation of the Orion 
nebula taken with the same instrumental
set-up as the galaxy spectra. The individual relative hydrogen line strengths
are adjusted to the Case B recombination spectrum values.  This spectrum, hereafter
referred to as the Balmer decrement spectrum, is then smoothed to the velocity dispersion
of the emission-contaminated galaxy and normalized to have the same H$\beta$ 
emission flux as that expected from the H$\alpha$ emission for the galaxy.  
The blue contaminated spectrum is then
normalized to match the flux level of the red spectrum by utilizing the $\sim$50\AA\
of overlap between the blue and red spectra at $\lambda$5500-5550.  Finally, the Balmer
decrement spectrum is subtracted from the normalized blue spectrum, which
yields the ``corrected'' Balmer absorption line strengths.  The Hn/Fe index is re-measured 
and the process is repeated, usually only once, until the corrected Hn/Fe
converges to a stable value.  

Errors in this correction could result from a lack of spatial correspondence of the
blue spectra and the H$\alpha$ spectra, which is the case for some of the galaxies.
If the emission distribution is not uniform (which is usually the case), then
spectra taken at different position angles will have different amounts of
emission contribution to the absorption spectrum.  However, in general the
emission found in E/S0 galaxies is highly concentrated (Buson \etal 1993), and the extraction of
our blue and red spectra was heavily weighted towards the central few arcseconds. Thus we
feel that any associated errors are minor.

\subsection{Measuring the H$\alpha$ Flux via the [OII]$\lambda$3727 Flux}

The method described above for correcting the emission contamination constitutes
our preferred technique,
since the corrections are based directly on the amount of H$\alpha$ emission 
observed on an individual 
galaxy basis.  However, we do not have red spectra, and hence the H$\alpha$ emission
information, for all of the galaxies in the sample.  Rather than leave the
line indices uncorrected for the remaining 60\% of the galaxies, we have opted to 
correct the indices by inferring the H$\alpha$ emission flux from the emission present
at the [OII]$\lambda$3727 doublet.  In this method, the assumption is that the excitation mechanism for 
emission in early-type galaxies shows little variation from one galaxy to another.

\begin{figure}
\plotone{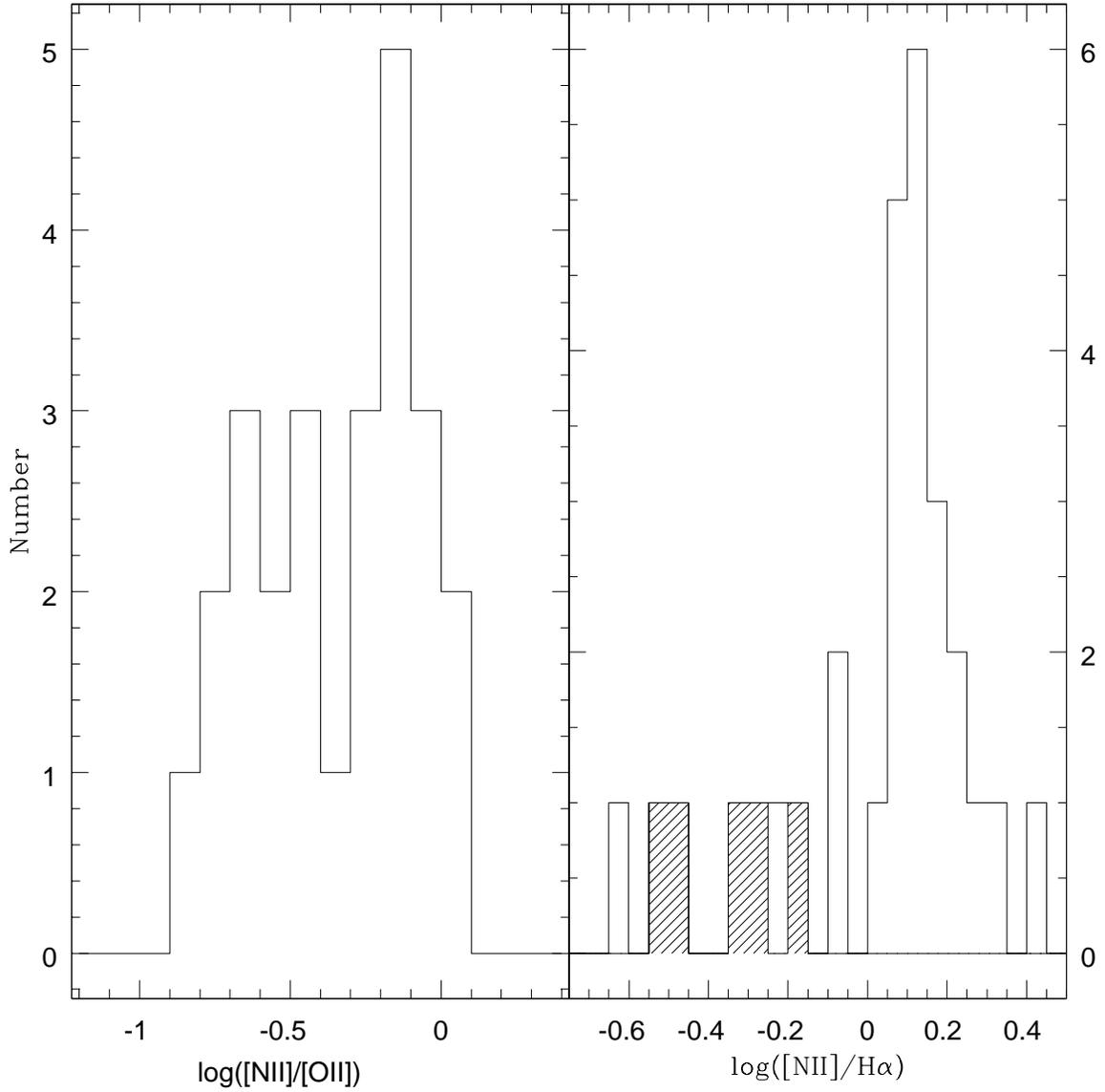}
\caption{Histograms of the line ratio log([NII]/[OII]) (left panel), and
log([NII]$\lambda$6584/H$\alpha$) (right panel) for those galaxies with 
measurable emission in their red spectra.  In the right hand histogram,
the shaded areas denote the low-luminosity Virgo galaxies; the
open regions are higher mass galaxies.}
\label{fig:histemis}
\end{figure}

Measuring the emission flux in the [OII]$\lambda$3727 line first requires subtracting
off the underlying continuum.  To do this, an emission-free template galaxy 
is chosen, VCC1279, and all galaxies are normalized to match the continuum
level near $\lambda$3727 for that galaxy.  The template galaxy is then subtracted
from each normalized galaxy spectrum, and the residual emission flux at [OII] 
is measured.  For the galaxies \emph{with} emission information at H$\alpha$, the red spectra
are re-normalized to match the overlap with their blue counterparts, and the flux
in the [NII]$\lambda$6584 line is measured.  A comparison between the two emission line fluxes
yields a median log([NII/OII]) ratio of -0.45 (Figure~\ref{fig:histemis}). 
Thus, for each galaxy \emph{without} information in the red, the [OII] emission flux can be
measured, and, from the median of the flux ratio, an estimate of the flux at [NII] can be  
made. Since the ultimate goal is to correct the Balmer line indices, the flux in [NII] 
must now be related to the expected flux in the Balmer lines.  For the galaxies in
our sample that have observed emission lines in their red spectrum,  we have
measured the ratio of the flux at [NII] compared to the flux at H$\alpha$ 
(Figure~\ref{fig:histemis}) and find a median log([NII]/H$\alpha$) value for the sample of
0.1.  Assuming that all galaxies have the same emission mechanism and the same ratio, 
(which is not true, of course, but only an approximation), the expected flux in the
H$\alpha$ emission line can be estimated.  Finally, using the Case B Balmer decrement,
the emission flux in H$\beta$ is determined.  

To translate the H$\beta$ emission flux estimated for each galaxy into spectral 
index corrections, we first evaluate the effects of a range of emission fluxes 
on two individual galaxies, VCC1279, an old galaxy, and VCC1912, a young galaxy.
The Balmer decrement spectrum is scaled to reflect the range of expected
H$\beta$ emission fluxes (0.5--5.0 x 10$^{-15}$) for the galaxy sample and subtracted from the two
template galaxies.  The index values are measured for each of the 
Balmer-subtracted galaxy spectra and the change in the index values as a function
of the removed emission flux is determined from a linear fit.  The coefficients
of these fits are given in Table~\ref{tab:e-lines}.  

\begin{figure}
\plotone{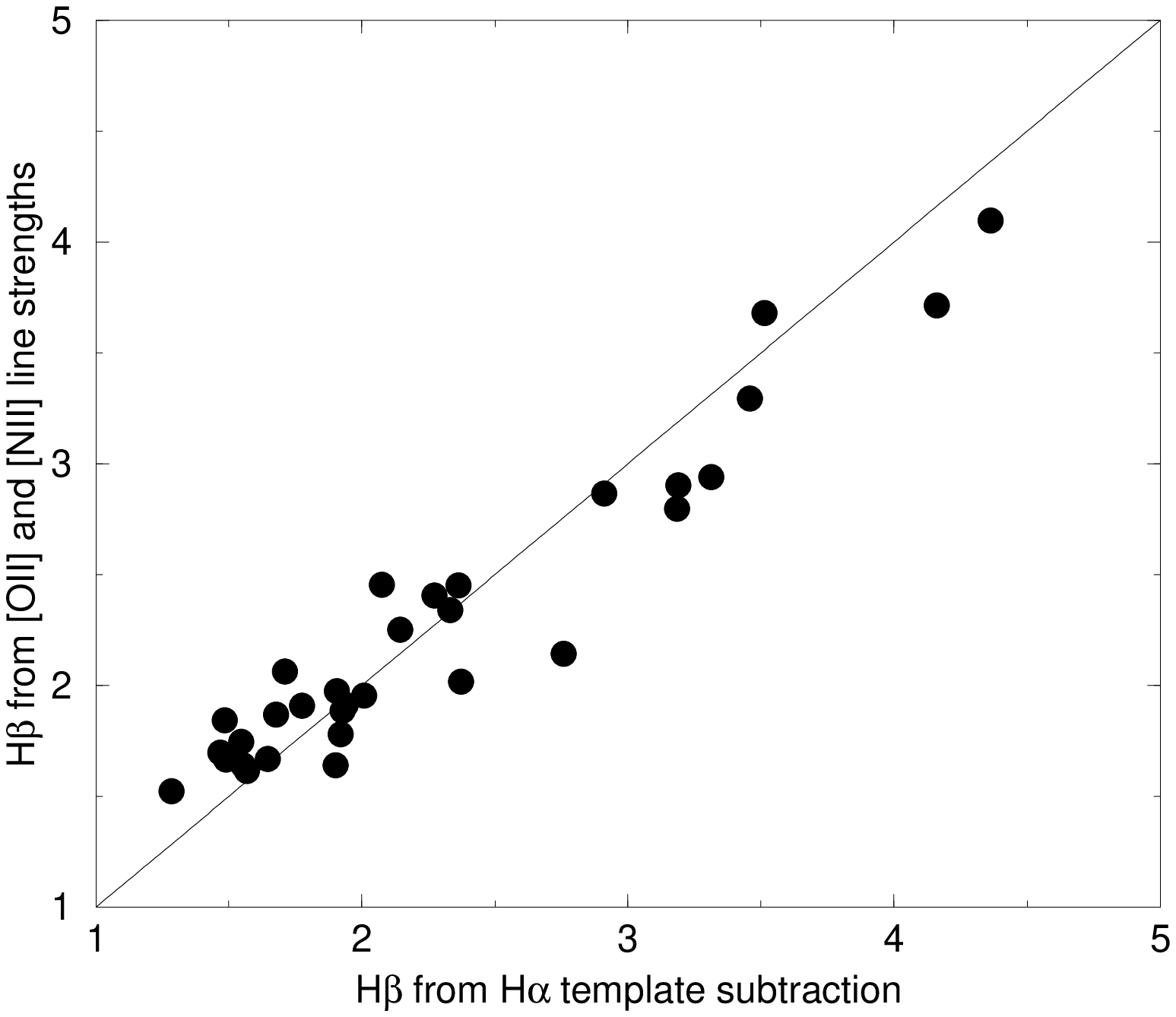}
\caption{Comparison of H$\beta$ corrections using the two different methods.  The rms 
scatter is 0.2\AA\ .}
\label{fig:HB_e}
\end{figure}

In summary, for each galaxy in the sample that
has not already been corrected for emission contamination, the flux in the OII
line is measured.  The OII flux is related to the NII flux, and in turn, the H$\alpha$
flux and the H$\beta$ emission flux.  Once the H$\beta$ flux is estimated, corrections to the line
indices are made using the linear relations in Table~\ref{tab:e-lines}.  For the
two main age indicators used in this work -- the Lick H$\beta$ index and the Rose
Hn/Fe index -- the mean index corrections are 0.15 and 0.012, respectively.
For a stellar population with an age of 8 Gyr, these corrections correspond
to an age difference of 0.5-1 Gyr, depending on the metallicity.  To gauge the 
reliability of using the OII emission line flux to correct the spectral indices,
we have compared the corrected index values for both this method and the more 
robust method using the H$\alpha$ emission flux from the red spectra (\S3.3.2).
Figure~\ref{fig:HB_e} shows that the results of the two methods agree quite well,
with an rms scatter about the unity line of 0.25.

\section{Correction for Non-Solar Abundance Ratios}
\label{sec:nsarcorrect}

As mentioned in \S4.5, the analysis of the integrated light of galaxies is
plagued by the fact that both the current models and the empirical stellar
spectral libraries are based on stars having solar abundance ratio patterns.
Below we describe how corrections are made for NSAR.  These corrections are
entirely based on the extensive work carried out by \cite{tr00a} and
by \cite{ku01}.

\subsection{Evidence for NSAR}

Shown in Figure~\ref{fig:nsar} are three index-index grids -- Fe 5270 vs.\ H$\beta$,
Mg~$b$ vs.\ H$\beta$, and Fe 5270 vs.\ Mg~$b$ -- with our galaxy data over-plotted. 
Of interest in these diagrams is the fact that the mean age and metallicity of the
galaxy sample changes from diagram to diagram.  Specifically, the galaxies appear
to be older and more metal-poor in the Fe5270-H$\beta$ diagram than they do in the 
Mg~$b$-H$\beta$ diagram.  The different metallicity (and age, to a lesser extent)
predictions can be most easily seen in the upper left panel of Figure~\ref{fig:nsar}
in which the bulk of the galaxies lie offset from the Mg-Fe population grids.
This discrepancy between the metal abundances derived from the two different indices
is further demonstrated in Fig.~\ref{fig:mgfemet}, and 
is now widely accepted to be caused by non-solar abundance ratio effects. Namely,
the models predict too little Mg at a given Fe-absorption line strength, since the
massive early-type galaxies have an enhancement of Mg (and other $\alpha$ elements)
relative to the solar-abundance stars which comprise the population synthesis models.
Recently, \cite{tr00a} have argued that rather than considering Mg and other
$\alpha$-elements to be enhanced in
massive early-type galaxies, one should consider the Fe-peak elements to be
underabundant.  

\subsection{Scaling the Galaxies to Solar-Abundance}

The overabundance of certain elements compared to Fe (or more appropriately, the 
underproduction of Fe compared to the $\alpha$ elements) has a profound effect on 
the use of age/metallicity index diagrams if the model predictions reflect only
solar abundance ratios.  
As noted above, the Mg-derived metal 
abundances will be greater than those derived using an Fe-index.  
In
addition, since the age-metallicity degeneracy has not been completely broken
(the age and metallicity tracks are not perpendicular in the model grids), the
age estimates derived using Mg and Fe line indices will differ as well.  Thus, 
trends in the intrinsic abundance ratios of the galaxies can lead to artificial 
trends in metallicity and/or age if the galaxies are not scaled to reflect the
same elemental abundance ratios.

Recently, \cite{tr00a} carried out an extensive investigation of the effects of 
NSAR in early-type galaxies and tabulated corrections for a selection of important
Lick indices based on the work of Tripicco \& Bell (1995) and the Worthey models.
We use their preferred Model 4 (enhancement in both C and O) corrections 
(given in their Table~5) and the method
of Kuntschner et al (2001) to correct a subset of our indices for the effects of
NSAR.  The first step in the correction process is to estimate the galaxy age using
the program described in \S4.4, using H$\beta$-Fe5270 model index grids 
(Figure~\ref{fig:nsar}). Galaxies which lie outside the grids are assigned an age
of 19 Gyr.  Each galaxy is then shifted in the Mg~$b$-Fe5270 index grid, along
the ratio vector defined from Table~5 of \citet{tr00a}, until it reaches the age
predicted by the H$\beta$ index.  The amount that the galaxy is shifted in the
Mg-Fe index grid determines the scaling factor for that galaxy, which is then used to
correct the other indices with $\delta$s listed in Table~5 of \citet{tr00a}, including
H$\beta$.  The galaxy age is then redetermined using the corrected index values
on the H$\beta$-Fe5270 index grid, and shifted further in the Mg-Fe diagram until
the corrected age estimate is reached.  The remaining indices are then corrected
for the new scaling factor and the entire process is reiterated one final time.
The mean changes in all NSAR-corrected indices, 
$\Delta$Index = Index({\it NSAR}) - Index({\it 0}),
are -0.038 for H$\beta$, -0.307 for Mg~${b}$,
and 0.178 for Fe~5270.

\subsection{Calculating the Abundance Ratios}

\begin{figure}
\plotone{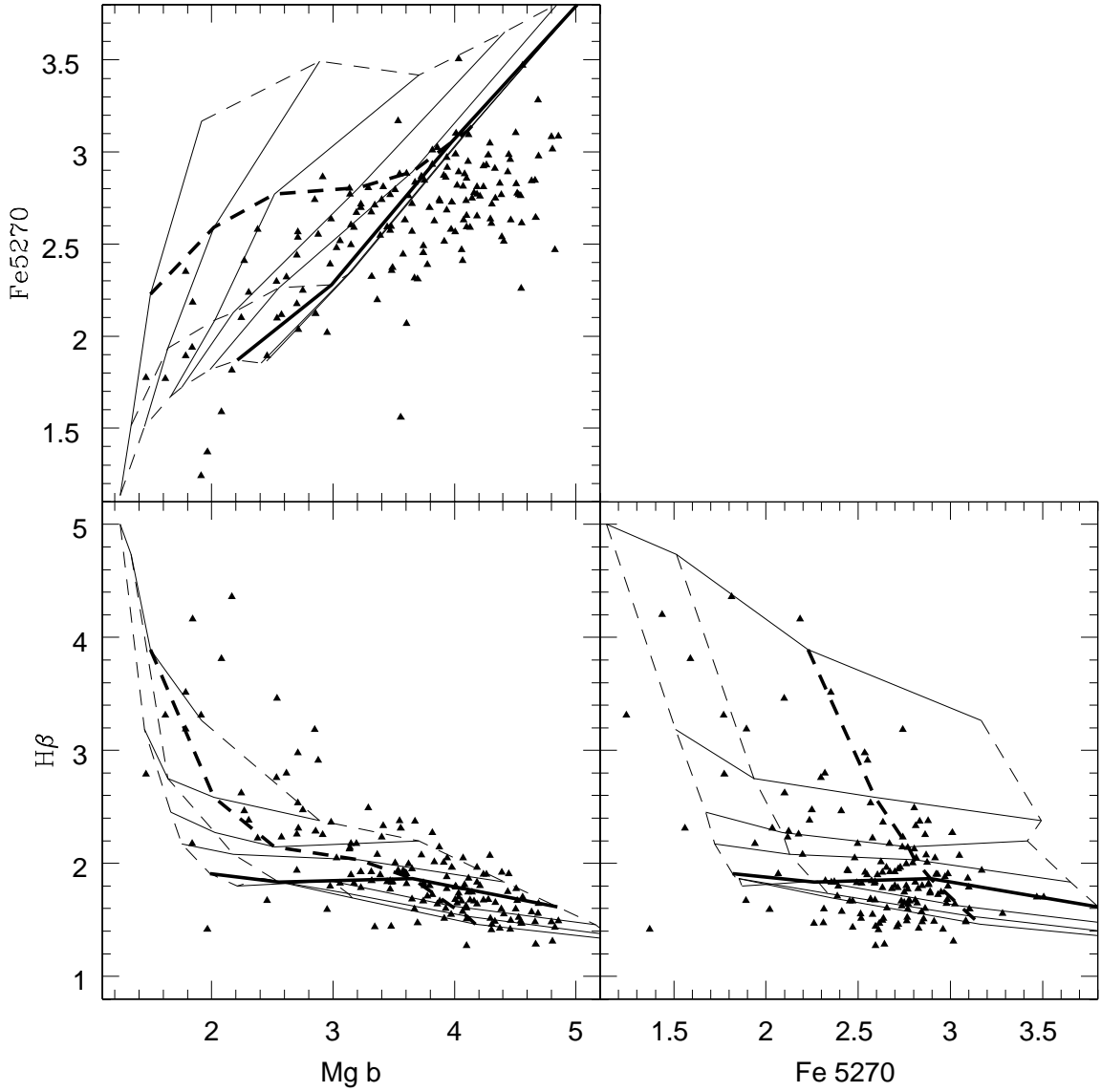}
\caption{Three indices which suggest the overabundance of Mg in early-type
galaxies.  The age and metallicity contours are the same as in previous figures.
Age increases from top to bottom in the two bottom panels, and increases
from left to right in the single top panel. Metallicity increases from
left to right in the two lower panels, and from bottom to
top in the upper panel.  The galaxy indices
have been corrected for
emission in H$\beta$ but not for NSAR.
Note that the Mg~$b$ index 
predicts younger ages and higher metallicities than the Fe index.}
\label{fig:nsar}
\end{figure}

\begin{figure}
\plotone{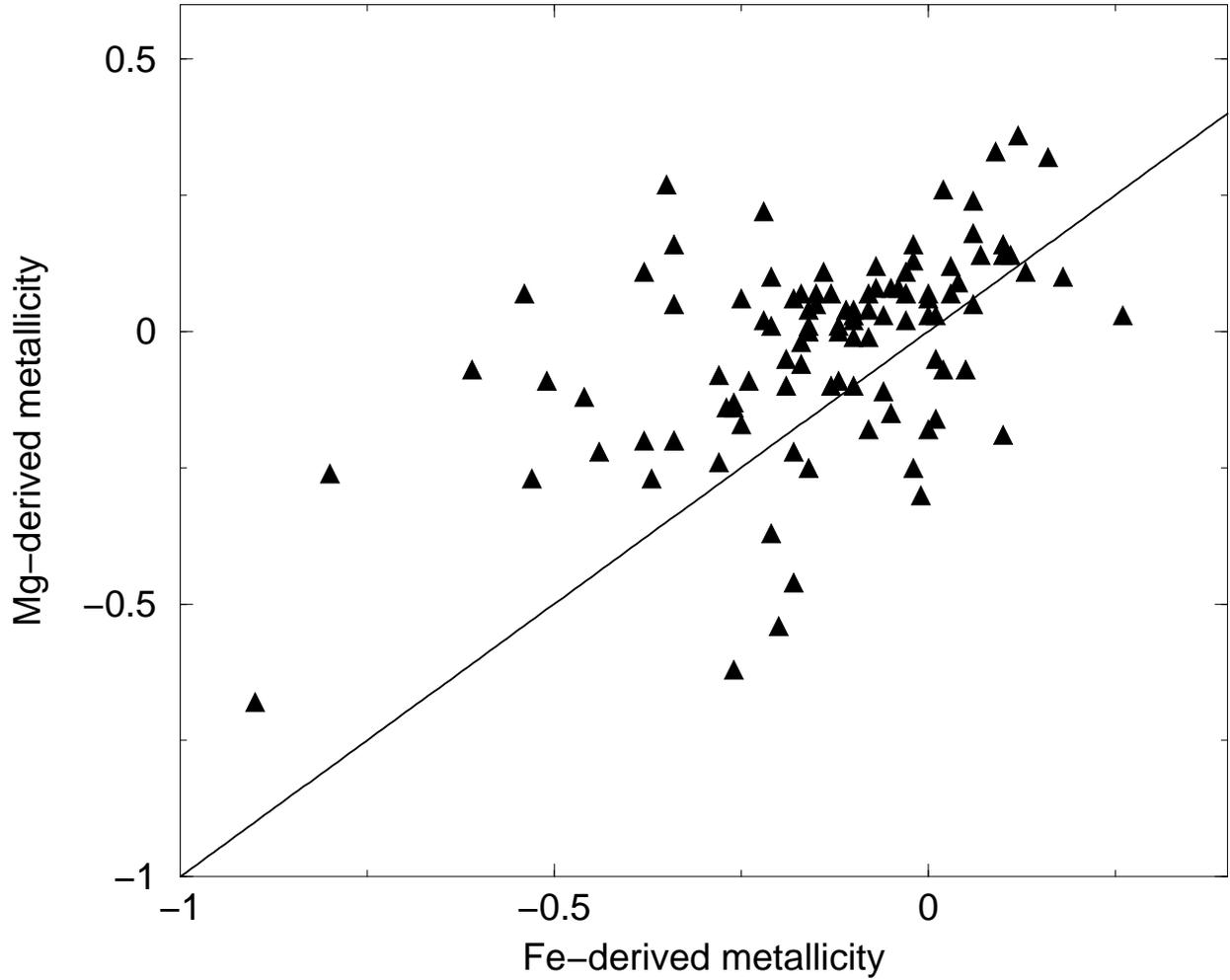}
\caption{The comparison between the metallicity derived from the Mg~$b$-H$\beta$ model
grids and that derived from the Fe 5270-H$\beta$ model grids. In both cases we have derived the abundances from spectral indices that have 
{\it not} been corrected for NSAR.  The solid line is the unity
relation. The Mg~$b$ index tends to yield higher metallicity values.}
\label{fig:mgfemet}
\end{figure}

A measure of the Mg/Fe enhancement ratio in the galaxies can be estimated by comparing
the metallicity derived from a non-NSAR-corrected H$\beta$-Fe diagram to that measured in a
non-corrected H$\beta$-Mg~$b$ diagram (e.g, Figure~\ref{fig:nsar}).  Determining the metallicity 
from these diagrams is complicated by the large number of galaxies that lie outside the model
grids whose metallicities must be extrapolated.  Recent studies have found a mean
enhancement ratio of Mg/Fe$\approx$+0.3 dex in luminous E's (e.g., Trager 2000a, 
Kuntschner 2001), yet the mean value for our sample is found to be half of this
value.  Because our sample is dominated by lower
mass systems which tend to have solar or only mildly enhanced abundance ratios,
the difference is not unexpected.

\newpage

\clearpage

\clearpage

\clearpage

\clearpage

\clearpage

\clearpage

\clearpage

\newpage

\begin{deluxetable}{lllcrrrrrrrlc}
\tablenum{1}
\tablecolumns{13}
\footnotesize
\tablecaption{Log of Galaxy Observations}
\tablewidth{0pt}
\tablehead{
\colhead{Name} & \colhead{Alt-Name} & \colhead{Env} & \colhead{N} & 
\colhead{S/N} & \colhead{cz$_{meas}$} & \colhead{cz$_{err}$}  & \colhead{cz$_{lit}$} 
& \colhead{$\sigma_{meas}$} & \colhead{$\sigma$$_{err}$} & \colhead{$\sigma_{lit}$} & \colhead{Morph} & 
\colhead{H$\alpha$} \\
\colhead{(1)} &
\colhead{(2)} &
\colhead{(3)} &
\colhead{(4)} &
\colhead{(5)} &
\colhead{(6)} &
\colhead{(7)} &
\colhead{(8)} &
\colhead{(9)} &
\colhead{(10)} &
\colhead{(11)} &
\colhead{(12)} &
\colhead{(13)} 
}
\startdata
A00368+25 & --- & F & 4 & 45 & 4675 & 26 & 4614 & 162 & 7  & --- & E & --- \\
A10025+59 & --- & F & 3 & 68 & 2916 & 23 & 2871 & 219 & 7  & --- & E & --- \\
A15565+64 & --- & F & 3 & 36 & 9260 & 32 & 9216 & 302 & 7  & --- & E & --- \\
A15572+48 & --- & F & 2 & 31 & 5990 & 66 & 6011 & 153 & 7  & 212 & S0 & --- \\
A16044+16 & --- & F & 2 & 42 & 11010 & 44 & 11015 & 229 & 7  & 216 & S0+ & u \\
A23174+26 & --- & F & 4 & 37 & 5915 & 43 & 5864 & 240 & 7  & --- & S0 &  y  \\
A23302+29 & --- & F & 6 & 37 & 5555 & 34 & 5595 & 206 & 14  & --- & S0 &  y  \\
IC0195   & --- & F & 3 & 45 & 3675 & 43 & 3648 & 127 & 7  & --- & S0+ & --- \\
IC0555   & --- & F & 3 & 29 & 6733 & 35 & 6732 & 211 & 8  & --- & S0 & --- \\
IC0598   & --- & F & 1 & 17 & 2246 & 10 & 2258 & 115 & 7  & --- & S0 & --- \\
IC1153   & --- & F & 3 & 44 & 6017 & 24 & 5879 & 224 & 7  & --- & S0 &  y  \\
IC1211   & --- & F & 4 & 28 & 5475 & 20 & 5563 & 209 & 12  & --- & E & --- \\
IC2382   & --- & F & 3 & 60 & 6033 & 41 & 6139 & 282 & 7  & --- & --- & --- \\
M32  & NGC221 & LG & 8 & 166 & -222 & 16 & -145 & 100 & 7  & 79 & E & --- \\
NGC183 & --- & F & 4 & 57 & 5289 & 64 & 5402 & 177 & 7  & --- & E & --- \\
NGC205 & --- & LG & 2 & 71 & -243 & 44 & -241 & 93 & 7  & 97 & dE & --- \\
NGC311 & --- & F & 3 & 51 & 5095 & 41 & 5065 & 263 & 7  & 286 & S0 & --- \\
NGC380 & --- & F & 3 & 48 & 4435 & 121 & 4426 & 261 & 9  & 275 & E &  n  \\
NGC384 & --- & F & 4 & 42 & 4844 & 31 & 4287 & 193 & 7  & --- & E & -- \\
NGC448 & --- & F & 2 & 31 & 1833 & 13 & 1917 & 103 & 7  & 90 & S0 & --- \\
NGC474 & --- & F & 3 & 53 & 2349 & 20 & 2372 & 159 & 7  & 169 & S0+ &  u  \\
NGC502 & --- & F & 2 & 61 & 2570 & 17 & 2489 & 129 & 7  & 74 & S0+ & --- \\
NGC516 & --- & F & 3 & 32 & 2432 & 23 & 2432 & 46 & 7  & --- & S0 & --- \\
NGC525 & --- & F & 4 & 46 & 2192 & 19 & 2146 & 105 & 7  & --- & S0 & --- \\
NGC584 & IC1712 & F & 3 & 73 & 1834 & 27 & 1802 & 210 & 7  & 230 & E & --- \\
NGC636 & --- & F & 2 & 46 & 1838 & 25 & 1860 & 176 & 7  & 166 & E & --- \\
NGC656 & --- & F & 4 & 49 & 3806 & 43 & 3916 & 76 & 7  & --- & S0 &  u  \\
NGC770 & --- & F & 4 & 51 & 2482 & 31 & 2458 & 51 & 7  & --- & E &  n  \\
NGC774 & --- & F & 4 & 48 & 4664 & 25 & 4595 & 173 & 7  & --- & S0 & --- \\
NGC821 & --- & F & 2 & 81 & 1762 & 22 & 1735 & 202 & 7  & 207 & E+ &  n  \\
NGC938 & --- & F & 4 & 60 & 4118 & 26 & 4099 & 186 & 7  & --- & E & --- \\
NGC990 & --- & F & 3 & 62 & 3476 & 54 & 3508 & 137 & 7  & 179 & E & --- \\
NGC1026 & --- & F & 5 & 51 & 4142 & 74 & 4179 & 158 & 19  & 191 & S0 &  n  \\
NGC1029 & --- & F & 3 & 41 & 3658 & 44 & 3635 & 145 & 7  & 159 & S0 &  u  \\
NGC1107 & --- & F & 4 & 35 & 3350 & 72 & 3424 & 176 & 12  & 256 & S0 & --- \\
NGC1153 & --- & F & 4 & 61 & 3116 & 33 & 3126 & 226 & 7  & --- & S0 &  u  \\
NGC1211 & --- & F & 2 & 44 & 3250 & 37 & 3171 & 114 & 7  & 174 & S0+ &  y  \\
NGC1298 & --- & F & 3 & 30 & 6514 & 26 & 6528 & 166 & 7  & 178 & E+ & --- \\
NGC1550 & --- & F & 2 & 46 & 3836 & 48 & 3714 & 295 & 7  & --- & S0 & --- \\
NGC1552 & --- & F & 3 & 38 & 4958 & 96 & 4924 & 232 & 7  & --- & S0 & --- \\
NGC2332 & --- & F & 3 & 39 & 5882 & 42 & 5836 & 252 & 9  & --- & S0 & --- \\
NGC2693 & --- & F & 3 & 69 & 4928 & 68 & 4942 & 267 & 25  & 326 & E3 & --- \\
NGC2778 & --- & F & 2 & 51 & 2082 & 25 & 2016 & 160 & 7  & 180 & E &  u  \\
NGC2880 & --- & F & 3 & 74 & 1608 & 21 & 1551 & 146 & 7  & 142 & S0 & --- \\
NGC2918 & --- & F & 2 & 39 & 6823 & 43 & 6810 & 233 & 7  & --- & E & --- \\
NGC2950 & --- & F & 2 & 91 & 1334 & 17 & 1337 & 168 & 7  & 183 & S0 & --- \\
NGC2954 & --- & F & 2 & 48 & 3743 & 20 & 3821 & 195 & 7  & 218 & E & --- \\
NGC3071 & --- & F & 3 & 28 & 6430 & 29 & 6422 & 150 & 7  & --- & --- &  u  \\
NGC3102 & --- & F & 4 & 22 & 3052 & 12 & 3066 & 134 & 7  & 166 & S0 & --- \\
NGC3156 & --- & F & 2 & 50 & 1375 & 37 & 1266 & 79 & 7  & 87 & S0 &  y  \\
NGC3193 & --- & F & 2 & 76 & 1318 & 17 & 1399 & 197 & 7  & 191 & E & --- \\
NGC3248 & --- & F & 2 & 54 & 1523 & 18 & 1466 & 114 & 7  & --- & S0 & --- \\
NGC3266 & --- & F & 2 & 49 & 1765 & 12 & 1616 & 139 & 7  & --- & S0+ & --- \\
NGC3489 & --- & F & 2 & 116 & 623 & 19 & 677 & 115 & 7  & 138 & S0+ & --- \\
NGC3524 & --- & F & 2 & 30 & 1365 & 7 & 1369 & 109 & 7  & --- & S0 & --- \\
NGC3599 & --- & F & 2 & 40 & 755 & 19 & 835 & 81 & 7  & 79 & S0 &  y  \\
NGC3648 & --- & F & 2 & 51 & 1948 & 20 & 1988 & 170 & 7  & --- & S0 &  y  \\
NGC3731 & --- & F & 4 & 28 & 3124 & 14 & 3212 & 171 & 7  & 176 & E & --- \\
NGC3757 & --- & F & 4 & 37 & 1161 & 11 & 1273 & 165 & 7  & --- & S0 & --- \\
NGC3818 & --- & F & 3 & 36 & 1664 & 26 & 1701 & 183 & 7  & 198 & E & --- \\
NGC3872 & --- & F & 2 & 64 & 3161 & 43 & 3186 & 296 & 7  & 256 & E & --- \\
NGC4283 & --- & F & 3 & 49 & 984 & 7 & 1076 & 130 & 7  & 105 & E & --- \\
NGC4296 & --- & F & 2 & 41 & 4095 & 29 & 4227 & 213 & 7  & 195 & S0 &  u  \\
NGC4308 & --- & F & 3 & 15 & 589 & 10 & 624 & 90 & 7  & 87 & E & --- \\
NGC4648 & --- & F & 3 & 17 & 1414 & 18 & 1474 & 210 & 7  & 224 & E & --- \\
NGC5342 & --- & F & 3 & 54 & 2249 & 17 & 2211 & 178 & 13  & 188 & S0 &  n  \\
NGC5370 & --- & F & 3 & 30 & 3081 & 35 & 3057 & 109 & 7  & 135 & S0+ &  y  \\
NGC5422 & --- & F & 5 & 54 & 1820 & 14 & 1782 & 176 & 8  & --- & S0 & --- \\
NGC5424 & --- & F & 2 & 34 & 6052 & 35 & 5951 & 194 & 7  & 189 & S0 &  y  \\
NGC5459 & --- & F & 3 & 35 & 5193 & 25 & 5261 & 200 & 10  & 217 & S0 & --- \\
NGC5481 & --- & F & 2 & 40 & 2132 & 14 & 2064 & 148 & 7  & 140 & E &  y  \\
NGC5500 & --- & F & 4 & 33 & 1988 & 25 & 1881 & 107 & 7  & --- & E & --- \\
NGC5576 & --- & F & 2 & 88 & 1511 & 14 & 1482 & 182 & 7  & 190 & E & --- \\
NGC5582 & --- & F & 2 & 53 & 1362 & 16 & 1448 & 179 & 7  & 150 & E &  n  \\
NGC5590 & --- & F & 2 & 31 & 3221 & 21 & 3242 & 173 & 7  & --- & S0 &  u  \\
NGC5596 & --- & F & 3 & 38 & 3143 & 20 & 3167 & 124 & 7  & 152 & S0 &  u  \\
NGC5603 & --- & F & 3 & 49 & 5664 & 20 & 5635 & 206 & 8  & --- & S0 &  y  \\
NGC5611 & --- & F & 2 & 56 & 2015 & 22 & 1957 & 152 & 7  & --- & S0 & --- \\
NGC5623 & --- & F & 3 & 40 & 3642? & 30 & 3356 & 273 & 17  & --- & E & y   \\
NGC5629 & --- & F & 2 & 35 & 4444 & 36 & 4498 & 258 & 7  & 261 & S0 & --- \\
NGC5631 & --- & F & 2 & 35 & 2067 & 17 & 1979 & 169 & 8  & 135 & S0+ &  y  \\
NGC5644 & --- & F & 4 & 44 & 7502 & 30 & 7562 & 272 & 7  & --- & S0 & --- \\
NGC5770 & --- & F & 2 & 52 & 1491 & 14 & 1464 & 104 & 7  & 124 & S0 & --- \\
NGC5813 & --- & F & 2 & 62 & 1930 & 23 & 1972 & 233 & 7  & 230 & E &  y  \\
NGC5831 & --- & F & 3 & 60 & 1703 & 22 & 1656 & 167 & 7  & 168 & E & --- \\
NGC5845 & --- & F & 2 & 56 & 1471 & 33 & 1456 & 229 & 7  & 244 & E & --- \\
NGC5869 & --- & F & 3 & 55 & 1934 & 14 & 2087 & 182 & 7  & --- & S0 &  n  \\
NGC5966 & --- & F & 3 & 43 & 4477 & 26 & 4474 & 236 & 7  & 174 & E &  y  \\
NGC5982 & --- & F & 5 & 66 & 3155 & 23 & 3017 & 256 & 7  & 256 & E & --- \\
NGC6003 & --- & F & 2 & 35 & 3937 & 12 & 4060 & 153 & 7  & 184 & S0 &  n  \\
NGC6017 & --- & F & 2 & 38 & 1682 & 22 & 1626 & 117 & 7  & --- & S0+ &  y  \\
NGC6030 & --- & F & 3 & 34 & 4407 & 19 & 4491 & 197 & 7  & --- & S0+ & --- \\
NGC6126 & --- & F & 4 & 37 & 9859 & 40 & 9759 & 226 & 9  & --- & S0+ & --- \\
NGC6127 & --- & F & 2 & 30 & 4601 & 23 & 4644 & 248 & 18  & 225 & E & --- \\
NGC6137 & --- & F & 3 & 36 & 9306 & 38 & 9308 & 303 & 14  & 293 & E & --- \\
NGC7194 & --- & F & 3 & 39 & 8102 & 101 & 8042 & 243 & 7  & --- & E & --- \\
NGC7280 & --- & F & 3 & 59 & 1900 & 17 & 1844 & 98 & 7  & --- & S0+ &  y  \\
NGC7391 & --- & F & 4 & 53 & 2982 & 93 & 3049 & 196 & 7  & 281 & E &  u  \\
NGC7411 & --- & F & 3 & 40 & 6796 & 48 & 6835 & 251 & 18  & --- & E & --- \\
NGC7454 & --- & F & 3 & 32 & 1992 & 20 & 2051 & 120 & 7  & 121 & E & --- \\
NGC7461 & --- & C & 4 & 49 & 4234 & 51 & 4272 & 159 & 7  & --- & S0 & --- \\
NGC7557 & --- & P & 2 & 21 & 3692 & 19 & 3754 & 81 & 7  & --- & S0+ & --- \\
NGC7611 & --- & P & 4 & 79 & 3331 & 51 & 3252 & 194 & 7  & --- & S0 &  n  \\
NGC7612 & --- & P & 3 & 41 & 3186 & 45 & 3213 & 173 & 7  & 196 & S0 & --- \\
NGC7623 & --- & P & 3 & 51 & 3738 & 24 & 3739 & 171 & 7  & --- & S0+ & --- \\
NGC7628 & --- & F & 3 & 45 & 4251 & 45 & 4202 & 148 & 7  & --- & E & --- \\
NGC7671 & --- & F & 4 & 50 & 3904 & 69 & 3875 & 254 & 8  & --- & S0 &  u  \\
NGC7698 & --- & F & 4 & 41 & 6939 & 53 & 7028 & 126 & 9  & --- & S0 &  n  \\
NGC7703 & --- & F & 4 & 41 & 4062 & 22 & 3936 & 146 & 7  & --- & S0 &  y  \\
NGC7707 & --- & F & 4 & 41 & 5544 & 34 & 5484 & 187 & 10  & --- & S0 & --- \\
NGC7711 & --- & F & 4 & 63 & 3999 & 59 & 4057 & 151 & 7  & 181 & S0 &  y  \\
UGC06604 & NGC3795B & F & 6 & 28 & 1437 & 9 & 1358 & 126 & 7  & --- & E & --- \\
UGC08876 & --- & F & 4 & 46 & 2074 & 12 & 2113 & 159 & 7  & --- & S0 & --- \\
UGC12454 & --- & F & 3 & 17 & 4892 & 48 & 4793 & 155 & 15  & --- & S0 & --- \\
UGC12472 & --- & F & 3 & 45 & 6406 & 75 & 6511 & 151 & 7  & --- & S0 &  y  \\
UGC9519 & --- & F & 2 & 18 & 1702 & 29 & 1692 & 95 & 7  & --- & S0 &  y  \\
VCC049 & NGC4168 & V & 2 & 23 & 2215 & 11 & 2284 & 182 & 16  & 186 & E & --- \\
VCC140 & IC3065 & V & 2 & 21 & 1013 & 26 & 1072 & 63 & 13  & --- & S0 &  n  \\
VCC218 & IC3100 & V & 4 & 13 & 454 & 55 & 533 & 87 & 17  & --- & dS0 & y   \\
VCC345 & NGC4261 & V & 4 & 22 & 2220 & 24 & 2238 & 308 & 7  & 326 & E & --- \\
VCC389 & IC0781 & V & 4 & 23 & 1369 & 25 & 1330 & 55 & 15  & --- & dS0 &  n  \\
VCC523 & NGC4306 & V & 3 & 31 & 1981 & 20 & 1508 & 56 & 7  & --- & S0+ &  n  \\
VCC538 & NGC4309A & V & 4 & 15 & 620 & 37 & 500 & 57 & 12  & --- & E &  n  \\
VCC634 & NGC4328 & V & 4 & 7 & 429 & 40 & 499 & 53 & 20  & 36 & dE & --- \\
VCC685 & NGC4350 & V & 3 & 9 & 1200 & 15 & 1241 & 112 & 7  & --- & S0 & --- \\
VCC698 & NGC4352 & V & 7 & 40 & 2070 & 7 & 2106 & 63 & 14  & 62 & S0+ &  n  \\
VCC731 & NGC4365 & V & 2 & 68 & 1222 & 18 & 1243 & 271 & 7  & 261 & E & --- \\
VCC751 & IC3292 & V & 4 & 17 & 697 & 36 & 710 & 68 & 8  & --- & dS0 &  n  \\
VCC758 & NGC4370 & V & 4 & 13 & 726 & 18 & 782 & 104 & 15  & --- & S0+ &  u  \\
VCC781 & IC3303 & V & 4 & 17 & -337 & 54 & -254 & 63 & 16  & --- & dS0 &  y  \\
VCC784 & NGC4379 & V & 2 & 55 & 939 & 20 & 1069 & 153 & 7  & 70 & S0+ & --- \\
VCC828 & NGC4387 & V & 3 & 44 & 472 & 11 & 561 & 105 & 7  & 112 & E & --- \\
VCC856 & IC3328 & V & 3 & 10 & 961 & 34 & 972 & 74 & 16  & 38 & dE & --- \\
VCC929 & NGC4415 & V & 3 & 18 & 839 & 20 & 910 & 65 & 10  & 41 & E & --- \\
VCC944 & NGC4417 & V & 4 & 73 & 820 & 12 & 843 & 164 & 7  & 84 & S0+ & --- \\
VCC966 & NGC4421 & V & 2 & 42 & 1610 & 14 & 1603 & 71 & 7  & --- & S0+ &  n  \\
VCC1010 & NGC4431 & V & 4 & 16 & 878 & 12 & 913 & 75 & 12  & 55 & S0+ &  y  \\
VCC1025 & NGC4434 & V & 3 & 54 & 987 & 13 & 1071 & 128 & 7  & 115 & E & --- \\
VCC1030 & NGC4435 & V & 4 & 83 & 771 & 13 & 801 & 197 & 7  & 168 & S0+ & --- \\
VCC1036 & NGC4436 & V & 4 & 24 & 1004 & 28 & 1135 & 83 & 12  & 40 & dS0 & --- \\
VCC1062 & NGC4442 & V & 4 & 104 & 530 & 17 & 532 & 197 & 7  & 217 & S0+ & --- \\
VCC1073 & IC0794 & V & 4 & 5 & 1860 & 41 & 1899 & 67 & 28  & 41 & dE & --- \\
VCC1125 & NGC4452 & V & 4 & 37 & 165 & 6 & 195 & 114 & 7  & --- & S0 & --- \\
VCC1146 & NGC4458 & V & 3 & 48 & 593 & 14 & 635 & 113 & 7  & 106 & E & --- \\
VCC1178 & NGC4464 & V & 3 & 58 & 1173 & 16 & 1243 & 147 & 7  & 121 & E & --- \\
VCC1183 & IC3413 & V & 3 & 18 & 1283 & 31 & 1387 & 81 & 7  & 69 & dS0 &  n  \\
VCC1199 & --- & V & 4 & 9 & 1201 & 21 & 900 & 69 & 19  & --- & E &  n  \\
VCC1231 & NGC4473 & V & 2 & 83 & 2141 & 23 & 2244 & 201 & 7  & 193 & E & --- \\
VCC1242 & NGC4474 & V & 4 & 54 & 1588 & 7 & 1610 & 112 & 7  & 87 & S0+ & --- \\
VCC1250 & NGC4476 & V & 2 & 47 & 2013 & 20 & 1978 & 69 & 7  & 69 & S0+ &  y  \\
VCC1279 & NGC4478 & V & 2 & 74 & 1402 & 20 & 1349 & 155 & 7  & 143 & E &  n  \\
VCC1283 & NGC4479 & V & 4 & 24 & 912 & 23 & 876 & 93 & 7  & --- & S0+ &  n  \\
VCC1297 & NGC4486B & V & 3 & 53 & 1485 & 21 & 1555 & 172 & 7  & 185 & E & --- \\
VCC1303 & NGC4483 & V & 3 & 52 & 846 & 15 & 875 & 97 & 7  & --- & S0+ & --- \\
VCC1318 & NGC4488 & V & 4 & 31 & 972 & 8 & 980 & 80 & 7  & --- & S0+ & --- \\
VCC1321 & NGC4489 & V & 3 & 51 & 895 & 18 & 967 & 75 & 7  & 51 & S0 & --- \\
VCC1422 & IC3648 & V & 4 & 24 & 1351 & 29 & 1372 & 54 & 8  & 36 & E+ &  n  \\
VCC1440 & IC0798 & V & 3 & 19 & 382 & 21 & 414 & 59 & 9  & --- & E & --- \\
VCC1475 & NGC4515 & V & 3 & 61 & 1011 & 21 & 1018 & 104 & 7  & 91 & E & --- \\
VCC1479 & NGC4516 & V & 3 & 8 & 917 & 9 & 958 & 118 & 7  & --- & S0+ & --- \\
VCC1488 & IC3487 & V & 3 & 17 & 1038 & 60 & 1157 & 108 & 38  & --- & E &  n  \\
VCC1521 & IC3499 & V & 4 & 36 & 1198 & 20 & 1212 & 80 & 7  & --- & S0 &  n  \\
VCC1535 & NGC4526 & V & 4 & 104 & 572 & 25 & 448 & 282 & 11  & 267 & S0+ & --- \\
VCC1537 & NGC4528 & V & 4 & 65 & 1342 & 8 & 1374 & 124 & 7  & 107 & S0+ & --- \\
VCC1545 & IC3509 & V & 3 & 12 & 2000 & 22 & 2050 & 65 & 10  & --- & E & --- \\
VCC1614 & IC3540 & V & 4 & 23 & 754 & 36 & 753 & 58 & 7  & --- & S0+ &  y  \\
VCC1630 & NGC4451 & V & 3 & 47 & 1109 & 11 & 1172 & 109 & 7  & --- & E & --- \\
VCC1827 & NGC4598 & V & 4 & 26 & 2020 & 23 & 1961 & 65 & 7  & 92 & S0 &  n  \\
VCC1871 & IC3653 & V & 4 & 33 & 402 & 21 & 603 & 117 & 7  & 51 & E & --- \\
VCC1903 & NGC4621 & V & 3 & 49 & 453 & 17 & 410 & 228 & 7  & 230 & E & --- \\
VCC1912 & IC0810 & V & 4 & 30 & -128 & 35 & -169 & 64 & 14  & --- & dS0 &  n  \\
VCC1938 & NGC4638 & V & 4 & 88 & 1135 & 11 & 1164 & 157 & 7  & 132 & S0 & --- \\
VCC1939 & NGC4636 & V & 2 & 76 & 938 & 15 & 938 & 203 & 13  & 207 & E & --- \\
VCC2000 & NGC4660 & V & 2 & 74 & 989 & 20 & 1083 & 187 & 7  & 185 & E & --- \\
VCC2048 & IC3773 & V & 4 & 24 & 1144 & 38 & 1095 & 79 & 14  & 40 & dS0 &  n  \\
\enddata
\tablecomments{\\
column (1): Galaxy name\\
column (2): Alternate NGC or IC name from NED\\
column (3): F=``field''; V=Virgo Cluster; P=Pegasus Cluster; LG=Local Group\\
column (4): Number of combined spectra\\
column (5): Signal-to-noise ratio of combined spectrum, measured near 4000 \AA\\
column (6-7): Measured heliocentric radial velocity\\
column (8): Heliocentric radial velocity from NED or from Falco \etal
(1999)\\
column (9-10): Measured velocity dispersion \& errors \\
columns (11): Velocity dispersions, from McElroy (1995), Simien \& Prugniel
(2002), Pedraz \etal (2002), and Geha \etal (2002). \\
column (12): Morphology from NED (NASA Extragalactic Database). Those
in turn originate either from de Vaucouleurs \etal (1991), or Bingelli \etal (1985).  An E+ or S0+ is our own note to indicate
 the presence of rings, shells, or other peculiarities based on the 
NED designation\\
column (13): H$\alpha$ detection.  A `y' indicates a detection of H$\alpha$, 
`n' indicates no detection, and `u' indicates uncertain detection.}
\label{tab:log}
\end{deluxetable}

\begin{deluxetable}{lccc}
\tablenum{2}
\tablewidth{0pt}
\tablecaption{List of Spectral Indices Measured}
\tablehead{ \colhead{Spectral Index} & \colhead{Ref\tablenotemark{a}} &
\colhead{Col\tablenotemark{b}} & \colhead{Unit}}

\startdata
H$\delta$/$\lambda$4045 & R94 & (2) & \\
H$\delta$/$\lambda$4063 & " & (4) & \\
SrII$\lambda$4077/$\lambda$4045 & " & (6) & \\
SrII$\lambda$4077/$\lambda$4063 & " & (8) & \\
H$\gamma$/$\lambda$4325 & " & (10) & \\
$\lambda$4289/$\lambda$4271 & " & (12) & \\
$\lambda$4384/$\lambda$4352 & " & (14) & \\
p[Fe/H] & " & (16) & \\
Ca II H/K& " & (18) & \\
$\lambda$3888/$\lambda$3859 & " & (20) & \\
p4220/p4209 & " & (22) & \\
EW(Fe4045) & " & (24)  & \AA \\
EW(Ca4226) & " & (26) & \AA \\
EW(H$\gamma$) & " & (28) & \AA \\
H$\delta$$_A$ & WO97 & (30) & \AA \\
H$\gamma$$_A$ & " & (32) & \AA \\
H$\delta$$_F$ & " & (34) & \AA \\
H$\gamma$$_F$ & " & (36) & \AA \\
Ca4227 & W94 & (38) & \AA \\
G4300 & " & (40) & \AA \\
Fe4383 & " & (42) & \AA \\
H$\beta$ & " & (44) & \AA \\
Fe5015 & " & (46) & \AA \\
Mg$~b$ & " & (48) & \AA \\
Fe5270 & " & (50) & \AA \\
CN1 & " & (52) & mag \\
CN2 & " & (54) & mag \\
Mg$_1$ & " & (56) & mag \\
Mg$_2$ & " & (58) & mag \\
Hn/Fe & CRC00 & (60) & \\
\enddata
\tablenotetext{a}
{
References: R94 (Rose, 1994); WO97 (Worthey \& Ottaviani, 1997); W94
(Worthey \etal ,1994); CRC00( Concannon, Rose, \& Caldwell 2000)
}
\tablenotetext{b}
{{Column number in Table 3}
}
\footnotesize
\label{tab:listind}
\end{deluxetable}

\begin{deluxetable}{lrrrrrrrrllll}
\rotate
\tabletypesize{\scriptsize}
\setlength{\tabcolsep}{0.02in}
\tablenum{3}
\tablecaption{Spectral Indices and Their Errors (Full Table Available as a Separate File)}
\tablewidth{0pt}
\tablehead{\colhead{Galaxy ID} &
\multicolumn{1}{l}{$\frac{H\delta\tablenotemark{a}}{\lambda4045}$} &
\multicolumn{1}{l}{$\frac{H\delta}{\lambda4063}$} &
\multicolumn{1}{l}{$\frac{SrII\lambda4077}{\lambda4045}$} &
\multicolumn{1}{l}{$\frac{SrII\lambda4077}{\lambda4063}$} &
\multicolumn{1}{l}{$\frac{H\gamma\tablenotemark{a}}{\lambda4325}$} &
\multicolumn{1}{l}{$\frac{\lambda4289}{\lambda4271}$} &
\multicolumn{1}{l}{$\frac{\lambda4384}{\lambda4352}$} &
\multicolumn{1}{l}{p[Fe/H]} &
\multicolumn{1}{l}{Ca II} &
\multicolumn{1}{l}{$\frac{\lambda3888\tablenotemark{a}}{\lambda3859}$} &
\multicolumn{1}{l}{$\frac{p4220}{p4209}$} \\
\colhead{} &
\multicolumn{1}{l}{EW(Fe)} &
\multicolumn{1}{l}{EW(Ca)} &
\multicolumn{1}{l}{EW(H$\gamma$)} &
\multicolumn{1}{l}{H$\delta$$_A$} &
\multicolumn{1}{l}{H$\gamma$$_A$} &
\multicolumn{1}{l}{H$\delta$$_F$} &
\multicolumn{1}{l}{H$\gamma$$_F$} &
\multicolumn{1}{l}{Ca4227} &
\multicolumn{1}{l}{G4300} &
\multicolumn{1}{l}{Fe4383\tablenotemark{b}} &
\multicolumn{1}{l}{H$\beta$\tablenotemark{a,b}}\\
\colhead{} &
\multicolumn{1}{l}{Fe5015} &
\multicolumn{1}{l}{Mg$~b$\tablenotemark{b}} &
\multicolumn{1}{l}{Fe5270\tablenotemark{b}} &
\multicolumn{1}{l}{CN1} &
\multicolumn{1}{l}{CN2} &
\multicolumn{1}{l}{Mg$_1$\tablenotemark{b}} &
\multicolumn{1}{l}{Mg$_2$\tablenotemark{b}} &
\multicolumn{1}{l}{Hn/Fe\tablenotemark{a}} &
\colhead{$\Delta$H$\beta$$_{emiss}$\tablenotemark{c}} &
\colhead{$\Delta$H$\beta$$_{NSAR}$\tablenotemark{d}} &
\colhead{} & \colhead{}
}
\startdata
A00368+25   &  0.925~~0.013&  0.940~~0.013&  1.012~~0.015&  1.028~~0.015&  0.995~~0.010&  1.020~~0.010&  0.876~~0.008&  1.100~~0.013&  1.166~~0.058&  1.109~~0.033&  1.002~~0.010\\
&  0.328~~0.040&  0.567~~0.031&  0.458~~0.027& -1.441~~0.392& -5.524~0.318&  0.852~~0.220& -1.463~~0.167&  1.245~~0.144&  5.595~~0.234&  4.575~~0.308&  1.632~~0.114\\
&  4.160~~0.250&  3.614~~0.129&  2.729~~0.135&  0.058~~0.009&  0.104~~0.011& -0.266~~0.004&  0.083~~0.006&  1.009~~0.013&  0.018& -0.053 & &\\
A10025+59   &  0.977~~0.009&  0.988~~0.010&  1.009~~0.011&  1.019~~0.011&  1.004~~0.007&  1.023~~0.007&  0.878~~0.006&  1.137~~0.009&  1.215~~0.041&  1.212~~0.024&  1.016~~0.007\\
&  0.393~~0.048&  0.592~~0.038&  0.479~~0.033& -2.156~~0.202& -6.376~0.164&  0.663~~0.114& -1.782~~0.086&  1.251~~0.074&  5.854~~0.120&  4.919~~0.159&  1.634~~0.059\\
&  5.014~~0.129&  3.911~~0.067&  2.998~~0.070&  0.093~~0.004&  0.138~~0.004& -0.240~~0.002&  0.110~~0.002&  1.065~~0.009&  0.033& -0.051 & &\\
A15572+48   &  0.893~~0.014&  0.919~~0.015&  1.007~~0.017&  1.036~~0.016&  0.962~~0.011&  0.983~~0.011&  0.876~~0.009&  1.093~~0.014&  1.136~~0.063&  1.047~~0.036&  0.988~~0.011\\
&  0.227~~0.056&  0.470~~0.044&  0.512~~0.039& -0.434~~0.394& -4.848~0.319&  1.010~~0.221& -0.924~~0.167&  1.201~~0.144&  5.210~~0.235&  4.181~~0.309&  1.859~~0.115\\
&  4.451~~0.251&  3.108~~0.130&  2.573~~0.136&  0.014~~0.005&  0.054~~0.007& -0.082~~0.002&  0.132~~0.003&  0.967~~0.014&  0.046& -0.021 & &\\
A16044+16   &  0.914~~0.019&  0.923~~0.020&  1.009~~0.023&  1.019~~0.022&  0.965~~0.015&  1.006~~0.015&  0.889~~0.012&  1.111~~0.019&  1.115~~0.085&  1.100~~0.049&  1.008~~0.014\\
&  0.262~~0.036&  0.499~~0.029&  0.514~~0.025& -1.095~~0.207& -4.869~0.168&  1.113~~0.116& -0.809~~0.088&  1.312~~0.076&  5.327~~0.123&  4.645~~0.163&  2.164~~0.060\\
&  4.862~~0.132&  2.331~~0.068&  2.274~~0.071&  0.051~~0.002&  0.091~~0.002& -0.987~~0.001& -0.262~~0.001&  0.993~~0.019&  0.000& -0.147 & &\\
A23174+26   &  0.969~~0.024&  0.970~~0.025&  1.030~~0.028&  1.031~~0.028&  1.008~~0.019&  0.983~~0.019&  0.838~~0.015&  1.094~~0.024&  1.278~~0.107&  1.242~~0.061&  1.047~~0.018\\
&  0.608~~0.066&  0.440~~0.052&  0.381~~0.046& -2.593~~0.408& -6.610~0.331&  0.096~~0.229& -1.801~~0.173&  1.221~~0.150&  5.890~~0.243&  5.433~~0.321&  1.394~~0.119\\
&  4.829~~0.260&  3.947~~0.135&  2.966~~0.141&  0.104~~0.012&  0.151~~0.014& -0.378~~0.005&  0.071~~0.007&  1.073~~0.023&  0.264& -0.073 & &\\
A2330+29    &  0.994~~0.028&  0.988~~0.030&  1.043~~0.034&  1.037~~0.033&  1.028~~0.023&  1.045~~0.022&  0.866~~0.019&  1.123~~0.029&  1.152~~0.127&  1.171~~0.072&  1.012~~0.022\\
&  0.413~~0.108&  0.428~~0.085&  0.322~~0.075& -2.348~~0.525& -6.263~0.426&  0.360~~0.295& -1.747~~0.223&  1.438~~0.192&  5.524~~0.313&  5.432~~0.412&  1.591~~0.153\\
&  4.847~~0.335&  4.231~~0.173&  3.247~~0.181&  0.108~~0.023&  0.156~~0.028& -0.248~~0.010&  0.158~~0.014&  1.064~~0.028&  0.282& -0.055 & &\\
\enddata
\tablenotetext{a}{Corrected for emission}
\tablenotetext{b}{Corrected for non-solar abundance ratios (NSAR)}
\tablenotetext{c}{Correction to the Lick H$\beta$ index due to emission}
\tablenotetext{d}{Correction to the Lick H$\beta$ index due to NSAR}
\label{tab:indices}
\end{deluxetable}

\begin{deluxetable}{lccccc}
\tablenum{4}
\tablewidth{0pt}
\tablecaption{Measured Intrinsic Scatter}
\tablehead{\colhead{Group} & \colhead{Num} & \colhead{H$\beta$\tablenotemark{a}}
 & \colhead{Hn/Fe}
& \colhead{Mg b\tablenotemark{a}} & \colhead{Fe 5270\tablenotemark{a}}}
\startdata
\cutinhead{Scatter for Uncorrected Indices}
log $\sigma>$2.0 & 100 & 0.201$\pm$0.021 & 0.023$\pm$0.004 & 0.447$\pm$0.027 & 0.263$\pm$0.019 \\
\footnotesize (median index error) & \nodata & ($\pm$0.078) & ($\pm$0.014) & ($\pm$0.088) & ($\pm$0.090) \\
log $\sigma<$2.0 &  38 & 0.383$\pm$0.093 & 0.078$\pm$0.023 & 0.567$\pm$0.104 & 0.179$\pm$0.025 \\
\footnotesize (median index error) & \nodata & ($\pm$0.155) & ($\pm$0.031) & ($\pm$0.175) & ($\pm$0.184) \\
\cutinhead{Scatter for Emission-Corrected Indices}
log $\sigma>$2.0 & 100 & 0.214$\pm$0.027 & 0.021$\pm$0.005 & 0.449$\pm$0.029 & 0.264$\pm$0.020 \\
log $\sigma<$2.0 &  38 & 0.600$\pm$0.107 & 0.084$\pm$0.021 & 0.572$\pm$0.100 & 0.193$\pm$0.028 \\
\cutinhead{Scatter for Emission and NSAR-Corrected Indices}
log $\sigma>$2.0 & 100 & 0.208$\pm$0.025 & 0.021$\pm$0.005 & 0.297$\pm$0.023 & 0.144$\pm$0.017 \\
log $\sigma<$2.0 &  38 & 0.581$\pm$0.107 & 0.084$\pm$0.021 & 0.546$\pm$0.092 & 0.282$\pm$0.045 \\
\enddata
\tablenotetext{a}{The Lick indices have $not$ been converted to magnitudes}
\label{tab:inscat}
\end{deluxetable}

\begin{deluxetable}{lccc}
\tablenum{5}
\tablecaption{Summary of Mean Ages by log $\sigma$}
\tablewidth{0pt}
\tablehead{\colhead{log $\sigma$ Range} & \colhead{Number} &
\colhead{Mean Age} & \colhead {RMS}}
\startdata
log $\sigma<$ 1.8           & 12 &  3.6 &  1.9  \\
1.8 $\le$log $\sigma<$  2.0 & 30 &  3.6 &  2.1  \\
2.0$\le$log $\sigma<$  2.2  & 48 &  7.4 &  4.2  \\
log $\sigma\ge$ 2.2           & 61 &  9.9 &  4.2  \\
\enddata
\label{tab:agesum}
\end{deluxetable}

\begin{deluxetable}{cccc}
\tablenum{6}
\tablewidth{0pt}
\tablecaption{Kolmogorov-Smirnoff Test for Ages of Different log $\sigma$ Groups
}
\tablehead{ \colhead{Group Combination} & \colhead{Likelihood} &
\colhead{Group Combination} & \colhead{Likelihood}
}
\startdata
1-2 & 8.82$\times$10$^{-1}$ & 2-3 & 9.42$\times$10$^{-6}$\\
1-3 & 2.92$\times$10$^{-4}$ & 2-4 & 4.59$\times$10$^{-9}$\\
1-4 & 2.32$\times$10$^{-5}$ & 3-4 & 5.62$\times$10$^{-4}$\\
\enddata
\footnotesize
\tablecomments{Groups by velocity dispersion are:  Group1 --- log $\sigma <$ 1.8
,\\
Group 2 --- 1.8 $\le$ log $\sigma <$ 2.0, Group 3 --- 2.0 $\le$ log $\sigma <$ 2.
2,
Group 4 --- log $\sigma >$ 2.2}
\label{tab:ks_table}
\end{deluxetable}

\begin{deluxetable}{lcccc}
\tablenum{7}
\tablewidth{0pt}
\tablecaption{Effects of Index Corrections on Age and [Fe/H] Determinations }
\tablehead{ \colhead{Corrections} &
\colhead{$<$Age$>$} & \colhead{$<$[Fe/H]$>$} &
\colhead{$<$Age$>$} & \colhead{$<$[Fe/H]$>$} }
\startdata
\multicolumn{5}{c}{\hskip 3cm Fe4383 vs Hn/Fe \hskip 2cm Fe5270 vs H$\beta$}\\
None           & 8.9$\pm$5.3 & 0.01$\pm$ 0.37 & 11.1 $\pm$ 6.3 & -0.15$\pm$ 0.17
 \nl
Emission       & 8.2$\pm$5.0 & 0.01$\pm$ 0.33 &  9.3 $\pm$ 6.2 & -0.12$\pm$ 0.16
 \nl
Emission + NSAR & \nodata & \nodata & 9.3 $\pm$ 6.0 & -0.03$\pm$ 0.17 \nl
\enddata
\label{tab:indcor}
\end{deluxetable}

\begin{deluxetable}{lrrrrrrrr}
\tablenum{8}
\tablecolumns{9}
\footnotesize
\tablecaption{Age and Metallicity Determinations\tablenotemark{a}}
\tablewidth{0pt}
\tablehead{
\colhead{Galaxy} & \multicolumn{4}{c}{Ages} & \multicolumn{4}{c}{Metallicities}\\
\colhead{(1)} & \colhead{(2)} & \colhead{(3)} & \colhead{(4)} & \colhead{(5)} 
& \colhead{(6)} & \colhead{(7)} & \colhead{(8)} & \colhead{(9)} 
}	

\startdata

 A00368+25 &  11.2$\pm$6.8 &   8.2 &  15.5 &  15.1 &  -0.25$\pm$0.22   &  -0.10 &  -0.31 &  -0.21\\
 A10025+59 &  19.2$\pm$0.2 &  10.5 &  12.2 &  12.2 &   1.00$\pm$0.19   &   0.00 &  -0.18 &  -0.02\\
 A15572+48 &   7.3$\pm$5.4 &   6.4 &   8.4 &   8.2 &  -0.25$\pm$0.24   &  -0.17 &  -0.28 &  -0.20\\
 A16044+16 &   7.5$\pm$4.1 &  14.9 &   2.7 &   4.0 &  -0.14$\pm$0.17   &  -0.44 &   0.19 &  -0.30\\
 A23174+26 &  13.3$\pm$0.0 &  11.7 &  \nodata &  \nodata &  -0.10$\pm$0.00   &  -0.05 &  \nodata &  \nodata\\
  A2330+29 &  11.6$\pm$9.4 &   7.1 &  12.2 &  12.0 &  -0.08$\pm$0.29   &   0.17 &  -0.07 &   0.08\\
    IC0195 &   7.9$\pm$3.4 &   8.4 &   6.8 &   7.6 &  -0.12$\pm$0.17   &  -0.15 &  -0.07 &  -0.14\\
    IC0555 &   7.4$\pm$4.6 &   7.0 &   7.1 &   7.9 &   0.03$\pm$0.24   &   0.06 &   0.04 &   0.04\\
     IC1153 &   8.5$\pm$5.3 &   9.3 &   5.8 &   8.3 &   0.05$\pm$0.18   &   0.00 &   0.14 &   0.03\\
    IC1211 &  11.7$\pm$1.8 &   4.7 &   6.7 &   4.5 &  -0.16$\pm$0.17   &   0.25 &   0.01 &   0.26\\
    IC598a &   3.0$\pm$0.0 &   2.8 &   2.1 &   2.3 &  -0.26$\pm$0.00   &  -0.13 &  -0.04 &  -0.08\\
       M32 &   2.6$\pm$0.2 &   3.2 &   2.9 &   2.9 &   0.27$\pm$0.05   &   0.10 &   0.23 &   0.10\\
   NGC183 &  17.1$\pm$0.9 &   9.4 &   9.6 &   9.5 &  -0.33$\pm$0.02   &  -0.04 &  -0.19 &  -0.04\\
   NGC205 &   0.9$\pm$0.0 &   1.2 &   1.3 &   1.3 &  -0.37$\pm$0.02   &  -0.67 &  -0.49 &  -0.73\\
   NGC448 &   8.8$\pm$5.9 &  14.3 &   7.3 &   8.8 &  -0.11$\pm$0.23   &  -0.33 &  -0.06 &  -0.28\\
    NGC474 &   7.9$\pm$3.3 &   6.4 &   7.4 &   7.9 &  -0.06$\pm$0.17   &   0.06 &  -0.04 &   0.03\\
    NGC502 &   7.7$\pm$1.0 &   6.6 &   9.0 &   8.8 &   0.03$\pm$0.03   &   0.09 &   0.00 &   0.07\\
    NGC516 &   2.0$\pm$0.6 &   3.6 &   2.7 &   3.1 &   0.27$\pm$0.21   &  -0.26 &   0.16 &  -0.27\\
    NGC525 &   6.3$\pm$3.5 &   6.2 &   6.3 &   6.6 &  -0.12$\pm$0.20   &  -0.10 &  -0.11 &  -0.11\\
    NGC584 &   7.4$\pm$2.5 &   6.6 &   8.4 &   8.9 &   0.03$\pm$0.10   &   0.07 &   0.00 &   0.04\\
    NGC636 &   8.8$\pm$3.9 &   6.0 &  10.5 &  10.1 &  -0.08$\pm$0.18   &   0.11 &  -0.13 &   0.04\\
    NGC656 &   4.8$\pm$3.3 &   6.7 &   8.1 &   8.1 &   0.10$\pm$0.26   &  -0.03 &   0.00 &  -0.05\\
    NGC770 &   8.0$\pm$4.3 &   7.5 &   6.8 &   7.1 &  -0.15$\pm$0.19   &  -0.12 &  -0.11 &  -0.11\\
    NGC774 &  16.3$\pm$1.0 &   7.8 &  11.3 &  10.2 &  -0.35$\pm$0.19   &   0.00 &  -0.28 &  -0.04\\
    NGC821 &  11.6$\pm$2.4 &   9.3 &   9.8 &  10.2 &  -0.05$\pm$0.08   &   0.09 &   0.02 &   0.07\\
    NGC938 &   9.9$\pm$7.0 &   6.5 &   7.1 &   7.3 &  -0.24$\pm$0.22   &   0.00 &  -0.16 &  -0.01\\
    NGC990 &  11.3$\pm$4.2 &   8.5 &   8.3 &   8.0 &  -0.20$\pm$0.13   &  -0.04 &  -0.11 &  -0.03\\
   NGC1026 &  10.9$\pm$6.0 &   8.5 &  \nodata &  \nodata &  -0.15$\pm$0.19   &   0.00 &  \nodata &  \nodata\\
   NGC1029 &   8.2$\pm$3.5 &   7.6 &  \nodata &  \nodata &   0.06$\pm$0.13   &   0.08 &  \nodata &  \nodata\\
   NGC1107 &   9.9$\pm$5.8 &   8.2 &  \nodata &  17.3 &  -0.04$\pm$0.21   &   0.07 &  \nodata &  -0.07\\
   NGC1153 &   9.4$\pm$4.5 &   9.9 &  \nodata &  \nodata &   0.06$\pm$0.17   &   0.03 &  \nodata &  \nodata\\
   NGC1211 &   4.0$\pm$1.7 &   4.9 &   4.0 &   4.2 &   0.11$\pm$0.18   &   0.01 &   0.11 &   0.02\\
   NGC1298 &   6.4$\pm$7.6 &   5.7 &   4.9 &   5.5 &  -0.07$\pm$0.40   &   0.00 &  -0.01 &   0.00\\
   NGC1552 &  13.8$\pm$0.0 &  12.5 &  \nodata &  \nodata &  -0.18$\pm$0.00   &  -0.13 &  \nodata &  \nodata\\
   NGC2778 &  10.6$\pm$3.4 &   6.5 &   9.8 &  10.0 &  -0.12$\pm$0.13   &   0.12 &  -0.09 &   0.07\\
   NGC2880 &   7.6$\pm$1.9 &   6.3 &   4.7 &   5.6 &  -0.11$\pm$0.10   &   0.02 &   0.02 &   0.04\\
   NGC2918 &  \nodata &   6.9 &   8.9 &  11.3 & \nodata   &   0.04 &   0.19 &  -0.04\\
   NGC2950 &   7.3$\pm$1.0 &   4.5 &   3.2 &   3.2 &   0.02$\pm$0.05   &   0.21 &   0.25 &   0.27\\
   NGC2954 &   3.9$\pm$1.6 &   2.8 &   2.6 &   2.9 &   0.10$\pm$0.16   &   0.30 &   0.25 &   0.30\\
   NGC3071 &  10.7$\pm$6.4 &   9.3 &  12.4 &  14.3 &  -0.15$\pm$0.22   &  -0.07 &  -0.17 &  -0.14\\
   NGC3102 &   7.6$\pm$13.2 &   5.8 &   3.7 &   4.0 &  -0.18$\pm$0.53   &   0.01 &   0.04 &   0.03\\
  NGC3156 &   0.8$\pm$0.0 &   1.2 &   0.7 &   1.2 &   0.05$\pm$0.21   &  -0.26 &   0.10 &  -0.24\\
   NGC3193 &   9.1$\pm$1.7 &   7.1 &  11.7 &  11.9 &  -0.01$\pm$0.08   &   0.10 &  -0.10 &   0.02\\
   NGC3248 &   3.2$\pm$0.9 &   3.2 &   2.3 &   2.6 &   0.15$\pm$0.09   &   0.16 &   0.30 &   0.17\\
   NGC3266 &  11.1$\pm$3.3 &   9.4 &  15.4 &  14.6 &  -0.25$\pm$0.08   &  -0.17 &  -0.31 &  -0.24\\
   NGC3489 &   1.8$\pm$0.1 &   1.9 &   1.5 &   1.7 &   0.19$\pm$0.06   &   0.15 &   0.32 &   0.18\\
   NGC3524 &   4.9$\pm$1.3 &   4.0 &   3.2 &   3.1 &  -0.09$\pm$0.11   &   0.05 &   0.05 &   0.07\\
   NGC3599 &   1.9$\pm$0.4 &   1.8 &   1.1 &   1.3 &   0.17$\pm$0.14   &   0.23 &   0.55 &   0.32\\
   NGC3648 &  13.8$\pm$3.1 &   6.4 &  \nodata &  17.6 &  -0.14$\pm$0.09   &   0.22 &  \nodata &   0.09\\
   NGC3731 &  13.9$\pm$2.0 &  12.7 &  16.7 &  \nodata &  -0.14$\pm$0.11   &  -0.09 &  -0.18 &  \nodata\\
   NGC3757 &  13.5$\pm$1.7 &   8.2 &  11.4 &  10.4 &  -0.25$\pm$0.06   &   0.00 &  -0.22 &  -0.05\\
   NGC3818 &   8.9$\pm$5.8 &   5.8 &  10.7 &  11.0 &  -0.01$\pm$0.23   &   0.16 &  -0.08 &   0.07\\
   NGC4283 &   7.2$\pm$3.3 &   5.9 &   6.7 &   6.7 &   0.09$\pm$0.14   &   0.18 &   0.11 &   0.16\\
   NGC4296 &   7.8$\pm$3.3 &   5.7 &   5.8 &   6.3 &   0.08$\pm$0.13   &   0.21 &   0.16 &   0.19\\
   NGC4308 &   5.7$\pm$9.7 &   6.4 &   6.8 &   6.8 &   0.00$\pm$0.59   &  -0.08 &  -0.06 &  -0.09\\
   NGC4648 &   4.1$\pm$0.6 &   2.8 &  \nodata &  13.2 &   0.16$\pm$0.23   &   0.39 &  \nodata &   0.16\\
   NGC5342 &  16.6$\pm$0.0 &   6.7 &  \nodata &  15.6 &  -0.36$\pm$0.20   &   0.07 &  \nodata &  -0.05\\
   NGC5370 &   6.5$\pm$3.0 &   4.1 &   6.6 &   5.2 &   0.00$\pm$0.18   &   0.19 &   0.00 &   0.16\\
   NGC5422 &   8.4$\pm$5.4 &   6.4 &  10.3 &  10.4 &   0.05$\pm$0.24   &   0.15 &  -0.01 &   0.08\\
   NGC5424 &  13.3$\pm$5.1 &   9.5 &  \nodata &  \nodata &  -0.22$\pm$0.12   &  -0.05 &  \nodata &  \nodata\\
   NGC5459 &  12.2$\pm$9.4 &   8.3 &  11.0 &  14.2 &  -0.25$\pm$0.26   &  -0.07 &  -0.23 &  -0.17\\
   NGC5481 &   8.5$\pm$2.8 &   8.4 &  \nodata &  \nodata &  -0.10$\pm$0.12   &  -0.09 &  \nodata &  \nodata\\
   NGC5500 &   5.6$\pm$4.3 &   6.2 &  10.0 &  10.7 &  -0.20$\pm$0.30   &  -0.26 &  -0.34 &  -0.33\\
   NGC5576 &   5.9$\pm$1.4 &   6.7 &   6.1 &   6.8 &   0.05$\pm$0.07   &  -0.01 &   0.04 &  -0.01\\
   NGC5582 &  10.5$\pm$4.6 &   4.8 &   8.8 &   8.5 &  -0.29$\pm$0.13   &   0.09 &  -0.24 &   0.05\\
   NGC5590 &  10.3$\pm$5.5 &   5.7 &  10.9 &   9.8 &  -0.20$\pm$0.18   &   0.10 &  -0.21 &   0.03\\
  NGC5596 &   2.6$\pm$0.2 &   7.0 &  10.5 &  15.3 &   0.41$\pm$0.12   &  -0.17 &   0.00 &  -0.26\\
  NGC5603 &   3.0$\pm$0.5 &   4.1 &   5.7 &   6.7 &   0.41$\pm$0.12   &   0.20 &   0.17 &   0.14\\
   NGC5611 &   6.4$\pm$1.1 &   7.1 &   8.9 &   8.7 &  -0.11$\pm$0.08   &  -0.18 &  -0.20 &  -0.20\\
   NGC5631 &   3.1$\pm$0.9 &   2.7 &   2.8 &   3.0 &   0.19$\pm$0.11   &   0.30 &   0.24 &   0.30\\
   NGC5770 &   5.6$\pm$1.4 &   4.8 &   3.2 &   3.6 &  -0.13$\pm$0.10   &  -0.06 &   0.03 &  -0.03\\
   NGC5831 &   3.0$\pm$0.6 &   4.4 &   6.3 &   7.1 &   0.35$\pm$0.11   &   0.13 &   0.12 &   0.08\\
   NGC5845 &   9.8$\pm$0.0 &   7.9 &  10.0 &  10.8 &  -0.01$\pm$0.00   &   0.08 &  -0.02 &   0.04\\
   NGC5865 &   7.4$\pm$3.7 &   6.8 &  10.6 &  10.6 &   0.08$\pm$0.15   &   0.12 &   0.00 &   0.06\\
  NGC5966  &   4.7$\pm$0.0 &   6.1 &   2.1 &   2.9 &   0.11$\pm$0.00   &   0.02 &   0.41 &   0.08\\
   NGC6003 &   4.9$\pm$3.4 &   7.8 &   6.5 &   7.9 &   0.11$\pm$0.23   &  -0.11 &   0.06 &  -0.09\\
   NGC6017 &   5.9$\pm$3.3 &   3.1 &   2.7 &   2.8 &  -0.17$\pm$0.24   &   0.17 &   0.09 &   0.19\\
   NGC6030 &   6.5$\pm$6.3 &  10.8 &  11.3 &  15.9 &   0.09$\pm$0.35   &  -0.17 &  -0.06 &  -0.22\\
   NGC6126 &   9.1$\pm$6.9 &   3.8 &  10.9 &   7.9 &   0.00$\pm$0.28   &   0.40 &  -0.06 &   0.27\\
   NGC7280 &   3.7$\pm$1.1 &   3.7 &   2.5 &   2.7 &   0.05$\pm$0.12   &   0.04 &   0.20 &   0.08\\
   NGC7391 &  12.4$\pm$5.8 &   7.5 &  \nodata &  15.9 &  -0.12$\pm$0.17   &   0.11 &  \nodata &   0.01\\
   NGC7454 &   4.9$\pm$2.3 &   6.7 &   4.2 &   5.6 &  -0.03$\pm$0.17   &  -0.21 &   0.03 &  -0.18\\
   NGC7461 &   8.7$\pm$4.9 &   7.2 &  \nodata &  \nodata &  -0.07$\pm$0.23   &   0.03 &  \nodata &  \nodata\\
   NGC7557 &   1.7$\pm$0.1 &   2.6 &   2.6 &   2.9 &   0.31$\pm$0.23   &  -0.15 &   0.11 &  -0.19\\
   NGC7611 &   2.8$\pm$0.5 &   3.8 &   3.7 &   4.1 &   0.27$\pm$0.13   &   0.08 &   0.16 &   0.07\\
   NGC7612 &   2.8$\pm$0.9 &   4.9 &  11.3 &  12.9 &   0.40$\pm$0.13   &   0.06 &  -0.02 &  -0.05\\
   NGC7623 &   5.9$\pm$2.1 &   6.4 &  11.0 &  11.5 &   0.09$\pm$0.10   &   0.06 &  -0.08 &  -0.04\\
   NGC7628 &   6.6$\pm$3.1 &   6.7 &   3.3 &   3.5 &  -0.01$\pm$0.17   &  -0.03 &   0.17 &   0.03\\
   NGC7698 &  10.0$\pm$6.9 &   6.6 &   4.8 &   5.7 &  -0.14$\pm$0.24   &   0.07 &   0.04 &   0.09\\
   NGC7703 &   3.5$\pm$2.2 &   4.5 &   7.0 &   6.9 &   0.10$\pm$0.27   &  -0.06 &  -0.11 &  -0.10\\
  NGC7707 &  19.1$\pm$0.0 &   7.8 &   8.8 &   8.9 &   1.00$\pm$0.24   &   0.04 &  -0.20 &   0.04\\
   NGC7711 &   6.2$\pm$2.4 &   4.1 &   5.3 &   5.0 &  -0.02$\pm$0.14   &   0.16 &   0.02 &   0.15\\
  UGC06604 &   6.9$\pm$5.9 &   6.8 &   7.3 &   7.5 &  -0.10$\pm$0.31   &  -0.10 &  -0.13 &  -0.12\\
   UGC0887 &  18.2$\pm$0.7 &  10.2 &  \nodata &  16.8 &  -0.33$\pm$0.31   &  -0.08 &  \nodata &  -0.16\\
 UGC12454  &  \nodata &  12.7 &   4.4 &   5.2 & \nodata   &  -0.21 &  -0.12 &  -0.08\\
 UGC12472 &   0.7$\pm$0.0 &   1.1 &   0.5 &   1.0 &   0.42$\pm$0.14   &  -0.22 &   0.10 &  -0.20\\
   UGC9519 &   1.8$\pm$1.1 &   1.8 &   0.9 &   1.4 &   0.07$\pm$0.47   &  -0.03 &   0.35 &   0.07\\
   VCC049 &   4.8$\pm$5.3 &   3.5 &   2.1 &   2.4 &   0.10$\pm$0.34   &   0.27 &   0.43 &   0.29\\
    VCC140 &   3.4$\pm$2.0 &   3.5 &   2.7 &   2.8 &  -0.26$\pm$0.22   &  -0.27 &  -0.16 &  -0.24\\
    VCC218 &   1.2$\pm$0.4 &   2.1 &   1.4 &   2.0 &   0.15$\pm$0.12   &  -0.70 &   0.08 &  -0.69\\
    VCC389 &   3.7$\pm$2.9 &   3.1 &   2.0 &   2.2 &  -0.24$\pm$0.33   &  -0.12 &   0.03 &  -0.02\\
    VCC523 &   1.9$\pm$0.4 &   1.8 &   1.8 &   1.9 &   0.02$\pm$0.25   &   0.05 &   0.04 &   0.04\\
    VCC538 &   5.6$\pm$0.7 &   3.2 &   1.9 &   1.9 &  -0.62$\pm$0.25   &  -0.29 &  -0.22 &  -0.18\\
    VCC634 &   2.7$\pm$0.0 &   4.8 &   0.9 &   1.7 &  -0.20$\pm$0.13   &  -0.66 &   0.40 &  -0.42\\
    VCC685 &  10.4$\pm$2.2 &   5.9 &  \nodata &  17.4 &   0.00$\pm$0.10   &   0.24 &  \nodata &   0.08\\
    VCC698 &   4.3$\pm$4.2 &   3.5 &   3.0 &   3.1 &   0.00$\pm$0.36   &   0.10 &   0.09 &   0.10\\
    VCC751 &   2.4$\pm$1.3 &   3.2 &   3.0 &   3.0 &   0.16$\pm$0.35   &  -0.10 &   0.07 &  -0.11\\
    VCC758 &   7.9$\pm$0.0 &   3.7 &   2.4 &   2.6 &  -0.18$\pm$0.12   &   0.21 &   0.21 &   0.25\\
   VCC781 &   1.6$\pm$0.1 &   1.0 &   2.0 &   1.3 &  -0.80$\pm$0.25   &  -0.01 &  -1.00 &   0.01\\
   VCC784 &  \nodata &  15.3 &  10.5 &  13.8 & \nodata   &  -0.30 &  -0.28 &  -0.29\\
    VCC828 &  15.4$\pm$1.7 &   8.0 &  11.7 &  11.9 &  -0.35$\pm$0.04   &  -0.04 &  -0.30 &  -0.13\\
    VCC856 &   3.3$\pm$0.4 &   5.2 &  \nodata &  \nodata &  -0.43$\pm$0.12   &  -0.85 &  \nodata &  \nodata\\
    VCC929 &   6.0$\pm$13.7 &   9.6 &  \nodata &  \nodata &  -0.25$\pm$0.65   &  -0.52 &  \nodata &  \nodata\\
    VCC944 &   9.7$\pm$3.9 &   8.4 &  13.4 &  13.2 &  -0.08$\pm$0.14   &   0.02 &  -0.16 &  -0.07\\
    VCC966 &   2.8$\pm$0.5 &   3.0 &   2.1 &   2.4 &   0.30$\pm$0.12   &   0.24 &   0.43 &   0.26\\
   VCC1010 &   8.0$\pm$0.2 &   5.3 &   5.4 &   5.3 &  -0.40$\pm$0.34   &  -0.14 &  -0.28 &  -0.14\\
   VCC1025 &   6.8$\pm$3.6 &   6.5 &   9.4 &   9.6 &  -0.07$\pm$0.19   &  -0.03 &  -0.15 &  -0.09\\
   VCC1030 &   4.3$\pm$1.5 &   4.1 &   8.2 &   7.9 &   0.09$\pm$0.15   &   0.12 &  -0.07 &   0.05\\
   VCC1036 &   4.6$\pm$6.3 &   3.7 &   7.2 &   6.6 &  -0.34$\pm$0.39   &  -0.21 &  -0.46 &  -0.26\\
   VCC1062 &  12.3$\pm$2.7 &   7.4 &  \nodata &  \nodata &  -0.05$\pm$0.07   &   0.18 &  \nodata &  \nodata\\
   VCC1073 &   4.2$\pm$0.0 &   2.8 &   3.0 &   3.0 &  -0.06$\pm$0.07   &   0.19 &   0.05 &   0.18\\
   VCC1125 &   5.0$\pm$7.0 &   6.2 &  10.2 &  10.0 &   0.02$\pm$0.45   &  -0.09 &  -0.16 &  -0.16\\
   VCC1146 &  13.0$\pm$6.5 &   8.8 &  17.2 &  15.1 &  -0.34$\pm$0.16   &  -0.18 &  -0.39 &  -0.25\\
   VCC1178 &  17.2$\pm$0.9 &  13.8 &  \nodata &  \nodata &  -0.33$\pm$0.16   &  -0.22 &  \nodata &  \nodata\\
   VCC1183 &   4.8$\pm$12.2 &   3.1 &   2.2 &   2.2 &  -0.44$\pm$0.59   &  -0.17 &  -0.17 &  -0.08\\
  VCC1199  &  \nodata &  \nodata &   5.1 &   6.8 & \nodata   &  \nodata &   0.25 &   0.13\\
   VCC1231 &   9.4$\pm$1.3 &   5.7 &  12.1 &  11.3 &   0.00$\pm$0.06   &   0.21 &  -0.07 &   0.10\\
   VCC1242 &   6.5$\pm$4.2 &   7.1 &   3.9 &   4.7 &  -0.04$\pm$0.26   &  -0.09 &   0.10 &  -0.05\\
   VCC1250 &   2.2$\pm$0.4 &   2.1 &   1.9 &   1.9 &  -0.06$\pm$0.17   &  -0.03 &   0.03 &  -0.01\\
   VCC1279 &   5.9$\pm$1.1 &   5.0 &   5.2 &   6.2 &   0.02$\pm$0.06   &   0.06 &   0.05 &   0.05\\
   VCC1283 &   5.8$\pm$8.9 &   7.5 &  \nodata &  \nodata &   0.01$\pm$0.54   &  -0.17 &  \nodata &  \nodata\\
   VCC1297 &  15.1$\pm$6.2 &  10.5 &  \nodata &  \nodata &  -0.20$\pm$0.52   &  -0.01 &  \nodata &  \nodata\\
   VCC1303 &   7.5$\pm$3.4 &   4.3 &   9.7 &   8.3 &  -0.09$\pm$0.18   &   0.17 &  -0.16 &   0.11\\
   VCC1318 &   1.9$\pm$0.7 &   1.7 &   1.6 &   1.6 &   0.01$\pm$0.37   &   0.14 &   0.09 &   0.15\\
   VCC1321 &   5.0$\pm$2.1 &   3.4 &   2.9 &   3.0 &  -0.14$\pm$0.18   &   0.10 &   0.04 &   0.10\\
   VCC1422 &   4.2$\pm$4.5 &   3.3 &   2.9 &   3.1 &  -0.41$\pm$0.44   &  -0.26 &  -0.31 &  -0.27\\
   VCC1440 &   3.9$\pm$5.2 &   4.8 &  \nodata &  \nodata &  -0.45$\pm$0.41   &  -0.59 &  \nodata &  \nodata\\
   VCC1475 &   8.2$\pm$3.7 &   7.3 &   6.5 &   7.0 &  -0.27$\pm$0.12   &  -0.21 &  -0.21 &  -0.21\\
   VCC1479 &   4.0$\pm$4.7 &   4.3 &   1.2 &   1.6 &  -0.03$\pm$0.54   &  -0.09 &   0.60 &   0.09\\
   VCC1488 &   1.2$\pm$0.0 &   0.9 &   1.5 &   1.3 &  -0.47$\pm$0.54   &  -0.20 &  -0.62 &  -0.24\\
   VCC1521 &   4.9$\pm$3.6 &   5.1 &   8.5 &   8.1 &  -0.10$\pm$0.27   &  -0.11 &  -0.24 &  -0.17\\
   VCC1537 &   6.3$\pm$1.5 &   6.8 &   7.0 &   6.7 &  -0.05$\pm$0.12   &  -0.10 &  -0.08 &  -0.09\\
   VCC1545 &   2.4$\pm$10.7 &   3.5 &  10.6 &  12.7 &   0.12$\pm$0.12   &  -0.25 &  -0.33 &  -0.37\\
   VCC1614 &   1.9$\pm$0.8 &   1.8 &   1.8 &   1.8 &  -0.46$\pm$0.43   &  -0.34 &  -0.40 &  -0.34\\
   VCC1630 &   3.1$\pm$1.4 &   6.4 &  11.4 &  14.5 &   0.29$\pm$0.20   &  -0.05 &  -0.07 &  -0.18\\
   VCC1827 &   2.4$\pm$1.1 &   2.3 &   1.9 &   2.0 &   0.05$\pm$0.33   &   0.08 &   0.13 &   0.08\\
   VCC1871 &   5.7$\pm$7.7 &   7.4 &   3.5 &   4.0 &   0.06$\pm$0.39   &  -0.09 &   0.20 &  -0.02\\
   VCC1903 &  17.3$\pm$1.1 &   6.0 &  16.3 &  12.4 &  -0.23$\pm$0.39   &   0.26 &  -0.21 &   0.16\\
   VCC1912 &   1.1$\pm$0.0 &   1.5 &   1.7 &   1.7 &   0.00$\pm$0.02   &  -0.34 &  -0.25 &  -0.40\\
   VCC1938 &   6.3$\pm$2.4 &   6.5 &   3.8 &   4.1 &   0.15$\pm$0.12   &   0.13 &   0.29 &   0.19\\
   VCC1939 &  10.3$\pm$4.7 &   6.4 &  \nodata &  \nodata &  -0.04$\pm$0.19   &   0.18 &  \nodata &  \nodata\\
   VCC2000 &  15.3$\pm$2.5 &  11.2 &  \nodata &  \nodata &  -0.14$\pm$0.06   &   0.01 &  \nodata &  \nodata\\
   VCC2048 &   3.0$\pm$3.2 &   4.7 &   3.0 &   3.7 &  -0.25$\pm$0.46   &  -0.56 &  -0.26 &  -0.53\\

\enddata
\tablenotetext{a}
{
column (2): age and error in Gyr determined from Fe4383 $vs$ Hn/Fe diagram\\
column (3): age determined from Fe5270 $vs$ Hn/Fe\\
column (4): age determined from Fe4383 $vs$ H$\beta$\\
column (5): age determined from Fe5270 $vs$ H$\beta$\\
column (6): [Fe/H] and error determined from Fe4383 $vs$ Hn/Fe\\
column (7): [M/H] determined from Fe5270 $vs$ Hn/Fe\\
column (8): [Fe/H] determined from Fe4383 $vs$ H$\beta$\\
column (9): [M/H] determined from Fe5270 $vs$ H$\beta$
}
\label{agemet_table}
\end{deluxetable}

\begin{deluxetable}{ccrrcrr}
\tablenum{9}
\tablecolumns{7}
\tablewidth{0pc}
\tablecaption{Mean Age and [Fe/H] for Different $\sigma$ Groups}
\tablehead{ \colhead{log $\sigma$ Group} &
\colhead{Index Pair} & \colhead{$<$Age$>$} & \colhead{$<$[Fe/H]$>$} &
\colhead{Index Pair} & \colhead{$<$Age$>$} & \colhead{$<$[Fe/H]$>$} }
\startdata
log $\sigma$$<$1.8 & $Fe4383$ vs $H\beta$ & 2.4 & -0.02 & $Fe5270$ vs $H\beta$ & 2.3 & -0.22\nl
1.8$\le$log $\sigma$ $<$ 2.0 & \nodata & 2.3 & 0.14 & \nodata & 2.2 & -0.12\nl
2.0$\le$log $\sigma$ $<$ 2.2 & \nodata & 6.1 &  0.06 & \nodata & 5.0 & -0.13\nl
2.2$\le$log $\sigma$ & \nodata & 13.5 & -0.03 & \nodata & 12.8 & -0.14\nl
\multicolumn{3}{c}{ }\nl
log $\sigma$$<$1.8 & $Fe4383$ vs $Hn/Fe$ & 3.0 & -0.28 & $Fe5270$ vs $Hn/Fe$ & 3.0 & -0.28\nl
1.8$\le$log $\sigma$ $<$ 2.0 & \nodata & 2.8 & -0.06 & \nodata & 3.0 & -0.20\nl
2.0$\le$log $\sigma$ $<$ 2.2 & \nodata & 5.8 & -0.07 & \nodata & 6.9 & -0.17\nl
2.2$\le$log $\sigma$ & \nodata & 8.8 & -0.06 & \nodata & 9.4 & -0.08\nl
\multicolumn{3}{c}{ }\nl
log $\sigma$$<$1.8 & $Fe4383$ vs $H\gamma_F$ & 2.6 & -0.21 & $Fe5270$ vs $H\gamma_F$ &  2.7 & -0.26\nl
1.8$\le$log $\sigma$ $<$ 2.0 & \nodata & 2.0 & 0.08 & \nodata & 2.5 & -0.15\nl
2.0$\le$log $\sigma$ $<$ 2.2 & \nodata & 5.2 & -0.04 & \nodata & 6.2 & -0.15\nl
2.2$\le$log $\sigma$ & \nodata & 10.6 & -0.11 & \nodata & 10.6 & -0.11\nl
\multicolumn{3}{c}{ }\nl
log $\sigma$$<$1.8 & $Fe4383$ vs $H\gamma/4325$ & 3.1 & -0.30 & $Fe5270$ vs $H\gamma/4325$ & 3.1 & -0.29\nl
1.8$\le$log $\sigma$ $<$ 2.0 & \nodata & 2.6 & -0.02 & \nodata & 2.9 & -0.19\nl
2.0$\le$log $\sigma$ $<$ 2.2 & \nodata & 6.6 & -0.11 & \nodata &  7.2 & -0.18\nl
2.2$\le$log $\sigma$ & \nodata & 9.3 & -0.41 & \nodata & 10.5 & -0.11\nl
\multicolumn{3}{c}{ }\nl
log $\sigma$$<$1.8 & $Fe4383$ vs $H\delta_F$ & 2.0 & -0.06 & $Fe5270$ vs $H\delta_F$ & 2.3 & -0.22\nl
1.8$\le$log $\sigma$ $<$ 2.0 & \nodata & 1.7 & 0.17 & \nodata & 2.0 & -0.11\nl
2.0$\le$log $\sigma$ $<$ 2.2 & \nodata & 2.5 &  0.20 & \nodata & 3.1 & -0.11\nl
2.2$\le$log $\sigma$ & \nodata &  5.4 &  0.07 & \nodata & 6.6 & -0.03\nl
\multicolumn{3}{c}{ }\nl
log $\sigma$$<$1.8 & $Fe4383$ vs $H\delta/4045$ & 2.6 & -0.19 & $Fe5270$ vs $H\delta/4045$ & 2.7 & -0.25\nl
1.8$\le$log $\sigma$ $<$ 2.0 & \nodata & 2.0 &  0.08 & \nodata & 2.4 & -0.14\nl
2.0$\le$log $\sigma$ $<$ 2.2 & \nodata & 3.1 &  0.09 & \nodata & 4.1 & -0.12\nl
2.2$\le$log $\sigma$ & \nodata & 4.1 &  0.14 & \nodata & 5.2 & 0.00\nl
\multicolumn{3}{c}{ }\nl
log $\sigma$$<$1.8 & $Fe4383$ vs $H8/3859$ & 3.4 & -0.32 & $Fe5270$ vs $H8/3859$ & 3.2 & -0.28\nl
1.8$\le$log $\sigma$ $<$ 2.0 & \nodata & 5.7 &  0.37 & \nodata & 4.0 & -0.22\nl
2.0$\le$log $\sigma$ $<$ 2.2 & \nodata & 8.8 &  0.39 & \nodata & 9.2 & -0.21\nl
2.2$\le$log $\sigma$ & \nodata & 13.8 & -0.17 & \nodata & 12.6 & -0.13\nl
\enddata
\label{sigs_table}
\end{deluxetable}

\begin{deluxetable}{llcccc}
\tablenum{10}
\tablecaption{A Comparison with Previous Ages and Metallicities}
\footnotesize
\tablehead{\colhead{Galaxy} & \colhead{VCC Name} & \colhead{Age$_{Lit}$ (Gyr)} &
\colhead{[Z/H]$_{Lit}$} & \colhead {Age$_{Meas}$} & \colhead{[Fe/H]$_{Meas}$} }
\startdata
\sidehead{\hskip 5cm Trager et al.\ (2000a)}
NGC 221 &  M32  &  3.0  & 0.0  &  2.6  & 0.28 \\
NGC 636 &       &  4.1  & 0.34  &  8.8  & -0.08 \\
NGC 821 &       &  7.7  & 0.22  & 11.5  & -0.05 \\
NGC 2778 &      &  5.4  & 0.29  & 10.5  & -0.19 \\
NGC 3818 &      &  5.6  & 0.36  &  8.8  & -0.01 \\
NGC 4478 & VCC1279 & 4.6 & 0.29 & 5.86 & 0.02 \\
NGC 4489 & VCC1321 & 2.5 & 0.14 & 4.9  &-0.14 \\
NGC 5831 &      &  2.6  & 0.54 &  3.0 & 0.35\\
NGC 7454 &      & 5.0  & -0.06 & 4.9 & -0.03 \\
\sidehead{\hskip 5cm Kuntschner et al.\ (2001)}
NGC 2778 &      & 12.0 & 0.12 & 10.5 & -0.19 \\
NGC 3489 &      &  1.1 & 0.22 &  1.8 &  0.19 \\
NGC 4434 & VCC1025 & 12.3 & -0.24 & 6.8 & -0.06 \\
NGC 4458 & VCC1146 &  4.6 &  0.03 & 13.0 & -0.33 \\
NGC 4473 & VCC1231 & 12.8 &  0.16 & 9.3 & 0.00 \\
NGC 4621 & VCC1903 & 10.9 &  0.34 & 17.2 & -0.22 \\
NGC 4660 & VCC2000 & 21.2 & -0.07 & 15.3 & -0.14 \\
NGC 5582 &      &  13.3 & -0.15 & 10.4 & -0.28 \\
NGC 5831 &      &  17.5 & -0.07 & 3.0 & 0.35 \\
\tablebreak
\sidehead{\hskip 5cm Vazdekis et al.\ (2001b)}
NGC 4464 & VCC1178 & 19 & -0.27 & 17.2 & -0.32 \\
NGC 4387 & VCC 828 & 16 & -0.15 & 15.4 & -0.35 \\
NGC 4473 & VCC1231 & 12 &  0.14 & 9.3 & 0.00 \\
NGC 4478 & VCC1279 & 9.3 & 0.08 & 5.8 & 0.02 \\
\enddata
\tablecomments{The Trager et al. (2000a) ages are taken from 2000,AJ,119,1645,
Table 6A using the values for Model 4.  The metallicities are their
[Z/H].  The Kuntschner et al.(2001) values are from private communication of ages
\& metallicities derived from the data in that paper.
The Vazdekis et al.\ determinations were interpolated from a plot in 2001,ApJ,55
1,L127}
\label{tab:lit}
\end{deluxetable}

\begin{deluxetable}{lcc}
\tablenum{11}
\tablecaption{Spectral Index Emission Correction Relations}
\tablewidth{0pt}
\tablehead{\colhead{Index} &
\colhead{Slope} & \colhead{Zero-Point} }
\startdata
H$\delta$/4045  & -2.97E+12 &  0.266E-3 \nl
H$\gamma$/4325  & -5.03E+12 &  0.333E-6 \nl
H$_{8}$/3859    & -1.35E+12 & -0.233E-3 \nl
Hn/Fe           & -3.11E+12 & -0.144E-3 \nl
H$\beta$        &  8.59E+13 &  0.666E-6 \nl
\enddata
\tablecomments{The parameters of the line index corrections as a function of the
H$\beta$ emission flux removed.}
\label{tab:e-lines}
\end{deluxetable}


\clearpage

\begin{thebibliography}{}
\bibitem[Akritas \& Bershady (1996)]{ab96} Akritas, M. G., \& Bershady, M. A. 
   1996, \apj, 470, 706
\bibitem[Baldwin, Phillips, \& Terlevich (1981)]{bpt81} Baldwin, J. A., 
   Phillips, M. M., \& Terlevich, R. 1981, \pasp, 93, 5
\bibitem[Benson \etal (2001)]{be01} Benson, A. J., Pearce, F. R., Frenk, C. S.,
   Baugh, C. M., \& Jenkins, A. 2001, \mnras, 320, 261
\bibitem[Bertelli \etal (1994)]{padova} Bertelli, G., Bressan, A., Chiosi, C.,
   Fagotto, F., \& Nasi, E. 1994, \aaps, 106, 275
\bibitem[Bessell \etal (1989, 1991)]{bessell} Bessell, M. S., Brett, J. M., 
   Wood, P. R., and Scholtz, M. 1989, \aap, 213, 209
\bibitem[Bessell \etal (1989, 1991)]{bess91} Bessell, M. S., Brett, J. M.,
   Wood, P. R., and Scholtz, M. 1991, \aaps, 89, 335
\bibitem[Brown \& Wallerstein (1992)]{bw92} Brown, J. A., \& Wallerstein, G.
   1992, \aj, 104, 1818
\bibitem[Burstein \etal (1984)]{bu84} Burstein, D., Faber, S. M., Gaskell, C.
   M., \& Krumm, N. 1984, \apj, 287, 586
\bibitem[Buson \etal (1993)]{bus93} Buson, L. M., Sadler, E. M.,
 Zeilinger, W. W., Bertin, G.,
 Bertola, F., Danzinger, J.,
 Dejonghe, H., Saglia, R. P., \&
 de Zeeuw, P. T. 1993, \aap, 280, 409
\bibitem[Caldwell (1984)]{cald84} Caldwell, N. 1984, \pasp, 96, 287
\bibitem[Carretta \& Gratton (1997)]{cg97} Carretta, E., \& Gratton, R. G. 1997,
   \aaps, 121, 95
\bibitem[Concannon, Rose, \& Caldwell (2000)]{crc00} Concannon, K. D., Rose, 
   J. A., \& Caldwell, N. 2000, \apj, 536, L19
\bibitem[Davis \etal (1985)]{da85} Davis. M., Efstathiou, G., Frenk, C. S., \&
   White, S. D. M. 1985, \apj, 292, 371
\bibitem[De Propris (2000)]{dep00} De Propris, R. 2000, \mnras, 316, L9
\bibitem[de Vaucouleurs \etal (1991)]{rc3} de Vaucouleurs, G., de Vaucouleurs,
   A., Corwin, H. G., Buta, R. J., Paturel, G., \& Fouque, P. 1991, ``Third
   Reference Catalogue of Bright Galaxies, Version 3.9''
\bibitem[Dopita \& Sutherland (1995)]{ds95} Dopita, M. A., \& Sutherland, R. S.
   1995, \apj, 455, 468
\bibitem[Faber \etal (1985)]{fab85} Faber, S. M., Friel, E. D., Burstein, D.,
   \& Gaskell, C. M. 1985, \apjs, 57, 711
\bibitem[Fabricant \etal (1998)]{fa98} Fabricant, D., Cheimets, P., Caldwell, 
   N., \& Geary, J. 1998, \pasp, 110, 79
\bibitem[Falco \etal (1999)]{fa99} Falco, E. E., Kurtz, M. J., Geller, M. J.,
   Huchra, J. P., Peters, J., Berlind, P., Mink, D. J., Tokarz, S. P., \&
   Elwell, B. 1999, \pasp, 111, 438
\bibitem[Ferland \& Netzer (1983)]{fn83} Ferland, G. J., \& Netzer, H. 1983, 
   \apj, 264, 105
\bibitem[Ferraro \etal (2001)]{fe01} Ferraro, F. R., D'Amico, N., Possenti, A.,
   Mignani, R. P., \& Paltrinieri, B. 2001, \apj, 561, 337
\bibitem[Gabel \& Bruhweiler (2002)]{gb02} Gabel, J. R., \& Bruhweiler, F. C. 
   2002, \aj, 124, 737
\bibitem[Geha \etal (2002)]{ge02} Geha, M., Guhathakurta, P., \& van der Marel, R. P.
   2002, \aj, 124, 3073
\bibitem[Gibson \etal (1999)]{gi99} Gibson, B. K., Madgwick, D. S., Jones, L.
   A., Da Costa, G. S., \& Norris, J. E. 1999, \aj, 118, 1268
\bibitem[Grundahl, Stetson, \& Andersen (2002)]{gsa02} Grundahl, F., Stetson, 
   P. B., \& Andersen, M. L. 2002, \aap, 395, 481
\bibitem[Gunn \& Stryker (1983)]{gs83} Gunn, J. E., \& Stryker, L. L. 1983,
   \apjs, 52, 121
\bibitem[Howell, Guhathakurta, \& Gilliland (2000)]{hgg00} Howell, J. H.,
   Guhathakurta, P., \& Gilliland, R. L. 2000, \pasp, 112, 1200
\bibitem[Huchra \etal (1983)]{hu83} Huchra, J. P., Davis, M., Latham, D., \&
   Tonry, J. 1983, \apjs, 52, 89
\bibitem[Jones (1999)]{jo99} Jones, L. A. 1999, Ph.D. Thesis, University of 
   North Carolina
\bibitem[Jones \& Worthey (1995)]{jw95} Jones, L. A., \& Worthey, G.
   1995, \apj, 446, L31
\bibitem[Jorgensen (1999)]{jor99} Jorgensen, I. 1999, \mnras, 306, 607
\bibitem[Kauffmann (1996)]{ka96} Kauffmann, G. 1996, \mnras, 281, 487
\bibitem[Kauffmann \& Charlot (1998)]{ks98}Kauffmann, G., \& Charlot, S. 
   1998, \mnras, 294, 705
\bibitem[Kauffmann \etal (1999a,b)]{ks99a} Kauffmann, G., Colberg, J., Diaferio,
   A. \& White, S. D. M. 1999a, \mnras, 303, 188
\bibitem[Kauffmann \etal (1999b)]{ka99b} Kauffmann, G., Colberg, J., Diaferio, 
   A. \& White, S. D. M. 1999b, \mnras, 307, 529
\bibitem[Kjaergaard (1987)]{kj87} Kjaergaard, P. 1987, \aap, 176, 210
\bibitem[Kuntschner (2000)]{ku00} Kuntschner, H. 2000, \mnras, 315, 184
\bibitem[Kuntschner \etal (2001)]{ku01} Kuntschner, H., Lucey, J. R., Smith,
   R. J., Hudson, M. J., \& Davies, R. L. 2001, \mnras, 323, 615
\bibitem[Kuntschner \etal (2002)]{ku02} Kuntschner, H., Smith, R. J., Colless, 
   M., Davies, R. L., Kaldare, R., \& Vazdekis, A. 2002, \mnras, 337, 172
\bibitem[Kurucz (1993, 1994)]{kur93} Kurucz, R.L. 1993, SYNTHE Spectrum 
   Synthesis Programs and Line Data (Kurucz CD-ROM No 18)
\bibitem[Kurucz (1994)]{ku94} Kurucz, R.L. 1994, Solar Abundances Model 
   Atmospheres for 0,1,2,4,8 km s-1 , (Kurucz CD-ROM No19)
\bibitem[Leitherer \etal (1996)]{cou96} Leitherer, C., \etal 1996, \pasp, 108, 
   996
\bibitem[Leonardi \& Rose (1996)]{lr96} Leonardi, A. J., \& Rose, J. A. 1996, 
   \aj, 111, 182
\bibitem[Leonardi \& Rose (2002)]{lr02} Leonardi, A. J., \& Rose, J. A. 2002,
   \aj, (submitted)
\bibitem[Leonardi \& Worthey (2000)]{lw00} Leonardi, A. J., \& Worthey, G. 2000,
   \apj, 534, 650
\bibitem[Liu \& Chaboyer (2000)]{lc00} Liu, W. M., \& Chaboyer, B. 2000, \apj, 
   544, 818
\bibitem[Maraston, \& Thomas (2000)]{mt00} Maraston, C., \& Thomas, D. 2000, 
   \apj, 541, 126
\bibitem[McElroy (1995)]{mce95} McElroy, D. B. 1995, \apjs, 100, 105
\bibitem[O'Connell (1976)]{oc76} O'Connell, R. W. 1976, \apj, 206, 370
\bibitem[Osterbrock (1989)]{ost89} Osterbrock, D. E. 1989, Astrophysics of 
   Gaseous Nebulae and Active Galactic Nuclei, (Mill Valley, CA: University 
   Science Books)
\bibitem[Pedraz \etal (2002)]{pe02} Pedraz, S., Gorgas, J., Cardiel, N., 
   Sanchez-Blazquez, P., \& Guzman, R. 2002, \mnras, 332, L59
\bibitem[Peletier (1989)]{pe89} Peletier, R. F. 1989, Ph.D. Thesis, University
   of Groningen
\bibitem[Phillips \etal (1986)]{ph86} Phillips, M. M., Jenkins, C. R., Dopita,
   M. A., Sadler, E. M., \& Binette, L. 1986, \aj, 91, 1062
\bibitem[Proctor \& Sansom (2002)]{ps02} Proctor, R. N., \& Sansom, A. E. 2002,
   \mnras, 333, 517
\bibitem[Rose (1984)]{ro84} Rose, J. A. 1984, \aj, 89, 1238
\bibitem[Rose (1985)]{ro85} Rose, J. A. 1985, \aj, 90, 1927
\bibitem[Rose (1994)]{ro94} Rose, J. A. 1994, \aj, 107, 206
\bibitem[Rose \& Deng (1999)]{rd99} Rose, J. A., \& Deng, S. 1999, \aj, 117, 
   2213
\bibitem[Salaris \& Weiss (1998)]{sw98} Salaris, M., \& Weiss, A. 1998, \aap, 
   335, 943
\bibitem[Salasnich \etal (2000)]{sa00} Salasnich, B., Girardi, L., Weiss, A., \&
   Chiosi, C. 2000, \aap, 361, 1023
\bibitem[Simien, F., \& Prugniel, Ph. (2002)]{sph02} Simien, F., \& Prugniel, Ph.  
  2002, \aap, 384, 371
\bibitem[Smith \etal (2000)]{sm00} Smith, Russell J., Lucey, John R., 
Hudson, Michael J., Schlegel, David J., \& Davies, Roger L. 2000, \mnras, 313. 469
\bibitem[Somerville \& Primack (1999)]{sp99} Somerville, R. S., \& Primack, J. 
   1999, \mnras, 310, 1087
\bibitem[Terlevich \& Forbes (2002)]{tf02} Terlevich, A. I., \& Forbes, D. A. 
   2002, \mnras, 330, 547
\bibitem[Terlevich \& Melnick (1985)]{tm85} Terlevich, R., \& Melnick, J. 1985,
   \mnras, 213, 841
\bibitem[Trager \etal (2000a)]{tr00a} Trager, S., Faber, S., Worthey, G., \&
   Gonzalez, J. 2000a, \aj, 119, 1645
\bibitem[Trager \etal (2000b)]{tr00b} Trager, S., Faber, S., Worthey, G., \&
   Gonzalez, J. 2000b, \aj, 120, 165
\bibitem[Tripicco (1989)]{tr89} Tripicco, M. J. 1989, \aj,97, 735
\bibitem[Tripicco \& Bell (1995)]{tb95} Tripicco, M. J., \& Bell, R. A. 1995,
   \aj, 110, 3035
\bibitem[VandenBerg \etal (2000)]{vdb00} VandenBerg, D. A., Swenson, F. J., 
   Rogers, F. J., Iglesias, C. A., \& Alexander, D. R. 2000, \apj, 532, 430
\bibitem[Vazdekis \etal (1997)]{va97} Vazdekis, A., Peletier, R. F., Beckman,
   J. E., \& Casuso, E. 1997, \apjs, 111, 203
\bibitem[Vazdekis \& Arimoto (1999)]{va99} Vazdekis, A., \& Arimoto, N. 1999, 
   \apj, 525, 144
\bibitem[Vazdekis \etal (2001a)]{va01a} Vazdekis, A., Salaris, M., Arimoto, N.,
   \& Rose, J. A. 2001a, \apj, 549, 274
\bibitem[Vazdekis \etal (2001b)]{va01b} Vazdekis, A., Kuntschner, H., Davies, R. 
   L., Arimoto, N., Nakamura, O., \& Peletier, R. 2001b, \apj, 551, L127
\bibitem[Worthey (1994)]{wo94} Worthey, G. 1994, \apjs, 95, 107
\bibitem[Worthey \etal (1992)]{wo92} Worthey, G., Faber, S. M., \& Gonzalez, J.
   J. 1992, \apj, 398, 69
\bibitem[Worthey \etal (1994)]{wet94} Worthey, G., Faber, S. M., Gonzalez, J. 
   J., \& Burstein, D. 1994, \apjs, 94, 687
\bibitem[Worthey \& Ottaviani (1997)]{worott} Worthey, G., \& Ottaviani, D. L. 
   1997, \apjs, 111, 377
\bibitem[Zoccali \etal (2001)]{zo01} Zoccali, M., \etal 2001, \apj, 553, 733
   
\end{thebibliography}
\end{document}